\documentclass[apjfonts, numberedappendix]{emulateapj}

\newcommand{\Ang}{~\mbox{\AA}}

\newcommand{\invHz}{~\mbox{Hz}^{-1}}
\newcommand{\erg}{~\mbox{erg}}
\newcommand{\s}{~\mbox{s}}

\newcommand{\sri}{~\mbox{sr}^{-1}}
\newcommand{\Myr}{~\mbox{Myr}}

\newcommand{\eV}{~\mbox{eV}}

\newcommand{\cMpch}{~h^{-1}~\mbox{Mpc}~\mbox{comoving}}
\newcommand{\cMpc}{~\mbox{Mpc}~\mbox{comoving}}
\newcommand{\Mpch}{~h^{-1}~\mbox{Mpc}}

\newcommand{\pc}{~\mbox{pc}}
\newcommand{\ckpch}{~h^{-1}~\mbox{kpc}~\mbox{comoving}}
\newcommand{\ckpc}{~\mbox{kpc}~\mbox{comoving}}
\newcommand{\cmsq}{~\mbox{cm}^{2}}

\newcommand{\cmsqi}{~\mbox{cm}^{-2}}
\newcommand{\cmci}{~\mbox{cm}^{-3}}

\newcommand{\kpc}{~\mbox{kpc}}
\newcommand{\K}{~\mbox{K}}

\newcommand{\traphic}{{\sc traphic}}

\newcommand{\yri}{~\mbox{yr}^{-1}}

\newcommand{\invs}{~\mbox{s}^{-1}}

\newcommand{\Msun}{~\mbox{M}_{\odot}}
\newcommand{\Msunyri}{~\mbox{M}_{\odot}~\mbox{yr}^{-1}}
\newcommand{\Msuni}{~\mbox{M}_{\odot}^{-1}}
\newcommand{\Zsun}{~Z_{\odot}}
\newcommand{\cms}{~\mbox{cm}~\mbox{s}^{-1}}

\newcommand{\kms}{~\mbox{km s}^{-1}}

\newcommand{\cmbfast}{{\sc cmbfast}}

\newcommand{\Msunpcinvsq}{~\mbox{M}_{\odot}~\mbox{pc}^{-2}}

\slugcomment{Submitted to ApJ}

\begin{document}

\shorttitle{The First Galaxies}
\shortauthors{PAWLIK ET AL.}

\title{The First Galaxies: Assembly under radiative feedback from the first stars}

\author{Andreas H. Pawlik\altaffilmark{1,2}, Milo\v s Milosavljevi\'c\altaffilmark{1}, and Volker Bromm\altaffilmark{1}}

\altaffiltext{1}{Department of Astronomy and Texas Cosmology Center, The University of Texas at Austin, TX 78712}
\altaffiltext{2}{Max Planck Institute for Astrophysics, Karl-Schwarzschild-Str. 1, 85748 Garching, Germany}

\begin{abstract}
We investigate how radiative feedback from the first stars affects the
assembly of the first dwarf galaxies. To this end we perform
cosmological zoomed smoothed particle hydrodynamics simulations of a
dwarf galaxy assembling inside a halo reaching a virial mass $\sim
10^9 \Msun$ at $z = 10$. The simulations follow the non-equilibrium
chemistry and cooling of primordial gas and the subsequent conversion
of the cool dense gas into massive metal-free stars. To quantify the
radiative feedback, we compare a simulation in which stars emit both
molecular hydrogen dissociating and hydrogen/helium ionizing radiation
with a simulation in which stars emit only molecular hydrogen
dissociating radiation, and further with a simulation in which stars
remain dark. Photodissociation and photoionization exert a strong
negative feedback on the assembly of the galaxy inside the main
minihalo progenitor. Gas condensation is strongly impeded, and star
formation is strongly suppressed in comparison with the simulation in
which stars remain dark. The feedback on the gas from either
dissociating or ionizing radiation implies a suppression of the
central dark matter densities in the minihalo progenitor by factors of
up to a few, which is a significant deviation from the singular
isothermal density profile characterizing the dark matter distribution
inside the virial radius in the absence of radiative feedback. The
evolution of gas densities, star formation rates, and the distribution
of dark matter becomes insensitive to the inclusion of dissociating
radiation in the late stages of the minihalo assembly, and it becomes
insensitive to the inclusion of ionizing radiation once the minihalo
turns into an atomically cooling galaxy. The formation of a
rotationally supported extended disk inside the dwarf galaxy is a
robust outcome of our simulations not affected by the inclusion of
radiation. Low-mass galaxies in the neighborhood of the dwarf galaxy 
show a large scatter in the baryon fraction which is driven by radiative 
feedback from sources both internal and external to these galaxies. Our 
estimates of the observability of the first galaxies
show that dwarf galaxies such as simulated here will be among the
faintest galaxies the upcoming {\it James Webb Space Telescope} will
detect. Our conclusions regarding the structure and observability of
the first galaxies are subject to our neglect of feedback from
supernovae and chemical enrichment as well as to statistical 
uncertainties implied by the limited number of galaxies in our simulations.

\end{abstract}

\keywords{cosmology: theory -- galaxies: formation -- galaxies: high-redshift -- stars: formation -- hydrodynamics -- radiative transfer}

\doublespace

\section{Introduction}
The birth of star-forming galaxies a few hundred million years
after the Big Bang marks an important milestone in the history of our universe. 
The spectacular images returned by the {\it Hubble Space Telescope} ({\it HST}) in the
past few years have already allowed us to probe into the first billion
year of the universe. The last few years have also seen the development of an observationally testable 
theory of the formation of the first galaxies. As ongoing and 
upcoming observations, such as with the {\it James Webb Space Telescope} ({\it JWST}), 
are about to push to ever earlier times, approaching the epoch of the very first galaxies, this theory is put to
ever more stringent tests (for a review see, e.g., \citealp{D:2012}).
\par
Both analytical arguments (e.g.,\citealp{Tegmark:1997};
\citealp{Naoz:2006}; \citealp{Tseliakhovich:2011}) and simulations
(e.g., \citealp{Abel:2002}; \citealp{Bromm:2002};
\citealp{Stacy:2011}) suggest that the first stars have formed at $z
\gtrsim 30$ inside dark matter minihalos with virial temperatures
$\gtrsim 10^3\K$, corresponding to halo masses $\sim 10^5-10^6
\Msun$ (for a review, see \citealp{Bromm:2004}). The metal-free primordial
gas inside these minihalos cools and condenses to reach the high
densities needed to form stars primarily through the radiative de-excitation
of rovibrationally excited molecular hydrogen (e.g.,
\citealp{Haiman:1996b}; \citealp{Abel:1997}). As the minihalos grow in
mass, their virial temperatures increase and, after 
reaching $\sim 10^4 \K$, become sufficiently large to collisionally
excite atomic hydrogen. This gives birth to the first atomically cooling galaxies
with typical masses $\sim 10^7 -10^8 \Msun$ at $z \gtrsim 15$
(e.g., \citealp{Oh:2002}; \citealp{Wise:2007b}; \citealp{Greif:2008};
for a review see \citealp{Bromm:2011}). The first 
atomically cooling galaxies then evolve into the first dwarf galaxies
with characteristic masses $\gtrsim 10^9 \Msun$ (e.g.,
\citealp{Mashchenko:2008}; \citealp{Wise:2012}). The aim of the current work is to investigate 
the assembly of such galaxies under the radiative feedback from the first stars. 
The mass-scale marked by the first dwarf galaxies is closely related to a 
number of key open issues, some of which are outlined below.  
\par 
\par
{\it Suppression of star formation by radiative feedback. } The
radiation emitted by the first stars has a profound effect on
subsequently forming stars and galaxies (for a comprehensive overview,
\citealp{Ciardi:2005}). Radiation in the Lyman-Werner (LW) bands
dissociates molecular hydrogen, the main coolant in 
minihalos (e.g., \citealp{Haiman:1997}). Hydrogen-ionizing radiation heats the
gas inside the first halos and the intergalactic medium (IGM). The
associated increase in pressure drives the gas outside halos with
virial temperatures $\lesssim 10^4 \K$, suppressing star formation in
both minihalos and the first atomic cooling halos (e.g.,
\citealp{Thoul:1996}; \citealp{Barkana:1999}). The
increased pressure in the IGM impedes the accretion of gas onto these
low-mass halos, an effect known as Jeans-filtering (e.g.,
\citealp{Shapiro:1994}; \citealp{Gnedin:1998};
\citealp{Okamoto:2008}). In contrast, star formation inside the first
dwarf galaxies should be more robust to this negative radiative
feedback as their deeper gravitational potentials allow them to hold
on to their gas more strongly.
\par
{\it The sources of reionization.} The first galaxies are thought to
have started the reionization of the universe, which is the
transformation of the cosmic hydrogen from its early neutral to its
present ionized state that occurred during the first billion year
after the Big Bang (for reviews see, e.g., \citealp{Barkana:2001};
\citealp{Furlanetto:2006}; \citealp{Meiksin:2009}). However, whether galaxies could sustain
reionization and drive it to completion is a question of significant
debate (e.g., \citealp{Bouwens:2012}; \citealp{Finkelstein:2012}).
The largest uncertainties are related to our poor knowledge of the
escape fraction, i.e., the fraction of ionizing photons that leave the
galaxies unabsorbed and are thus available to reionize the IGM, and of
the abundance and ionizing luminosities of low-mass galaxies too faint
to be detected in current surveys (e.g.,
\citealp{Robertson:2010}). The most recent determinations of the UV
luminosity density at $z \gtrsim 7$ suggest that a significant contribution
from faint, yet to be observed, low-mass galaxies is likely needed to sustain
reionization in the recombining gas (\citealp{Finkelstein:2012}; see \citealp{KuhlenFG:2012}
for a comprehensive discussion). Because of their increased robustness 
against stellar feedback, dwarf galaxies are among the low-mass galaxies
expected to be especially efficient sources of reionization 
(e.g.,\citealp{Choudhury:2007}; \citealp{Raicevic:2011}).
\par
{\it The origin of the Milky Way (MW) satellites.}  Dissipationless
simulations of dark matter subhalos around MW-like galaxies imply 
the existence of a large population of satellite galaxies only few of which are 
currently observed. A number of solutions have been offered to
explain this missing satellite problem (\citealp{Moore:1999};
\citealp{Klypin:1999}; for reviews see, e.g.,
\citealp{Bullock:2010}; \citealp{Ricotti:2010}; \citealp{Mayer:2010}),
including the suppression of star formation in low-mass halos by
stellar feedback and reionization (e.g., \citealp{Ricotti:2005};
\citealp{Salvadori:2009}; \citealp{Bovill:2009}; \citealp{Sawala:2012}; \citealp{Brooks:2012}), 
the transformation of low-mass halos by tidal forces and ram-pressure stripping upon
their entry in the MW virial region (e.g., \citealp{Mayer:2007}),
modifications of the cold dark matter structure formation paradigm
(e.g., \citealp{Lovell:2012}), and others (e.g., \citealp{Bovy:2012}; 
\citealp{Vera-Ciro:2012}; \citealp{Wang:2012}). However, current theories still struggle to
explain the abundance and properties of the observed MW satellites
across the luminosity range, from the recently discovered ultra-faint to
the long-known classical satellites (e.g, \citealp{Strigari:2008};
\citealp{Bovill:2011}; \citealp{Boylan:2011}). The most massive of
the simulated MW dark matter subhalos have progenitors with masses
$10^8-10^{10} \Msun$ at $z \gtrsim 6$ (e.g.,
\citealp{Boylan:2012}). This suggests an intimate relation
with the first dwarf galaxies, and renders investigations into these
objects a promising tool to understand the origin of structure in
galaxies such as the MW. 
\par
{\it The first disk galaxies.} Simulations of the first atomically
cooling galaxies, i.e., galaxies inside halos with masses $10^7-10^8\Msun$ at redshifts $z
\sim 15$, reveal a highly irregular morphology of the 
halo gas (e.g., \citealp{Wise:2007b}; \citealp{Greif:2008}; \citealp{Regan:2009}). On
the other hand, simulations of galaxies inside halos with larger
masses $\gtrsim 10^9 \Msun$ and at lower redshifts $z \lesssim 6-10$ often
find the halo gas organized in rotationally supported disks (e.g., \citealp{Mashchenko:2008};
\citealp{Pawlik:2011a}; \citealp{Romano:2011}; \citealp{Wise:2012}). These findings suggest a dwarf-size
mass scale $\sim 10^8-10^{10} \Msun $ for the transition to disk-like
morphologies, and the emergence of the first disk galaxies at $z
\gtrsim 6$. Physical processes to imprint such a scale include 
the turbulence generated by the cold inflow of gas along filaments 
which characterizes gas accretion by the first atomic cooling halos 
(e.g., \citealp{Wise:2007b}; \citealp{Wise:2008c}; \citealp{Greif:2008}), and stellar 
feedback (e.g., \citealp{Kaufmann:2007}). Whether the first
halos may host disks is an important open issue, affecting estimates
of, e.g., the escape of ionizing photons into the IGM (e.g.,
\citealp{Gnedin:2008}; \citealp{Conroy:2012}), or the ability of
massive black holes to accrete gas and grow (e.g.,
\citealp{Eisenstein:1995}; \citealp{Koushiappas:2004};
\citealp{Lodato:2006}; \citealp{Petri:2012}).
\par
{\it The faintest galaxies JWST will
see.} The faintest $z \gtrsim 6$ galaxies {\it HST} has so far revealed have
estimated stellar masses $\gtrsim 10^8 \Msun$ (\citealp{Labbe:2010}; \citealp{Finkelstein:2010}; \citealp{Curtis:2012}). Future observations
with upcoming telescopes such as the {\it JWST} will allow to 
search for galaxies down to still lower stellar masses and out to higher redshifts, thus promising to test
our theories of the formation of the first stars and
galaxies. However, most studies agree that even {\it JWST} will not be
sensitive enough to detect the stellar radiation emitted from inside
the minihalos and the first atomic cooling halos (e.g.,
\citealp{Haiman:1998}; \citealp{Oh:2001}; \citealp{Ricotti:2008}; 
\citealp{Johnson:2009}; \citealp{Zackrisson:2012}; \citealp{Rydberg:2012}). {\it JWST} may detect the stellar
radiation from some of these objects if they are gravitational lensed
(e.g., \citealp{Zackrisson:2012}). But most of the stellar light
collected by {\it JWST} from high redshifts is expected to come from 
dwarf galaxies more massive than the first atomically cooling galaxies 
(e.g., \citealp{Johnson:2009}; \citealp{Pawlik:2011a}).
\par
Motivated primarily by the exciting prospects for observations with
the upcoming {\it JWST}, we have previously presented cosmological simulations of a dwarf
galaxy assembling inside a halo reaching $10^9\Msun$ at $z = 10$
(\citealp{Pawlik:2011a}). The simulations were performed using the
Smoothed Particle Hydrodynamics (SPH) technique and achieved high
resolution by zooming in a select region around the
galaxy. Following the non-equilibrium chemistry and cooling of
primordial gas, the simulations tracked the evolution of the dwarf
galaxy starting from before its birth inside a minihalo. An intriguing
outcome was the formation of a rotationally supported extended disk
just prior to the final simulation redshift. However, 
our previous simulations did not account for star formation and the associated feedback. As explained above, 
stellar feedback has the potential to significantly affect the assembly of the gas 
inside low-mass halos.
\par
In this study we present a new set of simulations similar to our
previous simulations, but extending them by including star formation
and radiation. We focus on the radiative feedback from LW 
and ionizing radiation.\footnote{We will use the 
terms LW radiation and dissociating radiation interchangeably.} To judge the impact of radiative
feedback we will compare a simulation that includes both LW and
ionizing radiation with a simulation that includes only LW radiation
and further with a simulation in which no radiation is emitted. Note that the simulations 
do not account for supernova (SN) feedback or metal enrichment, 
a limitation we will discuss in Section~\ref{Sec:Discussion} below. The simulations
are designed primarily to address the impact of radiative feedback on the
assembly of the emerging dwarf galaxy, and on the formation of galactic disks
inside it. However, we will also briefly discuss the impact of
radiative feedback on the assembly of galaxies in the neighborhood of
the simulated dwarf galaxy. Our simulations enable us to provide an
improved estimate of the observability of the first galaxies with
{\it JWST}.
\par
The organization of this paper is as follows. In
Section~\ref{Sec:Numerics} (as well as in the appendix) we describe our numerical techniques, and
in Section~\ref{Sec:Simulations} we describe the set of simulations
that we have carried out. In
Sections~\ref{Sec:Formation} and \ref{Sec:Reionization} we present the results
of our simulations, subsequently discussing 
the assembly of the dwarf galaxy and the radiative feedback on the IGM and the neighboring galaxies. In Section~\ref{Sec:Prospects} we use 
the simulated star formation rates to estimate the observability of 
the first galaxies with {\it JWST}. 
In Section~\ref{Sec:Discussion} we discuss our results and also address some of the most important
limitations. In Section~\ref{Sec:Summary}, we summarize our work. 
\par
Throughout this work we assume the $\Lambda$CDM cosmological model with parameters $\Omega_{\rm{m}} =
0.258, \Omega_{\rm{b}} = 0.0441, \Omega_\Lambda = 0.742, \sigma_8 =
0.796, n_{\rm{s}} = 0.963$, and $h = 0.719$, which are consistent with
the most recent analysis of the observations with the {\it
Wilkinson Microwave Anisotropy Probe} satellite (\citealp{Komatsu:2011}). Distances are
expressed in physical (i.e., not comoving) units, unless noted
otherwise. We will make use of the species number density
fractions with respect to hydrogen $\eta_\alpha \equiv n_{\rm \alpha}
/ n_{\rm H}$, where $\alpha$ labels the chemical species.
\par
\section{Numerical Methods}
\label{Sec:Numerics}
In this section we describe the numerical techniques employed. The
simulations presented below are identical to the simulations described
in \cite{Pawlik:2011a}, except for the inclusion of star formation and 
dissociating and ionizing radiation. We will therefore
only briefly review the simulation techniques already used in
\cite{Pawlik:2011a}. We will focus on the description of the
techniques used to model the formation of stars and 
the dissociative and ionizing impact of stellar radiation on the gas. We remind the reader that the 
simulations do not account for SN feedback or 
metal enrichment, as discussed further in Section~\ref{Sec:Discussion}.
\par
\subsection{Gravity and Hydrodynamics}
\label{sec:gravity}
We use a modified version of the $N$-body/TreePM Smoothed Particle
Hydrodynamics (SPH) code {\sc gadget} (\citealp{Springel:2005};
\citealp{Springel:2001a}; \citealp{Schaye:2010}) to perform 
a suite of zoomed cosmological hydrodynamical simulations of 
the assembly of a halo that reaches a virial mass 
$M_{\rm vir} \approx 10^9 \Msun$ at redshift $z = 10$. The
simulations are initialized at redshift $z = 127$ in a box of size $3.125 \cMpch$. Initial particle positions and velocities are obtained by
applying the Zeldovich approximation (\citealp{Zeldovich:1970}) to
particles arranged on a Cartesian grid. We adopt a transfer function
for matter perturbations generated with \cmbfast\ (version 4.1;
\citealp{Seljak:1996}).
\par
We first perform a low-resolution simulation without star formation
down to redshift $z = 10$, and locate the most massive halo, with
virial mass $M_{\rm vir} \approx 10^9 \Msun$ and virial radius $r_{\rm
vir} =3.1 \kpc$. We then trace the particles found within $3r_{\rm vir}$ 
from the most bound particle of this halo back to their
locations at the start of the simulation. We hierarchically refine
the initial particle setup using a nested sequence of cubical patches
(``zooms'') centered on the traced
particles, inside which we increase the mass resolution by successive 
factors of 8 with each increasing level of zoom. In the zoom with the highest mass resolution, 
which entirely contains the traced particles and which we refer to as the refinement region, 
gas (dark matter) particles have masses $m_{\rm g} = 484 \Msun$ ($m_{\rm DM} = 2350 \Msun$). 
The simulations are performed with the gravitational forces softened over
a sphere of Plummer-equivalent radius $\epsilon = 0.1 \ckpch$ applied
to all particles. 
\par
The particle dynamical variables, such as 
position, velocity, and density, are evolved in time using 
individual particle gravito-hydrodynamical time steps determined by the smaller 
of the dynamical time step and the Courant time step (e.g.,
Equation~16 in \citealp{Springel:2005}), which is the standard 
{\sc gadget} time stepping scheme. We do not explicitly 
limit the particle gravito-hydrodynamical time steps by either the chemical time or the 
radiative cooling or radiative heating time. However, chemistry, radiative 
cooling and radiative heating described below are solved by subcycling 
the smallest gravito-hydrodynamical time step among all particles 
in the simulation on the relevant time scales (see Appendix~\ref{sec:coupling}).
\par
\subsection{Chemistry and Cooling}
\label{sec:cooling}
We assume that the gas is of primordial composition with hydrogen
mass fraction $X = 0.75$ and a helium mass fraction $Y = 1-X$. We use
a modified version of the implicit solver {\sc dvode}
(\citealp{Brown:1989}) to follow the non-equilibrium chemistry and
cooling of $\rm H_2$, $\rm D$, $\rm HD$, $\rm D^+$, $\rm H^+$, $\rm
H$, $\rm D$, and $\rm He$, and we include $\rm H^{-}$ and $\rm H_2^+$
assuming their collisional equilibrium abundances (\citealp{Johnson:2006};
\citealp{Greif:2010}). We consider all relevant radiative
cooling processes: cooling by collisional ionization, collisional excitation of atomic and 
molecular lines, the emission of free-free and recombination
radiation, and Compton cooling by the CMB. Once stars form and emit radiation, the chemical and thermal evolution of the gas is
also affected by the photodissociation of molecular hydrogen and deuterium, 
and photoionization of hydrogen and helium, as we describe in 
Sections~\ref{Sec:dissociation} and \ref{Sec:ionization} below. It should be kept in mind that 
at high gas densities, a Jeans floor employed to avoid artificial fragmentation
artificially increases the gas temperature and 
affects the distribution and dynamics of the gas (see Equation~1 in \citealp{Pawlik:2011a}).
\subsection{Star Formation}
\label{Sec:StarFormation}
Star formation is a complex astrophysical phenomenon, many
details of which remain to be understood (for a review see, e.g.,
\citealp{McKee:2007}). In the MW and in nearby galaxies, star
formation is observed to occur inside a hierarchy of clouds with
masses $\sim 10^4-10^6 \Msun$. The clouds are
transformed into stars at a rate $\dot{\rho}_\star =
\rho_{\rm cl}/\tau_\star$, where $\tau_\star = \tau_{\rm
ff}/\epsilon_{\rm ff} $ is the star formation time scale, $\tau_{\rm
ff} = [3 \pi / (32 G \rho_{\rm cl})]^{1/2}$ the free fall time at the
characteristic density $\rho_{\rm cl}$ of the star-forming clouds, and
$\epsilon_{\rm ff}$ the star formation efficiency per free fall
time. Star formation is observed to be a slow process with a typical
efficiency per free fall time of only $\epsilon_{\rm ff} \sim 0.01$, 
independent of the characteristic densities of the
star-forming clouds in the range $n_{\rm H, cl} \equiv \rho_{\rm cl} X
/ m_{\rm H} \sim 10-10^5 \cmci$ (\citealp{Krumholz:2007}).
\par
While gas masses $\sim 10^4-10^6 \Msun$ that characterize star-forming
clouds in the nearby universe are resolved, our simulations lack the
resolution and physical detail to follow the formation of individual
stars. Hence, we cannot exploit our simulations to estimate the star
formation rates (SFRs) inside individual star-forming clouds from
first principles.  Instead, we adopt a phenomenological model that
specifies how quickly gas turns into stars. The model is motivated by
and consistent with investigations of star formation in the nearby
universe. Prescriptions for treating star formation in
simulations of high-redshift galaxies are commonly calibrated with
relations obtained from the local universe. This practice is
necessitated by the current lack of direct observations of star
formation at high redshifts. The phenomenological approach is
sufficient to allow us to investigate the effect of stellar
dissociating and photoionizing radiation on the interstellar gas and
the IGM. 
\par
We restrict star formation 
to occur only in regions with gas densities exceeding a threshold density, $n_{\rm H} \ge n_{\rm SF}$, 
where we set $n_{\rm SF}\equiv 500 \cmci$. This is somewhat
lower than the densities  $\sim 10^4 \cmci$ of metal-free clouds on the verge of collapse to
form stars (e.g., \citealp{Bromm:2002};
\citealp{Abel:2002}). It is also lower than the similar densities
of metal-free clouds able to shield their ${\rm H}_2$ from
external dissociating radiation (e.g., \citealp{Chalence:2012}). 
Because our simulations focus on the assembly of galaxies
inside more massive halos, we must adopt, for reasons of computational viability, 
a lower resolution than employed in these previous works, preventing us 
from adopting still larger star formation threshold densities. Our choice for 
the star formation threshold density is close to the lowest 
threshold density $n_{\rm H}=1000 \cmci$ explored by \citealp{Muratov:2012} above which the time of the formation 
of the first star formed inside high-redshift minihalos
was found to be insensitive to an increase in the threshold density (see also, e.g., 
\citealp{Maio:2009}; \citealp{Wise:2012}). 
\par
We adopt a star formation time scale that depends solely on the threshold density for 
star formation $n_{\rm SF}$,
\begin{equation}
\tau_\star \equiv  \frac{\tau_{\rm ff} (n_{\rm SF})} {\epsilon_{\rm ff}} = 201.2 \Myr \left (\frac{n_{\rm SF}}{500 \cmci}\right)^{-1/2},
\label{Eq:StarFormationTimeScale}
\end{equation}
where we have set $\epsilon_{\rm ff} = 0.01$. Hence, the star formation time scale is independent of the 
gas density $n_{\rm H} \ge n_{\rm SF}$. This amounts to assuming that all star 
formation occurs inside clouds with characteristic density $n_{\rm H, cl} = n_{\rm SF}$, 
and is consistent with both our limited resolution and the observations in the nearby universe. A similar star 
formation recipe using a density-independent star formation time has been employed in, 
e.g., \cite{Kravtsov:2003}. However, we caution that the SFRs of 
simulated galaxies depend on the specific choice for the value of the star formation efficiency, at least 
unless star formation is self-regulated by feedback (e.g., \citealp{Ricotti:2002}; \citealp{Haas:2012}).
\par
Our numerical implementation of the star formation law is identical to
that of \cite{Schaye:2008}. The star formation law is interpreted
stochastically, and the probability that a star-forming gas particle
is turned into a star particle in a time interval $\Delta t$ is given
by $\min(\Delta t/\tau_\star, 1)$. Gas particles are converted
to star particles assuming a conversion efficiency of 100\%, i.e., the masses of the star particles are
identical to those of the gas particles from which they are
formed. The implied relatively large mass of star particles is consistent with our resolution, but sets a 
lower limit on the stellar mass fractions and hence ionizing and LW 
luminosities in the simulated galaxies to which the effects of radiative feedback
may be sensitive. We impose an upper limit $T_\star \equiv
\max(1.5\times 10^4 \K, 10^{0.5} \times T_{\rm floor})$ on the
temperature at which gas is allowed to form stars, where $T_{\rm
  floor}$ is the minimum temperature set by the Jeans floor.
\par

\subsection{Population Synthesis}
\label{sec:popsyn}
We interpret the star particles in our simulations as simple stellar
populations, i.e., instantaneous stellar bursts that are characterized
by an initial mass function (IMF), metallicity, and age. We compute the time-dependent hydrogen and helium ionizing
luminosities $Q_{\rm HI}$, $Q_{\rm HeI}$, and $Q_{\rm HeII}$, as well
as the luminosities $Q_{\rm LW}$ in the LW band with energies
$11.2-13.6 \eV$, of these star formation bursts using the population
synthesis models from \cite{Schaerer:2003}. The models assume a power-law IMF $p(m_\star)
\propto m_\star^{-\alpha}$ with the \cite{Salpeter:1955} exponent $\alpha
= 2.35$ but allow for a variation of the range of the stellar masses
sampled from the power-law distribution. We describe the stellar
bursts using the \cite{Schaerer:2003} zero metallicity models with
initial masses in the range $50-500\Msun$. The age of a
burst is the time difference between the simulation time at which the
star particle was created and the current simulation time. For 
reference, the luminosities at zero age main
sequence are $Q_{\rm HI} = 10^{47.98}$, $Q_{\rm
HeI} = 10^{47.80}$, $Q_{\rm HeII} = 10^{47.05}$, and $Q_{\rm LW} =
10^{46.96} \,\textrm{photons} \Msuni \invs$. We only consider star particles inside the 
refinement region, and we do not follow the propagation of radiation outside this region. 
\par

\subsection{Photodissociation}
\label{Sec:dissociation}
Molecular hydrogen $\rm H_2$ and deuterated hydrogen HD are
photodissociated upon absorption of radiation in the 
LW band belonging to the photon energy range $11.2\eV-13.6\eV$. The
rate of photodissociation of $\rm H_2$ is (e.g., \citealp{Abel:1997})
\begin{eqnarray}
k_{\rm LW} &=& 1.1 \times 10^8 \invs \frac{F_\nu(\nu_{\rm LW})}{\erg \invHz \invs \cmsqi}\nonumber\\
&=& 1.38\times 10^{-12} \invs J_{21} ,
\end{eqnarray}
where  $\nu_{\rm LW} \equiv 12.4 \eV$ is the
characteristic LW frequency, $F_{\nu}(\nu_{\rm LW})$ is the LW flux, $J_{21} = J_\nu /
(10^{-21} \erg \invHz \invs \cmsqi\sri)$ is the normalized
LW mean intensity, and $J_\nu = F_\nu(\nu_{\rm LW})/4\pi$. We set the rate for photodissociation of HD
identical to that of $\rm H_2$ (e.g., \citealp{Glover:2007};
\citealp{Wolcott:2011}).
\par
We write the LW intensity $J_{21}$ as the sum of the
cosmological LW background intensity $\bar{J}_{21}$ and the
local LW intensity $J_{21}^{\rm loc}$ produced by the
stellar populations represented by the star particles in the
refinement region of our simulations, $J_{21} = \bar{J}_{21} +
J_{21}^{\rm loc}$. The refinement region is too small to follow the
build-up of the cosmological LW background in a
self-consistent manner. We therefore treat the intensity of the
LW background as a free parameter, and approximate its 
evolution using
\begin{equation}
 \bar{J}_{21}(z) = \bar{J}_{21,0} \times 10^{-(z - z_0)/5},
\label{eq:lw}
\end{equation}
where $\bar{J}_{21,0}$ is the intensity of the LW background at $z_0$. 
Setting $\bar{J}_{21,0}= 1$ and $z_0 = 10$, Equation~(\ref{eq:lw}) provides a good fit to 
the LW background evolution presented in Figure 1 of \cite{Greif:2006}, and 
is also consistent with computations of the LW background in other works 
(e.g., \citealp{Wise:2005}; \citealp{Ahn:2009}). The use of a LW background that is
a fixed function of redshift does not allow us to capture the effects of self-regulation of star formation inside 
minihalos by LW feedback (e.g., \citealp{Ahn:2012}), at least until the intensity of 
LW radiation emitted by the stars inside the simulation box 
has become higher than the intensity of the background.
\par
For reasons of computational efficiency we determine the contributions
from the $N_\star$ star particles in the refinement region to the
intensity $J_{i}^{\rm loc} = \sum_{j = 1}^{N_\star} J^\star_{j}$ evaluated 
at the location of gas particle $i$ in the optically thin
approximation (e.g., \citealp{Wise:2008a}), i.e., 
\begin{eqnarray}
J_{j}^{\star} &=& \frac{h_{\rm P} \nu_{\rm LW}}{4 \pi \delta \nu_{\rm LW}} \frac{Q_{{\rm LW}, j} m^\star_j}{4 \pi r_{ij}^2}   \label{eq:lwlocal},
\end{eqnarray}
or, in units of $10^{-21} \erg \invHz \invs \cmsqi\sri$,  
\begin{eqnarray}
J^\star_{21, j}&\approx& 1 \left(\frac{Q_{{\rm LW}, j}}{10^{47}\Msuni \invs}\right) \left(\frac{m^\star_j}{500 \Msun}\right) \left(\frac{r_{ij}}{1 \kpc}\right)^{-2}.
\end{eqnarray}
Here, $Q_{{\rm LW}, j}$ is the photon luminosity of star particle $j$ per unit mass in the
LW frequency band, $\delta \nu_{\rm LW} =
(13.6\eV-11.2\eV)/ h_{\rm P}$ is the width of the LW
band, $h_{\rm P}$ Planck's constant, $m^\star_j$ the mass of star
particle $j$, and $r_{ij}$ the distance between the gas and the
star particle. 
\par
For simplicity, we approximate the time-dependent 
LW luminosities $Q_{\rm LW}$ computed in Section~\ref{sec:popsyn} 
by their zero age main sequence values, and we assume that the 
stars emit LW radiation for $3 \Myr$ (\citealp{Schaerer:2003}). 
We approximately account for radiative transfer (RT) effects by attenuating
the total LW intensity $J_{21}$ by a local self-shielding factor (Equation~10
in \citealp
{Wolcott:2011b} with $\alpha = 1.1$). The self-shielding factor depends on 
the column density of molecular hydrogen, which we compute in the local Jeans
approximation (e.g., \citealp{Shang:2010}).
\par
When computing the contribution to the LW intensity from 
local sources using Equation~(\ref{eq:lwlocal}), we ignore the absorption of LW radiation along 
the way to the absorbing gas particle. Near emitters, 
this is justified because molecular hydrogen is
efficiently collisionally dissociated inside the \ion{H}{2} regions
surrounding them (\citealp{Johnson:2007}; but see \citealp{Ricotti:2002}). Outside the \ion{H}{2} regions, 
this approximation is good as long as the intergalactic molecular hydrogen fraction does
not substantially exceed its primordial value, $\eta_{\rm H_2} \sim
10^{-6}$. In this case, the optical depth for absorption in
the LW band remains insignificant out to distances
$\gtrsim 1 \cMpc$ (e.g., \citealp[their Figure~12]{Ricotti:2001}; see
also \citealp[their Figure~7]{Glover:2003}; \citealp{Ahn:2009}) much
larger than the size of the refinement region. Finally, we also ignore the absorption of LW photons by 
atomic hydrogen series lines, which also is not significant on the small scales 
simulated here (e.g., \citealp{Haiman:2000}; \citealp{Ahn:2009}). Our 
implementation of photodissociation is similar to that in \citealp{Johnson:2012}, 
but we do not account for the photo-detachment of H$^-$.
\par
\begin{deluxetable*}{rcccccccc}
\tablecolumns{10}
\tablecaption{Simulation parameters.}
\tablehead{
  \colhead{Simulation} &
  \colhead{$m_{\rm gas}$ \tablenotemark{a}} &
  \colhead{$m_{\rm DM}$ \tablenotemark{b}} &
  \colhead{$\epsilon$ \tablenotemark{c}} & 
  \colhead{$n_{\rm SF}$\tablenotemark{d}} &
  \colhead{LW\tablenotemark{e}} &
  \colhead{RT\tablenotemark{f}} }

\startdata

\textit{LW+RT}      &     $ 4.84 \times 10^2$  &$ 2.35 \times 10^3$ &  $ 0.1$  & $ 500 $  &yes & yes      \\
\textit{LW}      &     $ 4.84 \times 10^2$  &$ 2.35 \times 10^3$ &  $ 0.1$  & $ 500 $    &yes & no      \\
\textit{NOFB}   &      $ 4.84 \times 10^2$  &$ 2.35 \times 10^3$ &  $ 0.1$    & $ 500 $   &no & no   \\

\enddata
\tablenotetext{a}{Gas particle mass in the refinement region ($M_\odot$).}
\tablenotetext{b}{Dark matter particle mass in the refinement region ($M_\odot$).}
\tablenotetext{c}{Gravitational softening radius ($h^{-1}\,\textrm{kpc comoving}$).}
\tablenotetext{d}{Star formation threshold density $n_{\rm SF}$ ($\textrm{cm}^{-3}$).}
\tablenotetext{e}{Photodissociation by stellar LW photons and the LW background.}
\tablenotetext{f}{Radiative transfer of stellar ionizing photons.}

 \label{tbl:params}
\end{deluxetable*}

\subsection{Photoionization}
\label{Sec:ionization}
In this section we describe our implementation of photoionization by
stellar radiation. This implementation involves two main
steps. First, we transport ionizing photons radially from the star
particles through the simulation box, and compute the fraction of the
photons absorbed by atomic hydrogen and helium. Second, we infer the
associated photoionization and photoheating rates. Here we will only give a brief overview. 
We present a detailed description of the
implementation of the RT and the computation of the photoionization 
and photoheating rates in the Appendix. There we will also discuss tests of this implementation. The coupling of 
the RT with the hydrodynamical evolution is achieved 
by passing the photoionization and photoheating rates, along with the photodissociation rates described in the previous section, 
to the non-equilibrium solver for the chemical and thermal evolution of the gas described in Section~\ref{sec:cooling}, 
and this coupling is described in Appendix~\ref{sec:coupling}.
\par
We transport the ionizing radiation emitted by the star particles
using the multi-frequency RT code \traphic\
(\citealp{Pawlik:2008}; \citealp{Pawlik:2011b}). \traphic\ solves the
time-dependent RT equation by tracing photon packets
emitted by source particles at the speed of light and in a
photon-conserving manner through the simulation box. The photon
packets are transported directly on the spatially adaptive,
unstructured grid traced out by the SPH particles, which allows one to
exploit the full dynamic range of the SPH simulations. A directed
radial transport of the photon packets from the sources is
accomplished despite the irregular distribution of SPH particles by
guiding the photon packets inside cones. A photon packet merging
technique renders the computational cost of the RT independent of the
number of ionizing sources. The transport of photons is discretized in
RT time steps $\Delta t_{\rm r}$, after each of which the chemical and
thermal evolution of the gas is advanced based on the number of
absorbed photons. 
\par
\begin{figure*}
\begin{center}
  \includegraphics[trim = 0 0 0 0mm, width = 0.24\textwidth]{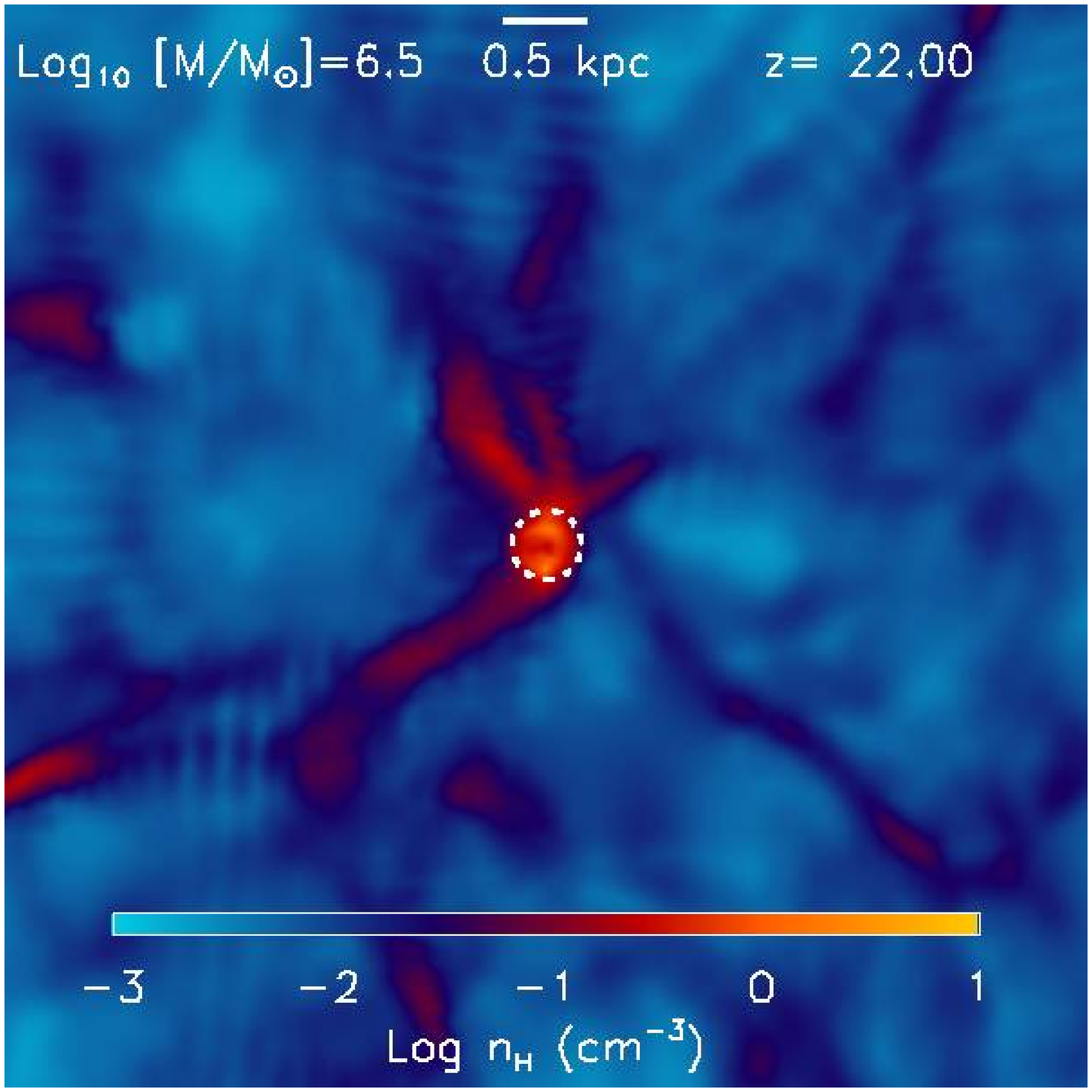}
\includegraphics[trim = 0 0 0 0mm, width = 0.24\textwidth]{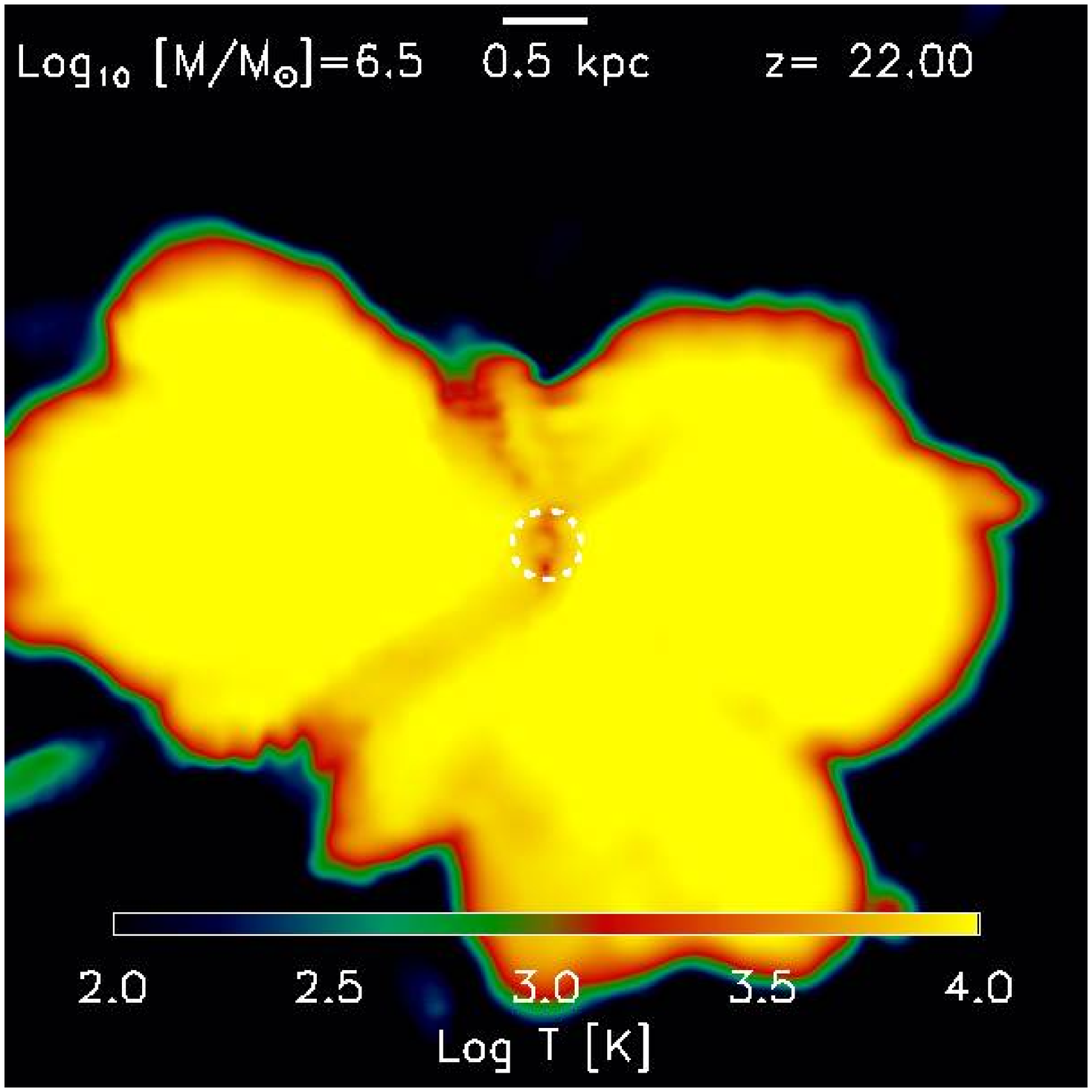}
\includegraphics[trim = 0 0 0 0mm, width = 0.24\textwidth]{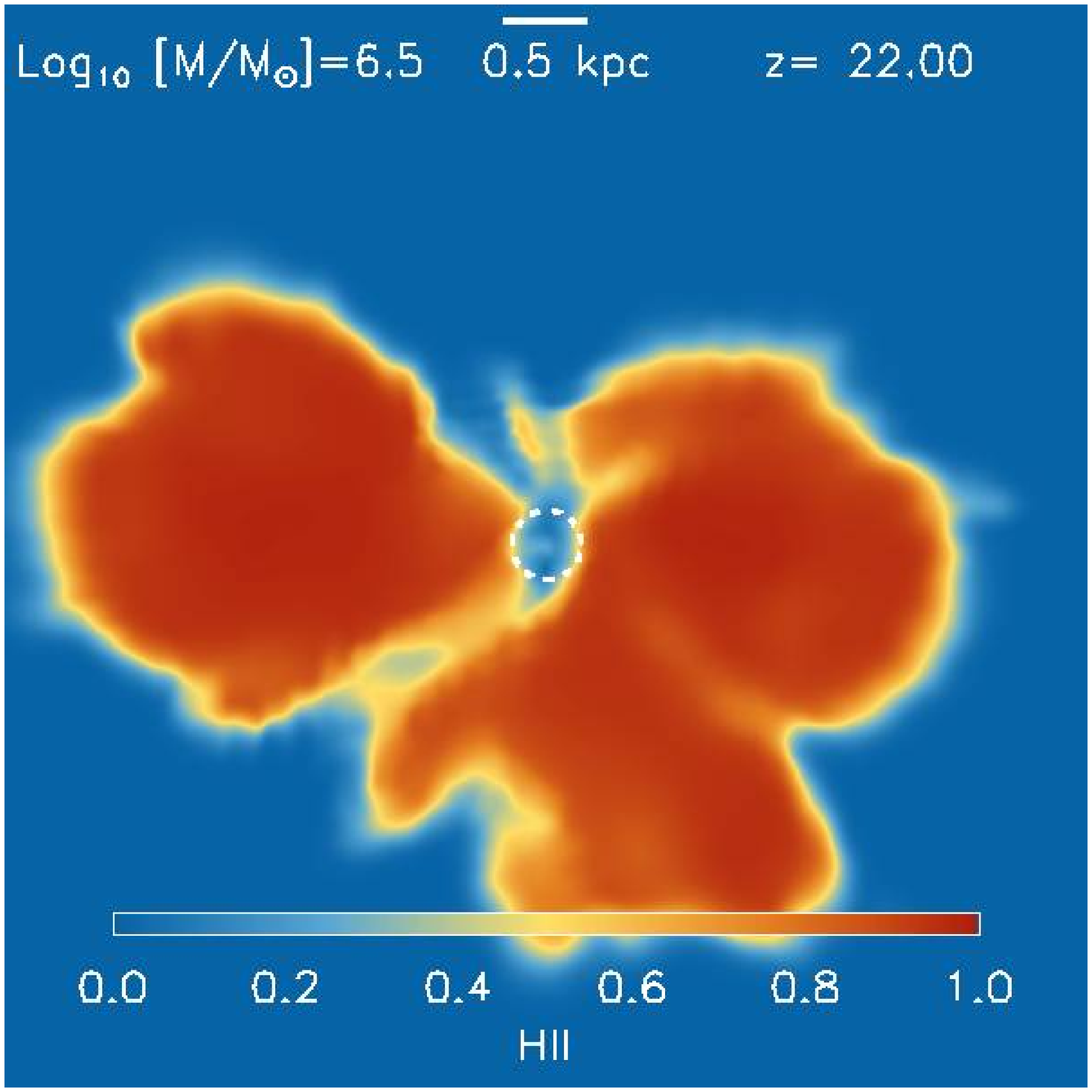}
\includegraphics[trim = 0 0 0 0mm, width = 0.24\textwidth]{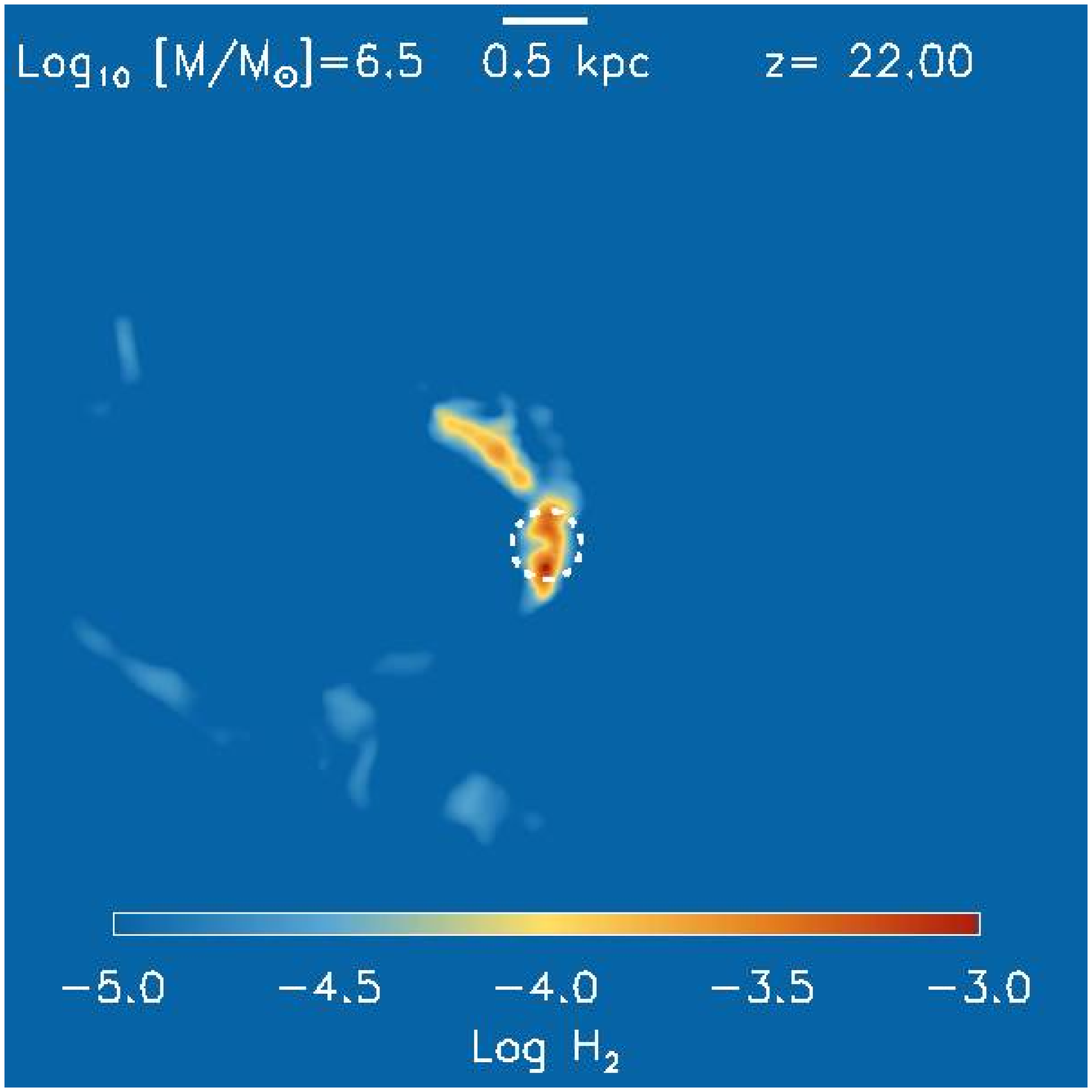}

\end{center}
\caption{Gas density, temperature, ionized hydrogen
  fraction, and molecular hydrogen fraction (from left to right) 
  averaged in a thin slice of thickness $0.2 \kpc$ through the center of the
  minihalo progenitor at $z = 22$, shortly after the first stellar
  burst has shut off. The relic photoionized region of diameter
  $4-6 \kpc$ shows the typical non-spherical ``butterfly'' shape (e.g., \citealp{Abel:1999b}). The
  dense gas inside the virial radius and along the filaments cools
  rapidly, allowing the residual free electrons to catalyze the
  formation of molecular hydrogen up to fractions $\eta_{\rm H_2} \sim 
  2\times 10^{-3}$ (\citealp{Oh:2002}). The molecular hydrogen fraction is
  also increased near the transition to the predominantly neutral IGM,
  a relict of the increased rate of molecular hydrogen formation
  inside the ionization fronts (\citealp{Ricotti:2002}). The dashed circle marks the virial radius. For purpose 
  of visualization, this figure has been obtained from a simulation identical to simulation {\it LW+RT} 
  but with an increased angular sampling, in which 
  the propagation of photons was not limited to a single inter-particle distance. 
  In Appendix~\ref{sec:parameters} we discuss that this change in parameters 
  leaves the evolution of the minihalo essentially unaffected.}
\label{Fig:FirstStar}
\end{figure*}

We make the following approximations specific to the current work.
Each star particle emits ionizing photon packets to its
neighboring SPH particles once per RT time step in a set of
tessellating emission cones centered around 8 different
directions. The effective angular sampling of the surrounding volume
is larger than implied by this number of directions thanks to the
splitting of photon packets among neighbors inside the same emission
cone, and because of the randomization of the emission directions at
each RT time step.  The photons are transported radially away from the
star particles by tracing them downstream inside transmission cones
with solid angle $4\pi / 128$. This angular resolution is sufficiently
high to track the delay of the ionization fronts around individual
halos by dense filaments, giving rise to the typical ``butterfly''
shape of ionized regions, as shown in Figure~\ref{Fig:FirstStar}. 
\par
To reduce the computational cost, we limit the propagation of photons to
at most a single inter-particle distance per RT time step, which
approximates the full time-dependent RT in the limit of small RT time
steps. Ionizing photons are transported using
a single frequency bin, and the absorption of the photons by neutral
hydrogen and neutral and singly ionized helium is computed in the grey
approximation. The grey approximation does not allow us to capture
effects of spectral hardening, and hence we may underestimate the effects
of photoionization and photoheating near and ahead of ionization
fronts, such as the enhanced formation of molecular hydrogen
(\citealp{Ricotti:2001}; \citealp{Ricotti:2002}). These 
approximations are discussed in further detail in the Appendix, in
particular in Appendix~\ref{sec:parameters}.
 
\section{Simulations}
\label{Sec:Simulations}
We employ a set of three simulations to study the effects of radiative
feedback on the assembly of high-redshift galaxies. Simulation
\textit{LW+RT} includes both star formation and the LW and ionizing
radiation emitted by the stars, as well as an imposed LW radiation
background, as described in Section~\ref{Sec:Numerics}. We will focus on discussing results of this simulation.
We will often present our results by comparing this simulation with
simulation \textit{LW}, which is identical except that the emission of
ionizing radiation is disabled, and with simulation \textit{NOFB}, in
which the emission of LW radiation is also disabled. In the last
simulation, gas forms stars, but these stars do not emit radiation,
and in addition, the intensity of the assumed LW background is set to
zero. This simulation is therefore identical to simulation {\it Z4}
reported in \cite{Pawlik:2011a}, except for the inclusion of star
formation. Important parameters of the simulations presented here are
summarized in Table~\ref{tbl:params}. The final redshift of the 
simulations is $z = 11$. We note that simulations {\it LW} and {\it
NOFB} are identical to the simulations used to estimate the
detectability of pair instability supernovae in \cite{Hummel:2011}.
\par
We use the friends-of-friends (FOF) halo finder, with linking parameter $b =
0.2$, built into the substructure finder {\sc subfind}
(\citealp{Springel:2001b}), to extract halos from our simulations. 
Given a FOF halo, we use {\sc subfind} to identify its most bound
particle and let it mark the halo center. We then
obtain the virial radius, defined as the radius of the
sphere centered on the most bound particle within which the
average matter density is equal to $200$ times the redshift-dependent 
critical density of the universe. The total mass inside this 
sphere defines the halo virial mass. We define the total SFR of a given halo
as the sum of the SFRs of the gas particles it contains. Because we employ a 
stochastic star formation recipe, the rate at which star-forming 
gas particles are converted to star particles may 
randomly fluctuate around this SFR.
\par
We make use of the following relation between virial temperature 
$T_{\rm vir}$ and virial mass $M_{\rm vir}$ (e.g., Equation~3.12 in \citealp{Loeb:2010}; see also \citealp{Barkana:2001}),
\begin{equation}
T_{\rm vir} = 1.04 \times 10^4 \K \left( \frac{M_{\rm vir}}{10^8 \Msun}\right)^{2/3} \left(\frac{\mu}{0.6}\right) \left(\frac{1+z}{10}\right).
\label{Eq:VirialTemp}
\end{equation}

\section{Formation and Evolution of the Dwarf Galaxy}
\label{Sec:Formation}
In this section we present the formation history of the simulated
dwarf galaxy. We start with discussing the radiative
feedback from LW and ionizing radiation on the assembly of gas and
the formation of stars inside the dark matter halo hosting the galaxy
(Section~\ref{Sec:Assembly}), and on the assembly of the dark matter
halo (Section~\ref{Sec:Profiles}). We finally proceed to investigate the
robustness of the gaseous disks forming at the center of the dwarf galaxy halo
under the radiative feedback (Section~\ref{Sec:Disks}).

\subsection{Baryon Assembly in the Dwarf Galaxy Halo}
\label{Sec:Assembly}

\begin{figure*}
\begin{center}
\includegraphics[trim = 0 0 0 0mm, width = 0.49\textwidth]{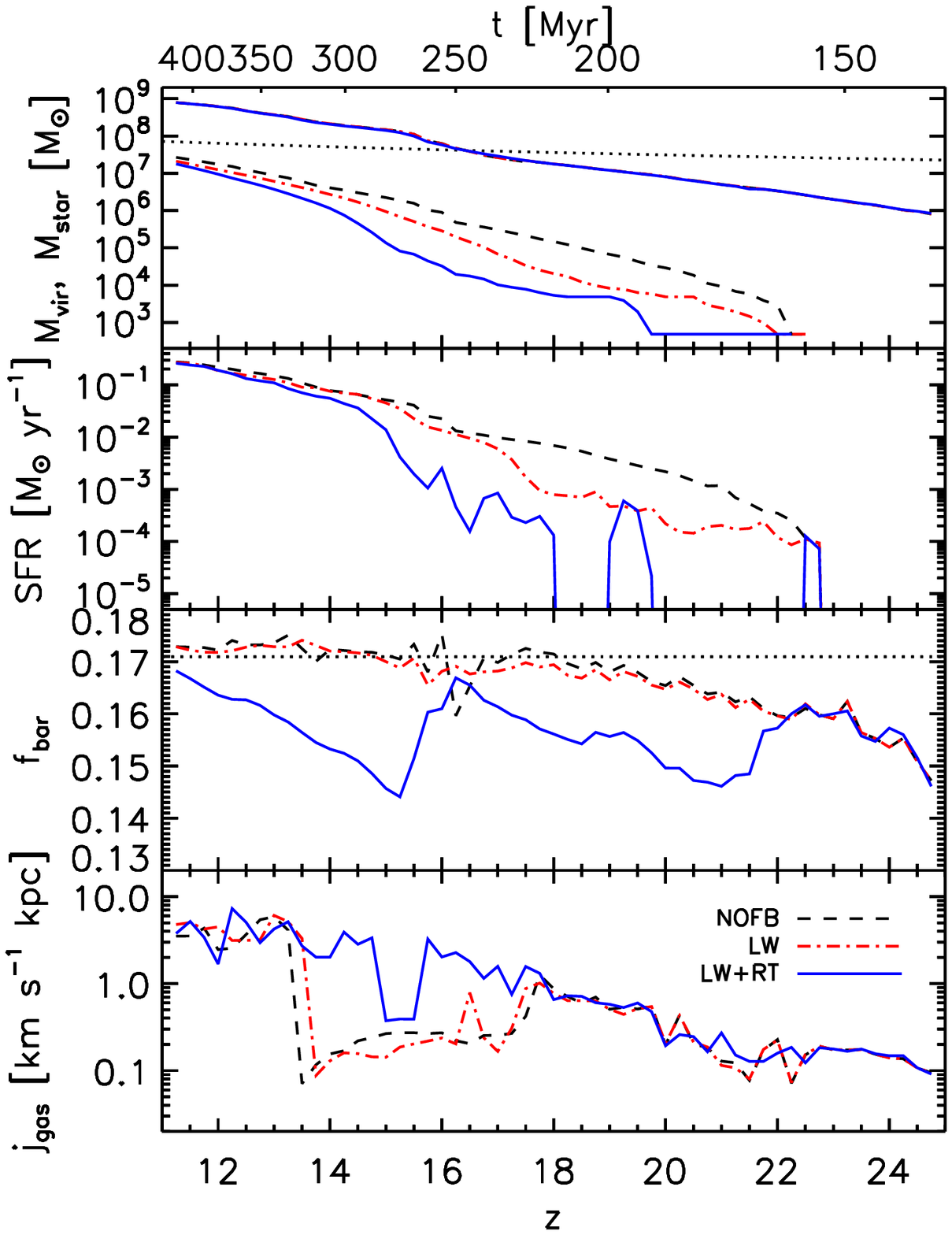}
\includegraphics[trim = 0 0 0 0mm, width = 0.49\textwidth]{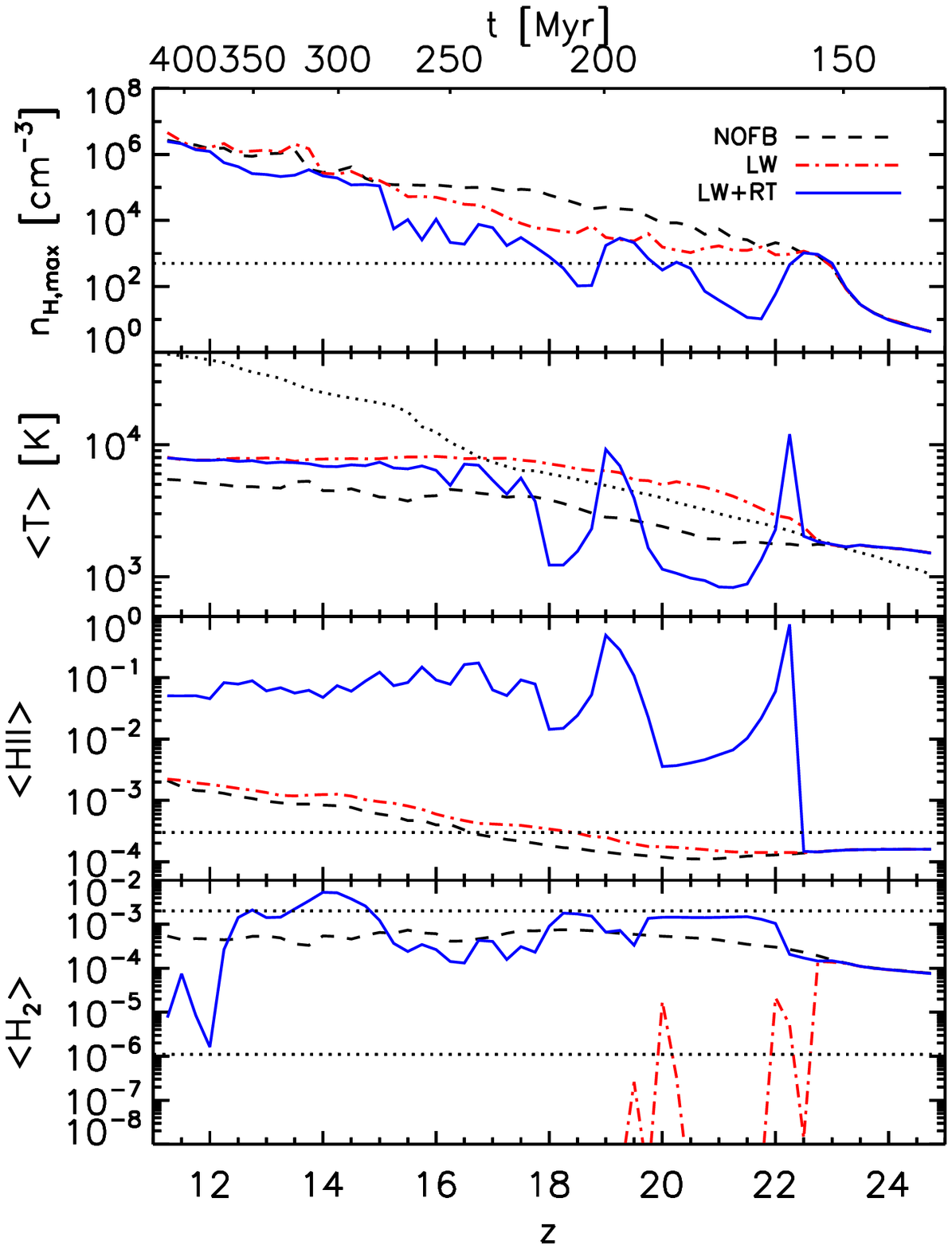}
\end{center}
\caption{Assembly history of the dwarf galaxy. The blue solid curves in each panel show the evolution of the indicated property of the 
  dwarf galaxy in simulation {\it LW+RT} which included both LW and ionizing radiation. For comparison, the red dash-dotted and the black dashed curves 
  show, respectively, the evolution of the same properties in simulation {\it LW} that included only LW but not ionizing radiation and 
  in simulation {\it NOFB} in which star particles remained dark and the intensity of the LW background was set to zero. The left-hand panel shows, from top to bottom: (a) virial mass $M_{\rm vir}$ (upper set of three curves which are nearly on top of each other) and stellar mass $M_{\star}$
 (lower set of three curves), and the mass corresponding to a virial temperature $T_{\rm
 vir} = 10^4 \K$ (dotted; Equation~\ref{Eq:VirialTemp} with $\mu = 0.6$);
 (b) SFR; (c) baryon fraction, and the
 universal baryon fraction $\Omega_{\rm b}/\Omega_{\rm m} \approx 0.17$ (dotted);
 (d) specific angular momentum $j_{\rm gas}$ of the gas. The right-hand panel
 shows, from top to bottom: (a) the maximum gas density $n_{\rm H,
 max}$; (b) the mass-weighted mean temperature $\langle T \rangle $,
 and the virial temperature corresponding to the virial mass of the 
 dwarf galaxy (in the {\it NOFB} simulation, dotted); (c)
 the mass-weighted mean ionized hydrogen fraction $\langle \eta_{\rm HII}
 \rangle $, and the initial ionized hydrogen fraction, $\eta_{\rm HII} =
 3\times 10^{-4}$ (dotted); (d) the mass-weighted mean molecular
 hydrogen fraction $\langle \eta_{\rm H_2} \rangle $, the initial molecular
 hydrogen fraction, $\eta_{\rm H_2} = 1.1\times 10^{-6}$ (lower dotted),
 and the freeze-out molecular fraction expected in fossil \ion{H}{2}
 regions, $\eta_{\rm H_2} \approx 2\times 10^{-3}$ (upper dotted; \citealp{Oh:2002}). All quantities are
 computed considering only matter inside the redshift-dependent
 virial radius $r_{\rm vir}$. The inclusion of photodissociation and photoheating implies as 
 strong negative feedback on the condensation of gas and the formation of stars in 
 the minihalo progenitor, but has little effect after the minihalo evolved 
 into an atomically cooling galaxy.}
\label{Fig:AssemblyHistory}
\end{figure*}

Figure~\ref{Fig:AssemblyHistory} shows the formation history of the 
dwarf galaxy in simulation {\it LW+RT}, including both photodissociating and photoionizing radiation (blue curves). For comparison, the figure also shows the
formation histories of the dwarf galaxy in the simulation in which
star particles were sources of LW but not of photoionizing radiation
({\it LW}; red curves), and in the simulation in which star particles
remained dark and the intensity of the LW background was set to zero ({\it NOFB}; black curves). 
The formation history is obtained by using {\sc subfind} to locate the dwarf halo progenitor that contains most of the 
50 most-bound particles of the dwarf halo at the final simulation redshift $z = 11$, and then repeating this 
procedure to find the progenitor of this progenitor and so on, tracing the halo assembly back to $z = 25$. The 
quantities displayed in Figure~\ref{Fig:AssemblyHistory} are obtained from the properties of
the particles inside the virial radius. The figure
shows that the dark matter halo hosting the emerging dwarf galaxy
grows by about three orders of magnitude in mass in about 300 million
years, consistent with expectations (\citealp{Pawlik:2011a}; their
Figure~1). 
\par
The evolution of the galaxy in simulation {\it LW+RT} 
proceeds in several main phases which will be discussed 
below: (1) the assembly of a dark matter
minihalo with mass $\sim 10^6 \Msun$ and the accretion and
condensation of gas inside it, leading to the formation 
of stars just below $z = 23$, (2) the subsequent accretion of gas under
feedback from star formation inside the dark matter minihalo, (3) the
evolution of this minihalo into an atomically cooling halo at $z
\approx 16$, during which the properties of the galaxy become
insensitive to the inclusion of LW radiation, (4) the ensuing growth
into a dwarf halo, during which the properties of the galaxy become
robust against feedback from photoionization, and, finally, (5) the
formation of two nested rotationally supported gaseous 
disks below $z \lesssim 13.5$. Our discussion of these phases will be 
complemented by comparisons with the evolution of the dwarf galaxy in 
simulations {\it LW} and {\it NOFB}. 
\par
During the first phase, and as the dark matter halo grows in mass and
accretes gas, both the baryon fraction, which is initially slightly
smaller than the cosmic baryon fraction $\Omega_{\rm b} / \Omega_{\rm
m} \approx 0.17$, and the central gas densities increase. By $z = 25$, the molecular
hydrogen fraction has significantly departed from its initial
value, enabling the minihalo gas to cool efficiently. The average gas temperature inside the virial
radius is initially slightly higher than the virial temperature (dotted
curve). Note that in simulation {\it NOFB}, as the molecular hydrogen fraction builds up, 
this relation then reverses and  the virial temperature becomes higher than the average gas temperature 
(e.g., \citealp{Oshea:2007}). At $z \approx 23$, the central 
gas densities become larger than the threshold density for 
star formation. A single gas particle is turned
into a star particle, triggering the emission of LW and ionizing
radiation from the associated stellar burst. At that time, the halo
has reached a mass of $ \sim 2 \times 10^6 \Msun$.
\par
Photoionization from the radiation emitted by the first
stellar burst almost instantly increases the average gas temperatures
to $\gtrsim 10^4 \K$. The associated increase in thermal pressure
pushes the gas away from the minihalo center, and removes a fraction
of it from inside the virial radius. As a consequence, the baryon
fraction is reduced to about $f_{\rm bar}= 0.15$. The reduction in the
baryon fraction is consistent with but slightly smaller than that found in
previous simulations of the assembly of minihalos under feedback from
star formation (e.g., \citealp{Wise:2008a}; \citealp{Wise:2012}). This may be 
because the minihalo simulated here is fed by dense filaments, and the inflow
of gas along these filaments provides a strong obstacle for
photoheating to drive the gas beyond the virial radius, and it 
replenishes the photoevaporated regions with fresh gas (e.g., \citealp{Abel:2007}). Nevertheless, the
reduction in the central gas mass is sufficient to drive the central gas densities below the threshold
density for star formation, and the stellar burst shuts itself
off. Figure~\ref{Fig:FirstStar} shows 
images of the gas density, temperature, and ionized and molecular 
hydrogen fraction around the minihalo at the end of the 
stellar burst.
\par

After the first stellar burst is shut off, the average mass-weighted 
molecular hydrogen fraction approaches $\sim 2 \times 10^{-3}$, as expected inside the
relic \ion{H}{2} region (\citealp{Oh:2002}). The increased fraction of
molecular hydrogen enables the gas to cool quickly, and the central
gas densities increase. A few tens of Myr after the end of the
first burst, the gas has become sufficiently cold and dense
for another stellar burst to be ignited (e.g., \citealp{Oshea:2005}; 
\citealp{Alvarez:2006}). Photoionization heating from 
this second burst again lowers the gas densities, but this time there is no strong 
decrease in the baryon fraction. The accreting minihalo is thus 
massive enough to retain most of the gas inside the virial
region. Because the negative feedback from photoheating is now less strong, this second
starburst is more extended in time than the first one,
and it involves the conversion of several gas particles to star
particles. However, the combined feedback from the stellar clusters
represented by the star particles eventually shuts off star
formation. But already a few tens of Myr later the central gas
densities have again increased above the SF threshold density, and the
galaxy continues to form stars. 
\par
By redshift $z \approx 16$, the halo has reached virial temperatures
$T_{\rm vir} \gtrsim 10^4 \K$. Consequently, a significant fraction of
the atomic hydrogen is collisionally excited, and its radiative
de-excitation endows the gas with an additional channel to lose its
thermal energy. Feedback from
photoionization heating continues to keep the gas inside the halo at relatively low densities. However, the galaxy now
forms stars continuously, albeit at a rate significantly
smaller than in the absence of radiative feedback. The comparison with
the quickly rising SFRs in simulation {\it LW} that included emission
of LW radiation but not that of ionizing photons shows that the feedback from the
photodissociation of molecular hydrogen by LW radiation alone becomes
inefficient in preventing the gas from forming stars as the halo mass
approaches the atomic cooling limit, in good agreement with previous
works (e.g., \citealp{Ahn:2007}; \citealp{Wise:2007a};
\citealp{Oshea:2008}). The transformation of the minihalo
into an atomic cooling halo is accompanied by a significant decrease of the
baryon fraction by $\approx 20\%$ at $z \approx 15-16$ and in the
specific angular momentum of the gas $j_{\rm gas} \equiv J_{\rm gas} / M_{\rm gas}$, 
where $J_{\rm gas}$ is the total gas angular momentum with respect to the motion of the 
most-bound particle and  $M_{\rm gas}$ the total gas mass. 
\par
\begin{figure*}
\begin{center}
\includegraphics[trim = 0 0 0 0mm, width = 0.49\textwidth]{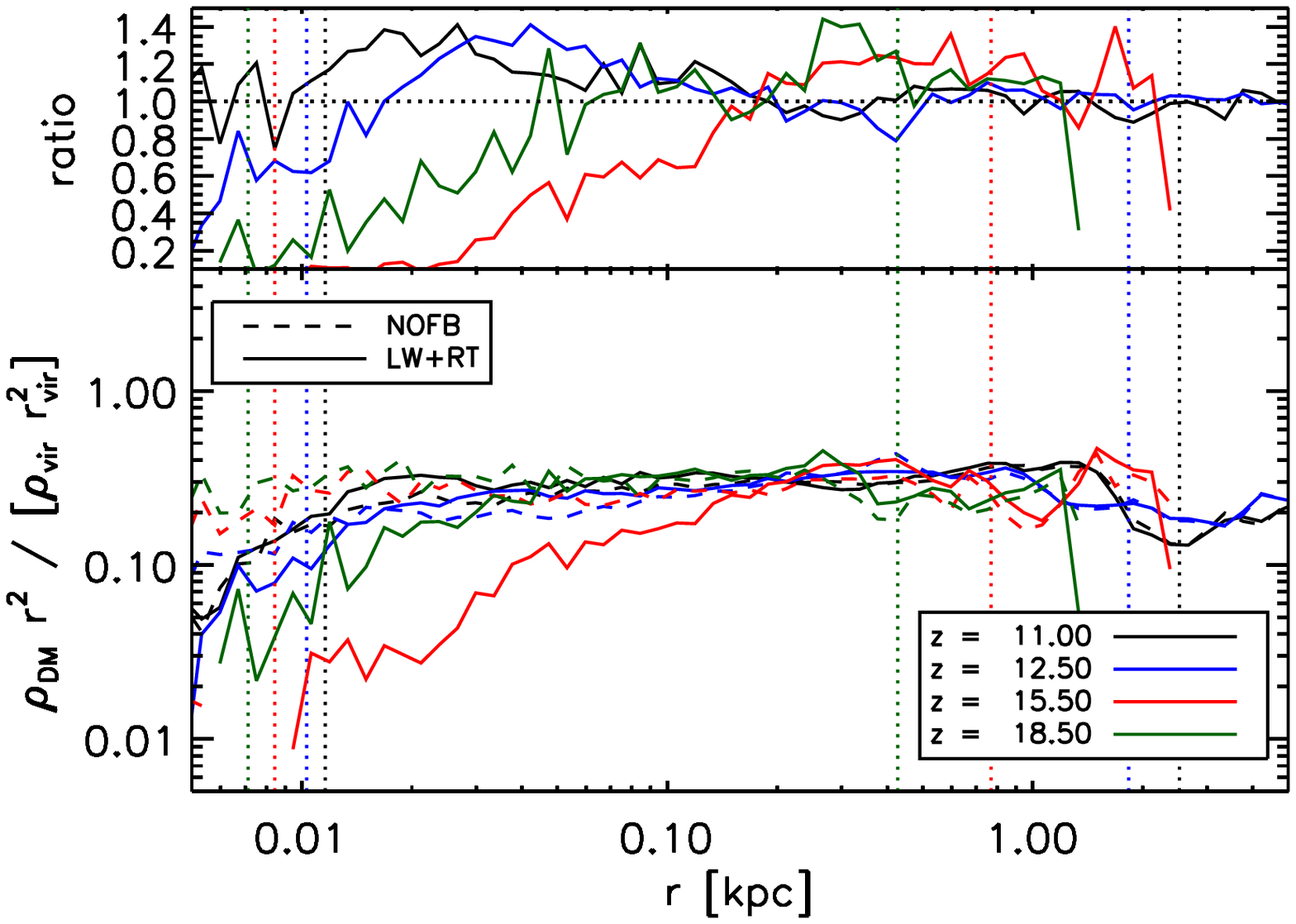}
\includegraphics[trim = 0 0 0 0mm, width = 0.49\textwidth]{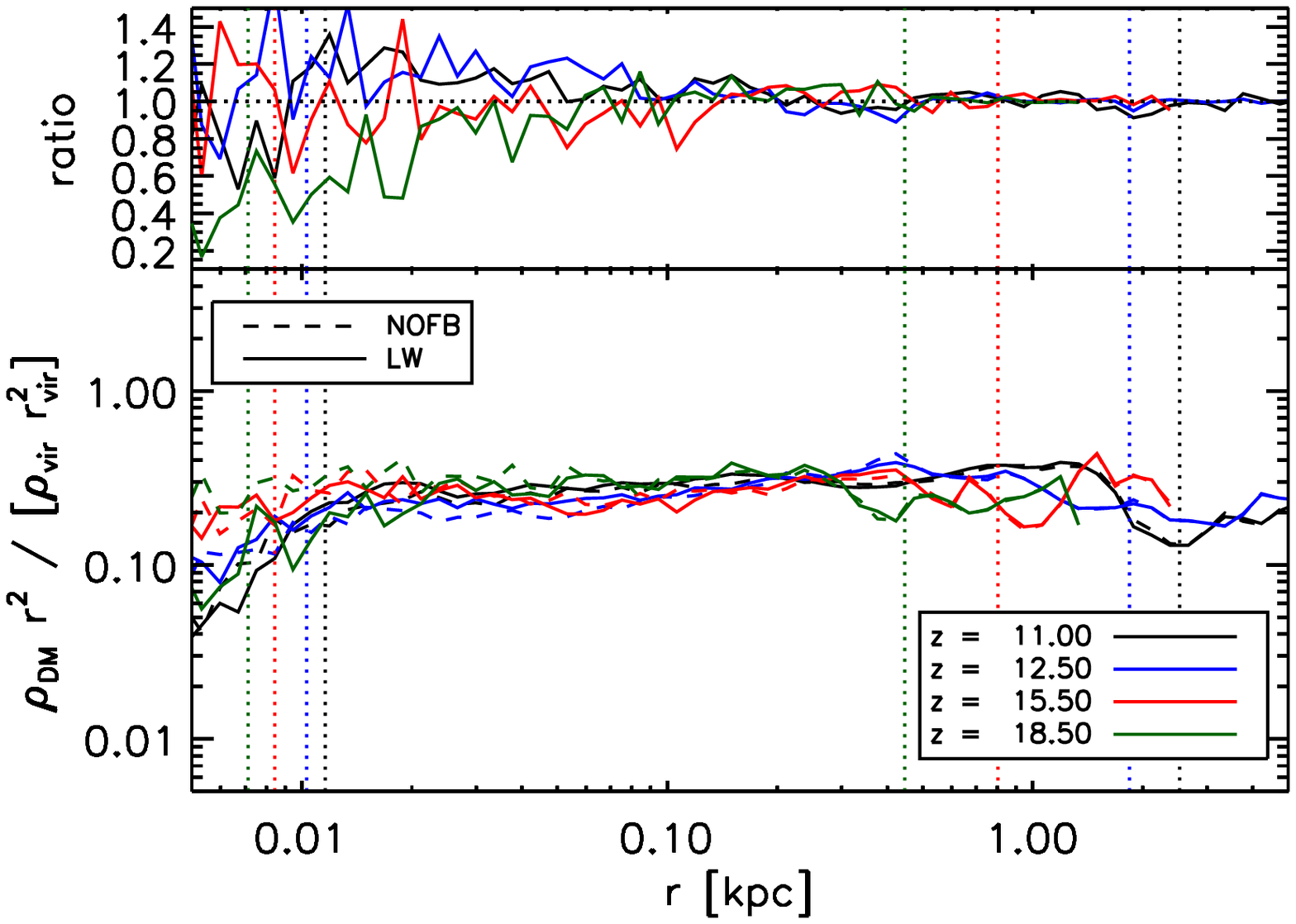}
\end{center}
\caption{Dark matter density profiles of the dwarf galaxy in simulation {\it LW+RT}, which included both LW and ionizing radiation
(bottom left; solid curves), and in simulation {\it LW}, which
included only LW radiation (bottom right; solid curves). Curves of
different colors show the density profile at different redshifts as
indicated in the legend. The density profiles have been normalized by
a singular isothermal density profile to reduce the
dynamic range. In each panel,
the set of dotted vertical lines on the left marks the
redshift-dependent gravitational softening length $\epsilon$, and the set of dotted
vertical lines on the right marks the redshift-dependent virial radius. 
For comparison, both bottom panels also
show the dark matter density profile of the dwarf galaxy in the
simulation without radiation ({\it NOFB}; dashed curves). The top panel in the
left-hand (right-hand) figure shows the ratio of the dark matter
density profile in the {\it LW+RT} simulation ({\it LW} simulation)
and the dark matter density profile in simulation {\it NOFB}. The horizontal dotted line marks a ratio of 1, corresponding to the outcome
in which the radiative feedback has no effect on the dark matter density profile. The inclusion of LW or 
ionizing radiation leads to a reduction of the central dark matter densities in the minihalo 
progenitor. As the minihalo turns into an atomically cooling galaxy and radiative feedback on the distribution of baryons 
becomes inefficient, the dark matter density profile approaches a singular isothermal profile in all simulations.}
\label{Fig:Profiles}
\end{figure*}

Below $z \lesssim 15$, photoionization heating becomes increasingly
inefficient at reducing the density of the halo gas, and the gas then
condenses up to $\gtrsim 10^5 \cmci$. As a result, the expansion of \ion{H}{2} regions 
is impeded by the increased recombination rates. 
The properties of the galaxy then become rapidly insensitive to the 
inclusion of ionizing radiation. The average mass-weighted ionized fraction, which has been increasing until then,
remains roughly constant at around $\langle \eta_{\rm HII} \rangle \lesssim 0.1$. As a result of the large
densities, the
SFRs approach those seen in the simulation without radiation. The average mass-weighted 
molecular hydrogen fraction reaches values in excess of
$\langle \eta_{\rm H_2} \rangle \gtrsim 2 \times 10^{-3}$ as ${\rm H_2}$ forms efficiently not only in fossil \ion{H}{2} 
regions (e.g., \citealp{Oh:2002}; \citealp{Johnson:2007}) 
but also in the ionization fronts of the \ion{H}{2} regions under irradiation from the
ionizing stars (\citealp{Ricotti:2002}; see also,
\citealp{Shapiro:1987}; \citealp{Jeon:2012}).
\par
A major merger at redshift $z \approx 15$ is accompanied by a strong
increase in the specific angular momentum of the halo gas. The gas
then settles in a disk, which is essentially in place at $z \approx 13.5$. 
The merger and the formation of this disk are shown in 
Figure~\ref{Fig:DisksDensities}, which display a sequence of 
snapshots of the gas density inside the forming dwarf galaxy. 
The disk is surrounded by a second larger-scale disk at $z \approx 12$. 
The evolution and the properties of these two disks will be discussed in more detail in
Section~\ref{Sec:Disks} below. After the formation of the disks, 
the difference between virial and average gas 
temperatures that has amplified since the halo has become an atomic cooling halo 
continues to increase. In the atomic cooling halo 
this difference arises primarily because a fraction of the gas does not 
shock-heat to the virial temperature upon accretion onto the halo. Instead,  
the gas can efficiently cool inside the dense filaments that reach into the
virial region and supply the halo center with gas
(\citealp{Pawlik:2011a}). A qualitatively similar difference between
average and virial temperatures has been seen in other 
simulations of high-redshift low-mass galaxies 
(e.g., \citealp{Oshea:2007}; \citealp{Wise:2007b}; \citealp{Greif:2008}). 
At $z = 11$, the final simulation redshift, the SFRs reach $\sim 0.2
\Msunyri$, consistent with the SFRs found in previous simulations of 
high-redshift low mass galaxies (e.g., \citealp{Wise:2009};
\citealp{Razoumov:2010}; \citealp{Yajima:2011};
\citealp{Wise:2012}). 
\par

\subsection{Effect of Radiative Feedback on the Dark Matter Halo}
\label{Sec:Profiles}

Radiative feedback also affects the properties of the 
dark matter halo hosting the emerging dwarf galaxy. Figure~\ref{Fig:Profiles} compares the evolution of the dark matter
density profile of the galaxy in simulation {\it NOFB} with that 
in simulation {\it LW+RT} (left panel), and 
with that in simulation {\it LW} (right panel). The
bottom panels show the dark matter density profiles for
each pair of simulations at four representative redshifts. The 
profiles are scaled by dividing by a singular isothermal profile
$\rho_{\rm iso} (r) = \rho_{\rm vir} r^2_{\rm vir} / r^2$. In the last
expression, $r_{\rm vir}$ is the virial radius of the halo, $\rho_{\rm
vir} = 200 \rho_{\rm crit}$, and $\rho_{\rm crit}$ is the critical
density. The top panels show the ratios of the density
profiles of the simulations including radiation and the simulation without radiation 
shown in the bottom panels, which helps to illustrate the 
effects of stellar feedback. In simulation {\it NOFB}, i.e., in the absence of radiative feedback 
(dashed curves), the dark matter density profile is approximately singular isothermal at all redshifts.
\par

The central dark matter densities in the simulations that included ionizing and/or
LW radiation (solid curves in each of the bottom panels) are
initially significantly lower, by up to factors $\sim 5$,
than those in the simulation without radiation. The dark matter
density profiles in these simulations thus do not follow a singular isothermal
shape but show a spatially resolved dark matter ``core'', extending to radii
significantly larger than the gravitational softening scale (left-most
vertical lines). The reduction in the central dark matter densities
is more distinct and exists down to lower redshifts in simulation {\it LW+RT} than in simulation {\it LW}. 
The difference in the central dark matter densities originates in the difference in the
distribution of the gas inside the assembling dwarf galaxy. In
simulation {\it LW}, gas cannot cool and condense as efficiently as in 
simulation {\it NOFB} because molecular hydrogen, the main coolant
in low-mass primordial galaxies, is photodissociated by LW
radiation. In simulation {\it LW+RT}, the central gas densities are,
on average, further reduced as photoionization heating drives the gas
away from the halo center. The radiative feedback on the distribution 
of baryons implies a significant change in the gravitational potential
and, in turn, in the gravitational pull on the dark matter, which
hence remains less centrally concentrated. 

\par
The comparison of the dark matter density profiles in the simulations with and
without feedback demonstrates that the ability of gas to cool and
condense to high densities is crucial for establishing the singular isothermal
density profile seen in the simulation without feedback 
(e.g., \citealp{Wise:2008b}; \citealp{Zemp:2012}). That gas condensation can lead to a more
concentrated dark matter density distribution is well known and is
usually described using the framework of halo contraction models
(\citealp{Blumenthal:1986}; \citealp{Gnedin:2004};
\citealp{Gnedin:2011}). One may then expect the reverse 
of this process, i.e., the removal of gas from the halo 
center, to lead to a reduction in the concentration of 
the dark matter. Indeed, previous works have demonstrated the ability of 
SN explosions to lower the central dark matter densities in
dwarf galaxies with masses $\gtrsim 10^9\Msun$ at $z \lesssim 10$ 
(e.g., \citealp{NavarroEke:1996}; \citealp{Mashchenko:2006};
\citealp{Mashchenko:2008}; \citealp{Governato:2012}). 
\par
Here we have shown that radiative feedback can have a qualitatively
similar effect on the dark matter distribution in high-redshift minihalos. 
The reduction in the central dark matter densities due to radiative
feedback is similar to that seen in the adaptive mesh refinement
simulations of \cite{Wise:2008b}. In our simulation {\it
LW+RT}, the centrally suppressed dark matter profiles shown in the
left panel of Figure~\ref{Fig:Profiles} can be approximated by a
\cite{Navarro:1997} profile and concentration parameter $\sim 2$ at
z = 18.5 and 15.5. The effect of radiative feedback on the dark matter
profiles of high-redshift low mass halos was also investigated by
\cite{Ricotti:2003}. The shape of dark matter profiles in a
cosmological simulation including gas dynamics and radiative feedback
was found to be similar, on average, to the shape in a simulation that
was identical except that it only treated the dynamics of the dark
matter. Our finding that radiative feedback can 
counteract the effects of gas condensation 
on the dark matter density profile is consistent with the results in \cite{Ricotti:2003} 
and \cite{Wise:2008b}. However, \cite{Ricotti:2003} 
also points out that there are significant statistical variations in the shape of the dark matter
profile of halos at fixed mass and redshift in the simulation that included gas physics. 
We therefore caution to overinterpret our conclusions drawn from the investigation of a single
dwarf halo.
\par

\subsection{The Disks}

\label{Sec:Disks}

\begin{figure*}
\begin{center}
\includegraphics[trim = 0 0 0 0mm, width = 0.24\textwidth]{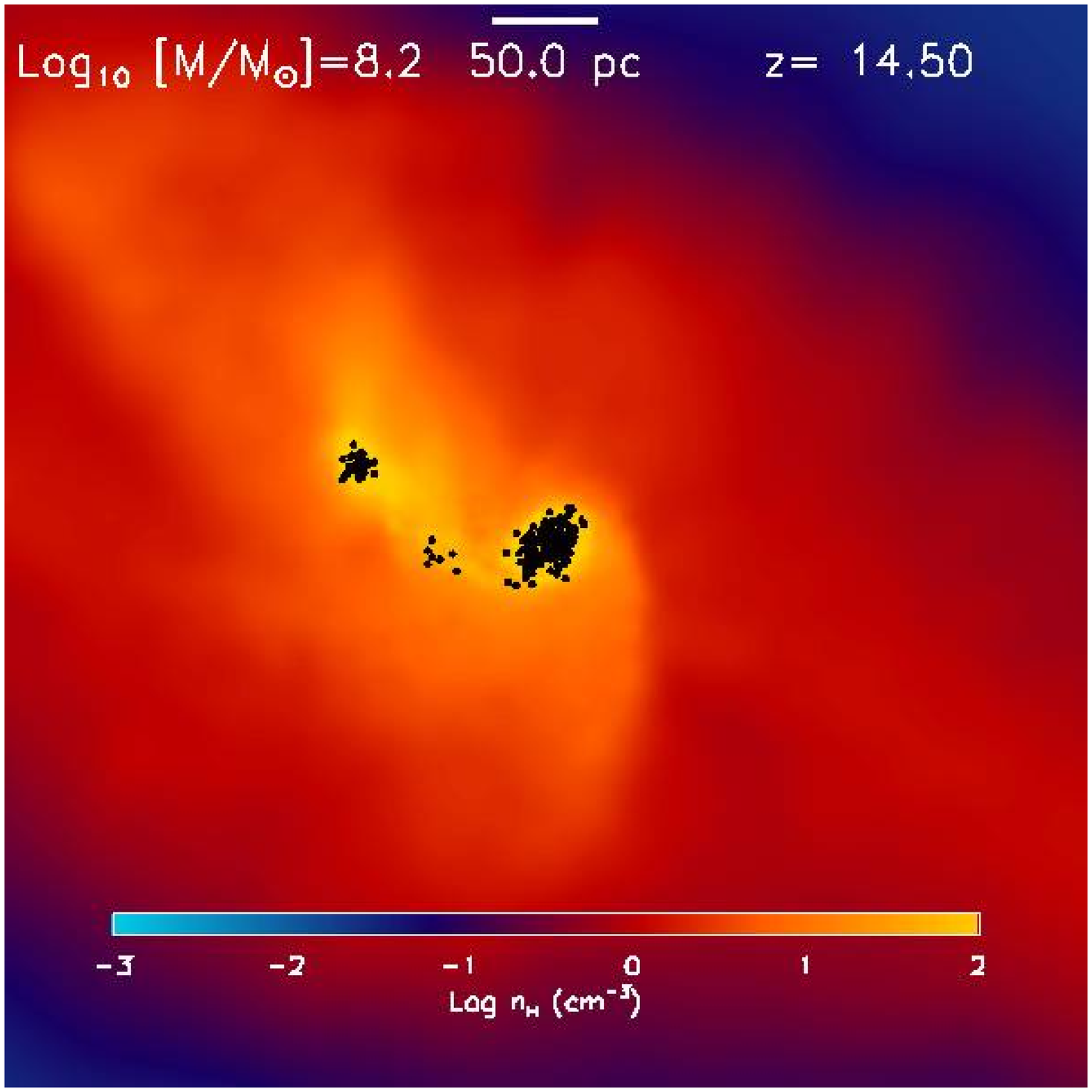}
\includegraphics[trim = 0 0 0 0mm, width = 0.24\textwidth]{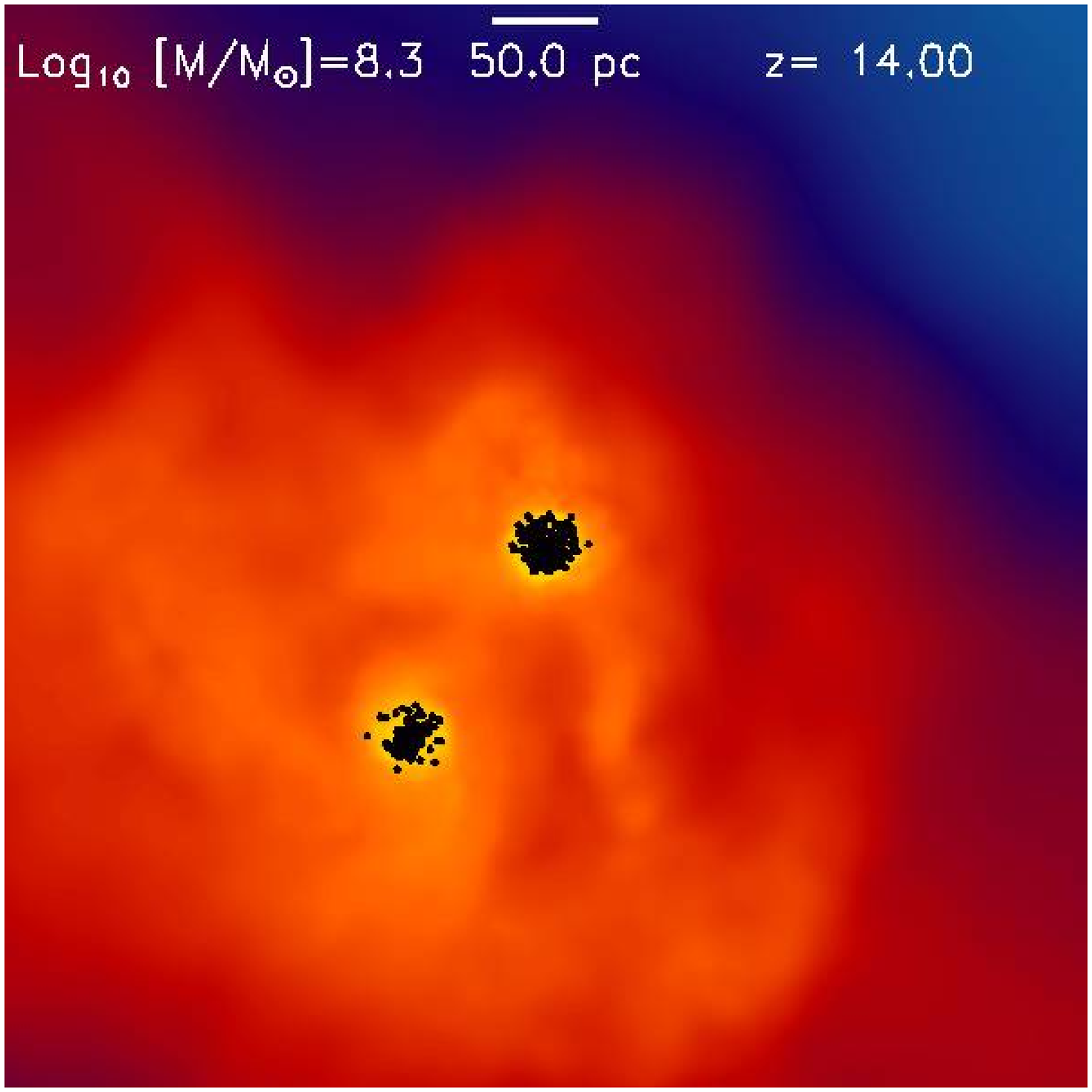}
\includegraphics[trim = 0 0 0 0mm, width = 0.24\textwidth]{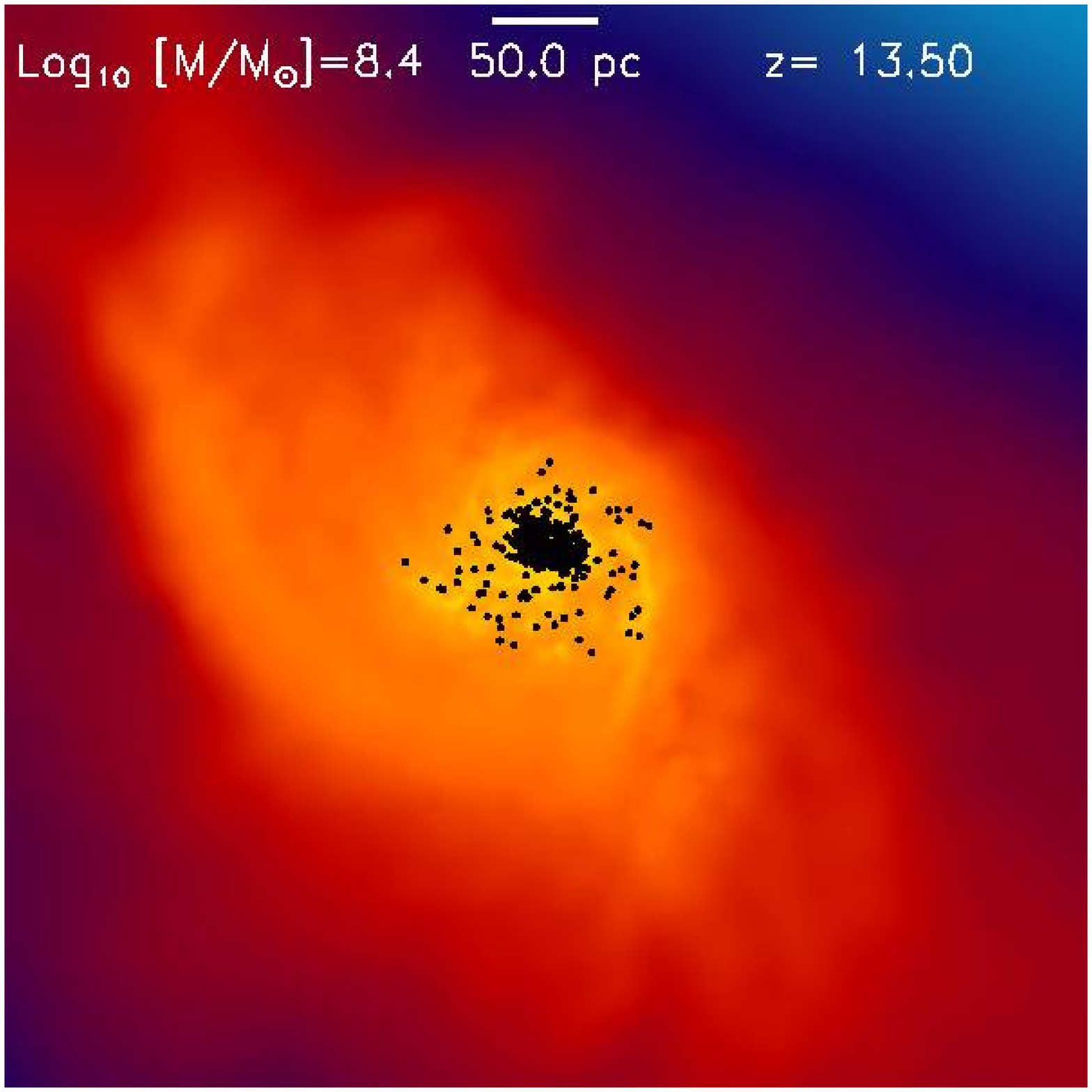}
\includegraphics[trim = 0 0 0 0mm, width = 0.24\textwidth]{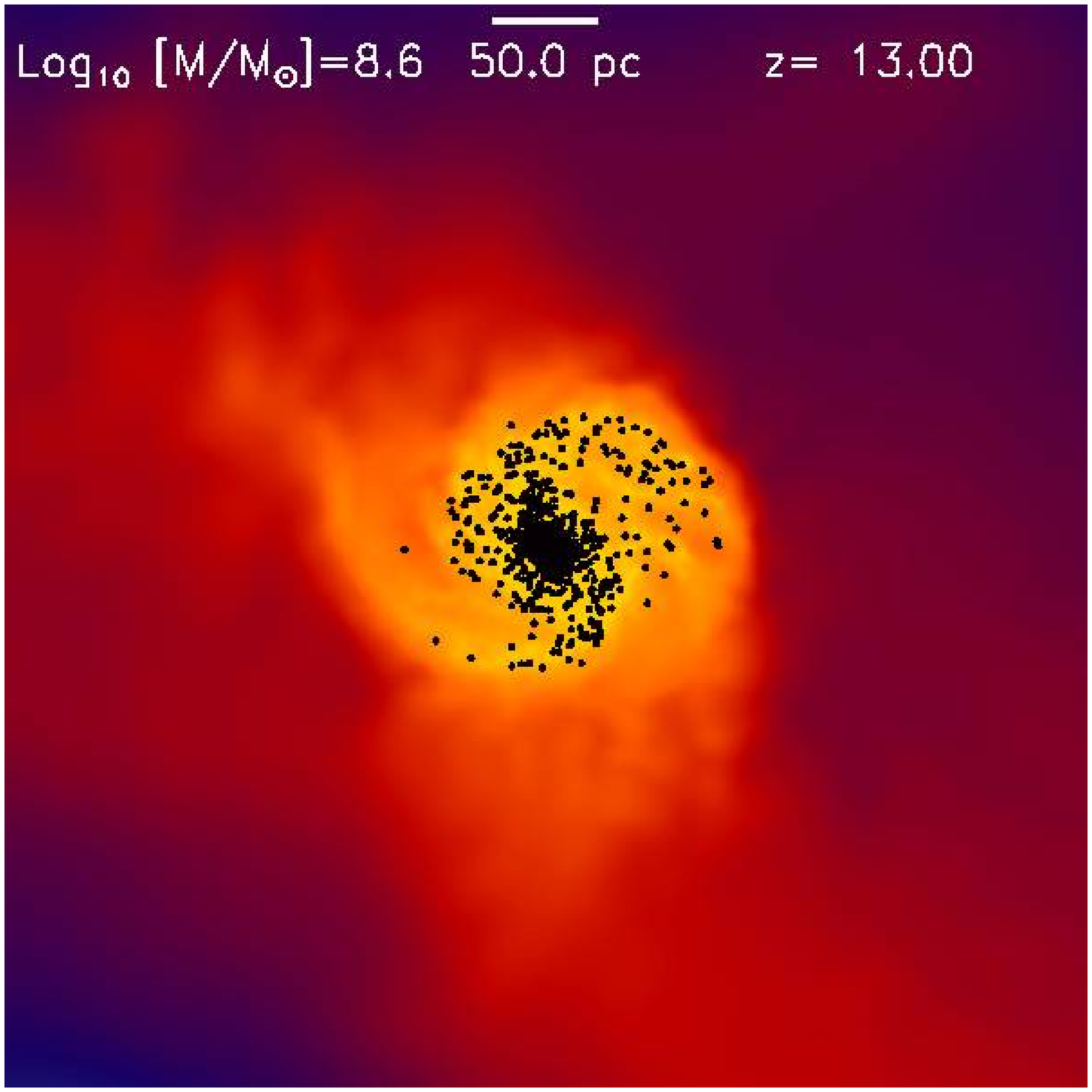}\\
\includegraphics[trim = 0 0 0 0mm, width = 0.24\textwidth]{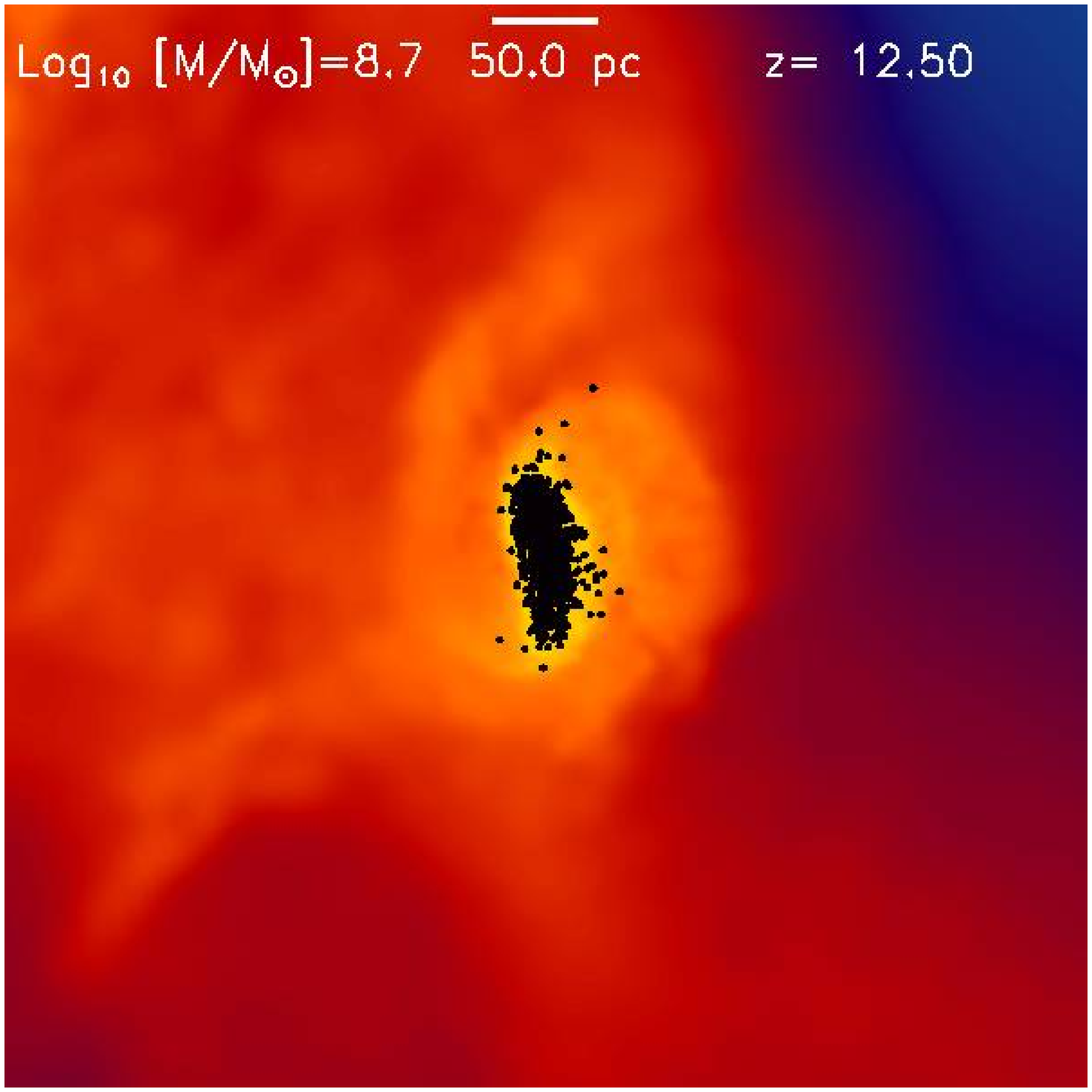}
\includegraphics[trim = 0 0 0 0mm, width = 0.24\textwidth]{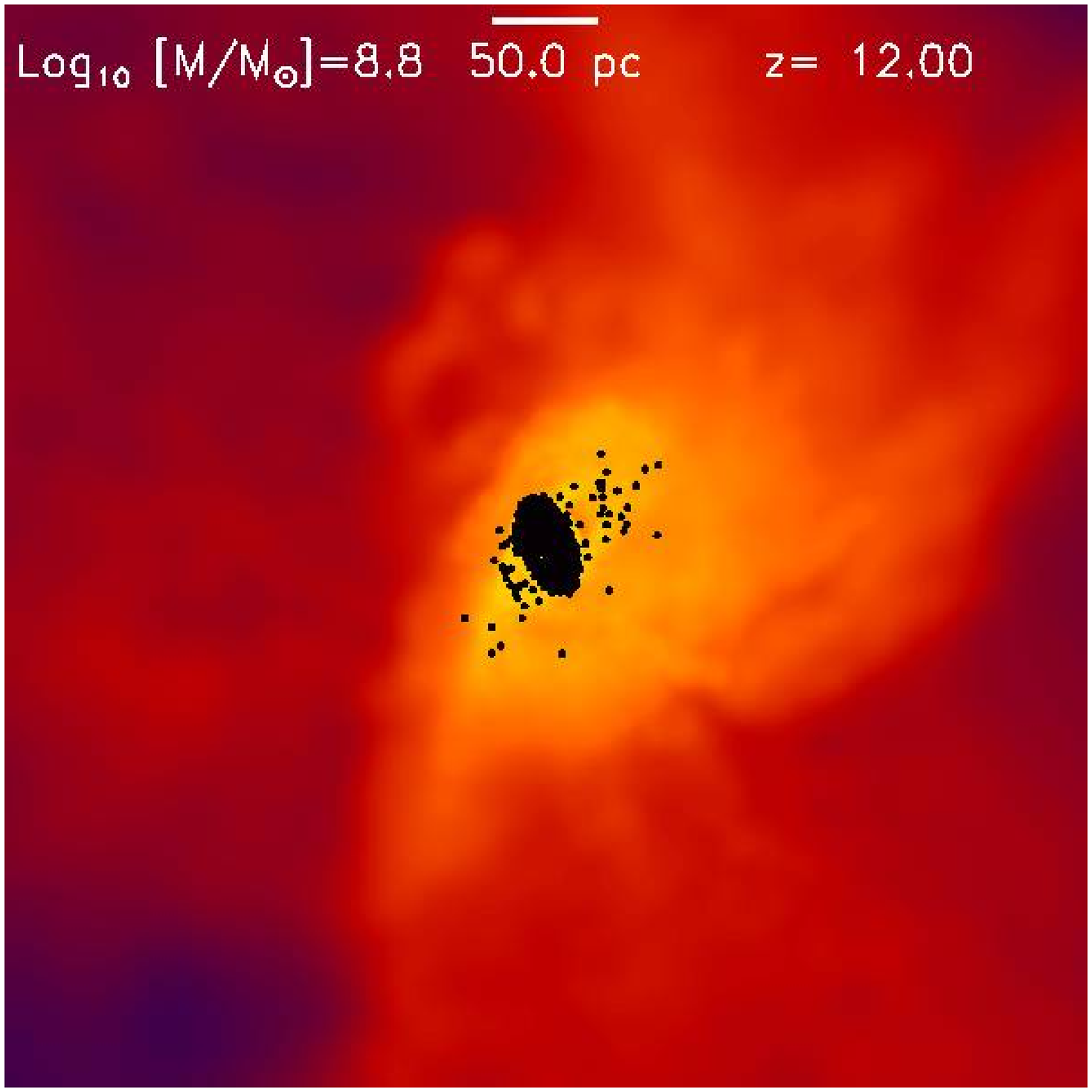}
\includegraphics[trim = 0 0 0 0mm, width = 0.24\textwidth]{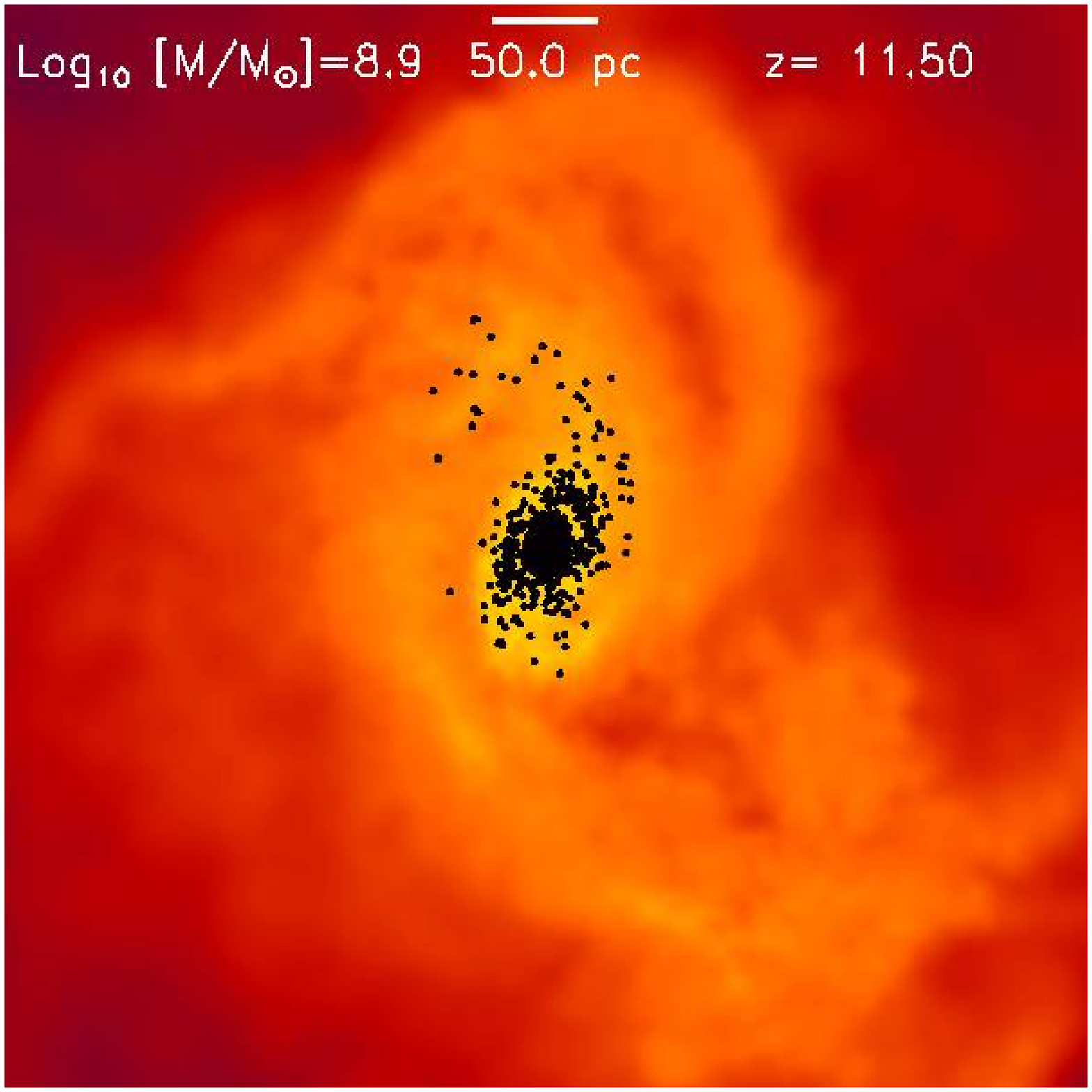}
\includegraphics[trim = 0 0 0 0mm, width = 0.24\textwidth]{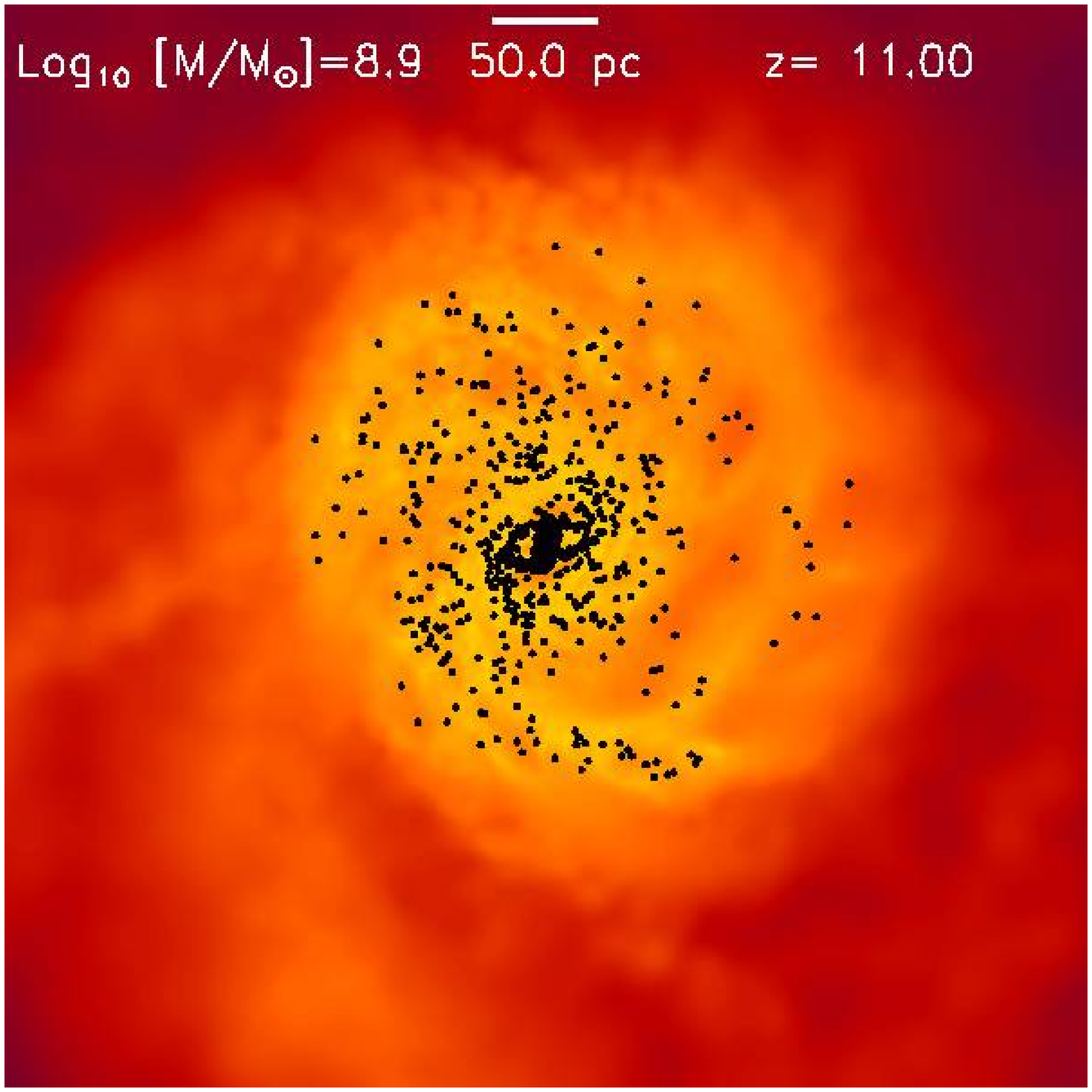}\\
\caption{Assembly history of the disks in simulation {\it LW+RT} which includes both LW and ionizing radiation. The panels present face-on
views of the assembling dwarf galaxy and show the evolution of gas
densities in the redshift range $11.0 \le z \le 14.5$ in cubical
slices of linear size 0.5 kpc. The locations of star particles emitting LW and ionizing
radiation is indicated with black dots. This set of panels can be
directly compared with that in Figure~8 of \cite{Pawlik:2011a}, which
shows the gas densities in a simulation identical to 
simulation {\it LW+RT} presented here except that it did not include star formation or radiation. 
Radiative feedback creates locally confined low-density cavities (best visible in the bottom right panel), 
but it is not sufficiently strong to prevent the assembly of the disks inside the atomically cooling dwarf galaxy. \label{Fig:DisksDensities}}
\end{center}
\end{figure*}

Figure~\ref{Fig:DisksDensities} shows
a sequence of snapshots of the gas density distribution centered on the dwarf galaxy in simulation {\it LW+RT}. At the final simulation
redshift, the gas at the halo center is organized in a massive central, 
spherical clump, an inner compact disk, and an outer extended disk. The panels
can be compared with Figure~8 of \cite{Pawlik:2011a}, which shows the gas
densities in our earlier simulation identical to simulation {\it LW+RT} discussed here except that it did not include star
formation or radiation. The comparison shows that the 
radiative feedback is weak and the 
inclusion of star formation and LW and ionizing radiation 
does not prevent the formation of the disks seen in that earlier simulation. 
\par
The sequence of events leading to the formation of the two disks is
very similar with and without the inclusion of radiation, and the
reader may therefore refer to \cite{Pawlik:2011a} for additional
details. A major merger at redshift $z \lesssim 15$ channels gas in
the halo center. This leads to the formation of the first gaseous disk
by $z = 13.5$. The disk subsequently develops spiral arms. The spiral
arms exert torques that imply an outward transport of angular momentum
in the disk gas. As a consequence, the disk shrinks in size. At $z
\approx 12$, a sequence of minor mergers replenishes the halo center
with gas, producing the second gaseous disk. This
disk remains spatially extended until the end of the simulation. The
second disk surrounds the first disk, and the two have orientations
tilted with respect to each other. This tilt is an interesting
consequence of the hierarchical assembly of the emerging dwarf galaxy
(e.g., \citealp{Roskar:2010}; \citealp{Pawlik:2011a}). The formation
redshifts of the disks are similar in all simulations and
hence insensitive to the inclusion of radiation.

\begin{figure*}
\begin{center}
\includegraphics[trim = 60 0 100 0mm, width = 0.32\textwidth]{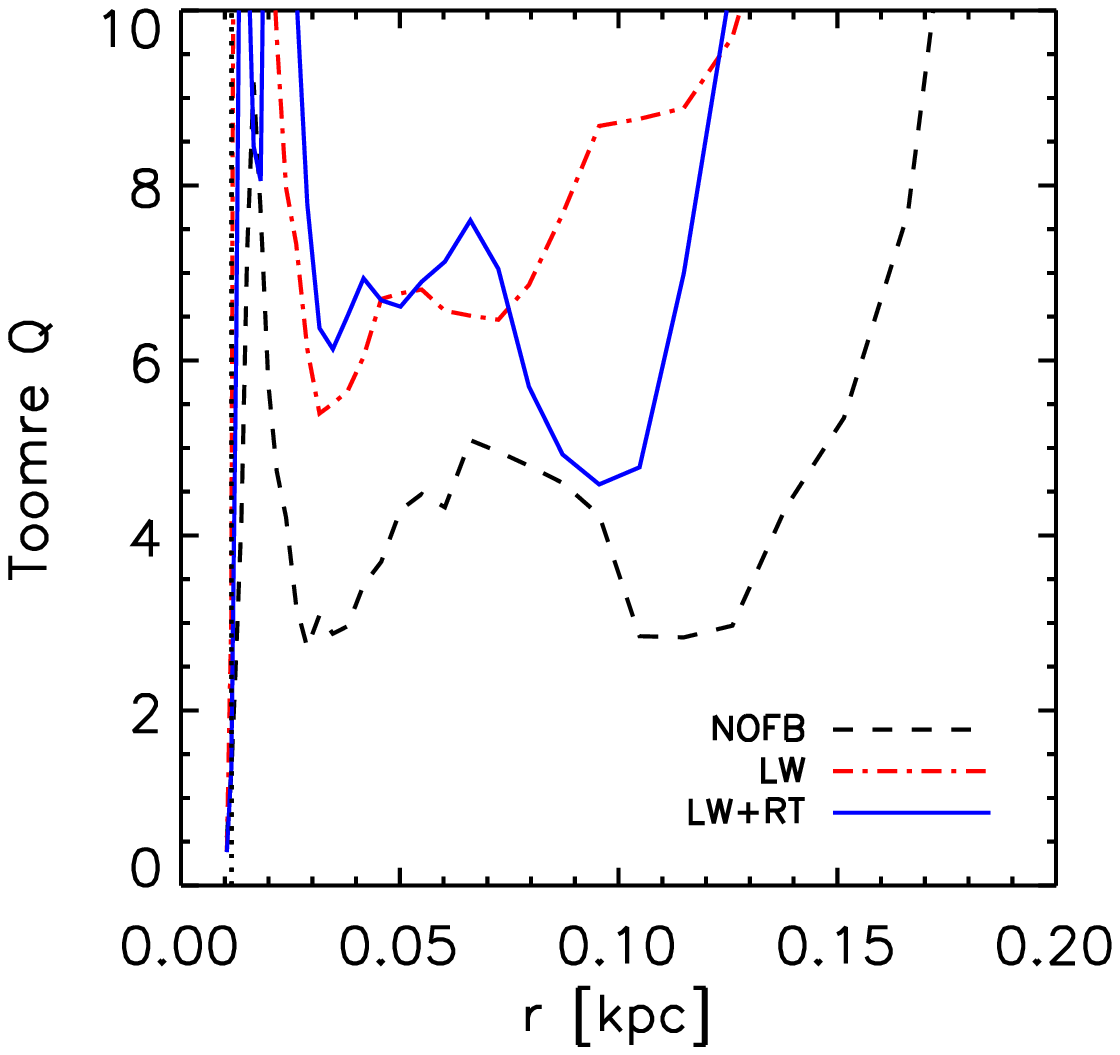}
\includegraphics[trim = 60 0 100 0mm, width = 0.32\textwidth]{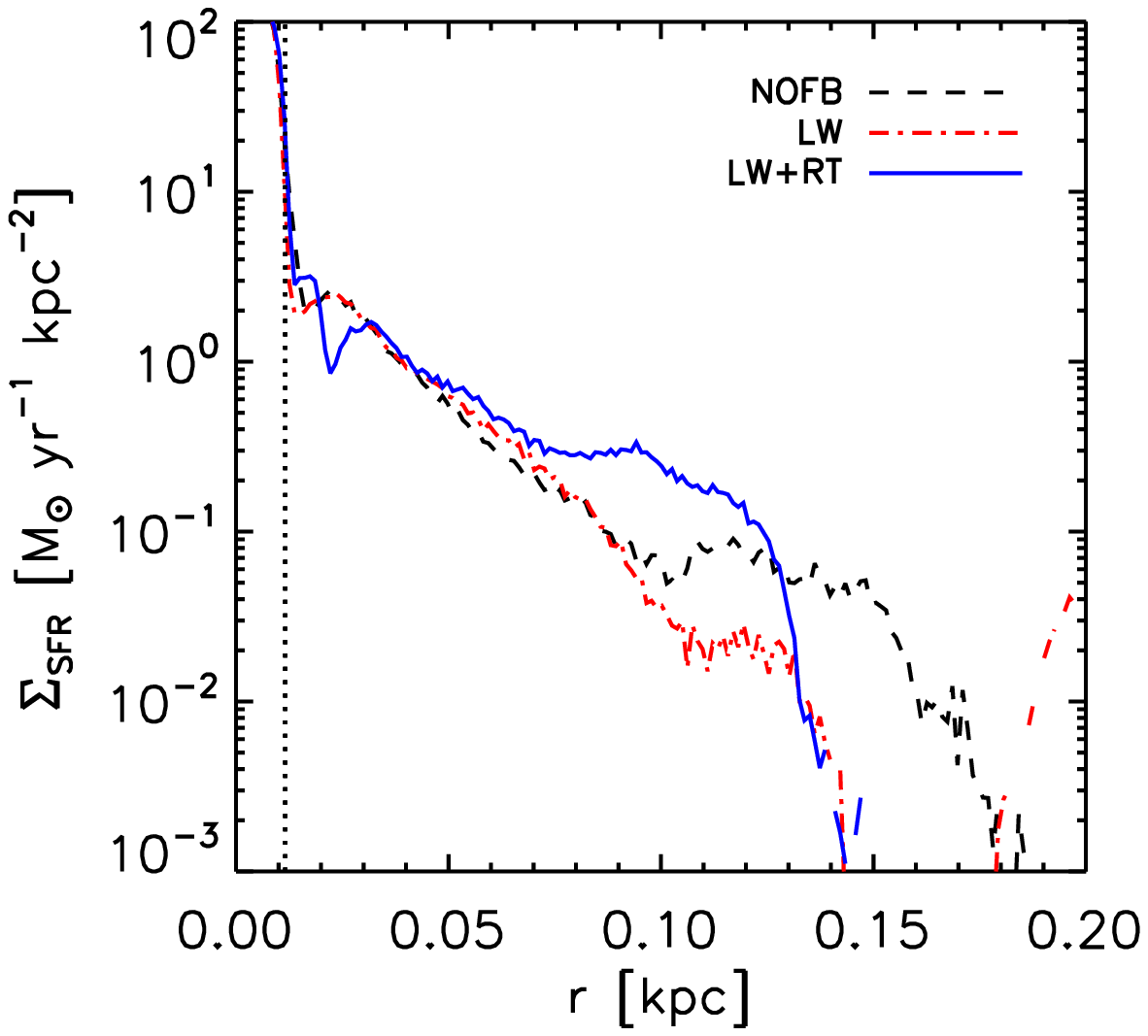}
\includegraphics[trim = 60 0 100 0mm, width = 0.32\textwidth]{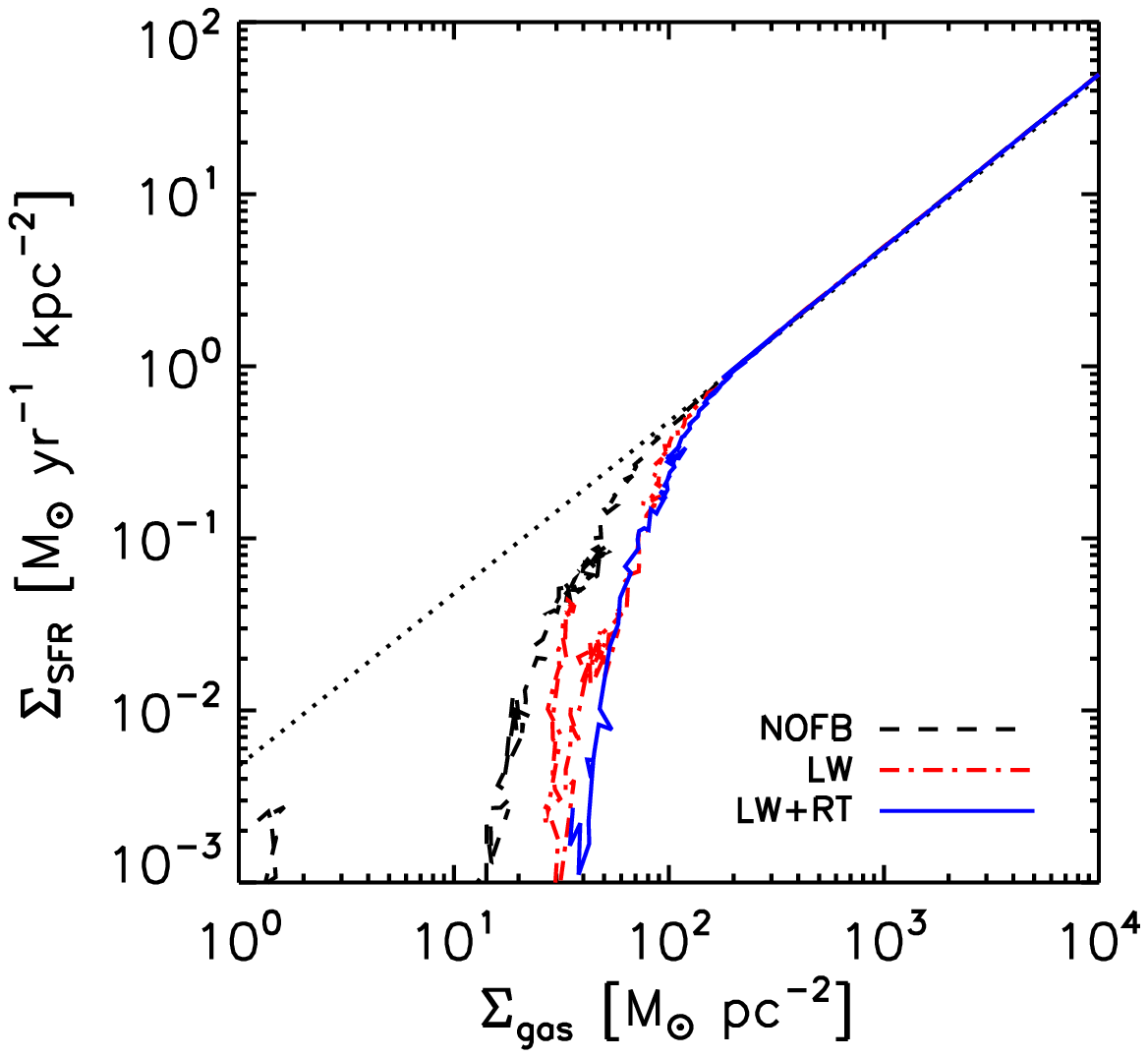}
\end{center}
\caption{Azimuthally averaged properties of the disks at $z = 11$ in the three
 simulations without radiation ({\it NOFB}; black dashed curves), LW radiation
 ({\it LW}; red dash-dotted curves), and LW and ionizing radiation ({\it
 LW+RT}; blue solid curves). {\it Left:}  Toomre $Q$ parameter of the disk gas. A value $Q\gtrsim 1$ implies stability against gravitational fragmentation. 
{\it Middle:} 
 SFR surface densities $\Sigma_{\rm SFR}$
 as a function of distance $r$ from the most bound particle. 
 {\it Right:}
 Star formation
 rate surface densities $\Sigma_{\rm SFR}$ as a function of gas
 surface densities $\Sigma_{\rm gas}$. The
 dotted line shows the relation expected from the star formation recipe,
 $\Sigma_{\rm SFR}= \tau_\star^{-1} \Sigma_{\rm gas}$. The vertical dotted lines
 in the left and middle panels mark the softening scale $\epsilon$.}
\label{Fig:DiskProfiles}
\end{figure*}

\par 
Photoheating creates low-density regions in the
disk gas, leaving behind a disk morphology more complex than 
in the absence of photoheating (compare, e.g., 
the bottom right panel of Figure~\ref{Fig:DisksDensities} with the corresponding panel in Figure~8 of \citealp{Pawlik:2011a}). These
regions remain, however, locally confined, and the disks remain, as a whole, intact and 
relatively unaffected by photoheating. This insensitivity of the gas to photoheating is
consistent with the fact that at the time of formation of the disks,
i.e., at $z \lesssim 15$, the galaxy halo has already entered the atomic cooling regime. 
Hence, its gravitational potential is sufficiently deep
to confine photoheated gas in its interior. A positive feedback loop
which consists of high gas densities confining the photoheated \ion{H}{2} regions,
allows the gas to collapse to increasingly higher densities, further
confining the photoheated gas. 
\par
The left panel of Figure~\ref{Fig:DiskProfiles} shows the 
Toomre parameter, averaged in annular bins, of the disk gas at the final simulation redshift, $Q = c_{\rm s} \kappa/ (\pi G \Sigma)$,
where $c_{\rm s}$ is the adiabatic sound speed, $\kappa = (4 \Omega^2 +
r d\Omega^2/dr)^{1/2}$ the epicyclic frequency, and $\Omega$ 
the angular velocity of the disk gas (\citealp{Toomre:1964}). In all three simulations $Q \gtrsim 1$ at radii
larger than the gravitational softening length, implying that the disk
configuration is stable against fragmentation. Note that the inclusion of LW and
ionizing radiation increases disk stability by increasing the sound
speed. The inclusion of ionizing radiation further helps to preserve the disks because
photoheating evaporates the gas from the low mass halos merging with
the dwarf galaxy. In the absence of feedback, on the other hand,
baryon-rich low mass halos that pass through the disks can disturb the
disks significantly (e.g., Figure~3 in \citealp{Pawlik:2011a}).
\par
The middle and right panels of Figure~\ref{Fig:DiskProfiles} show spherically averaged profiles of
the SFR surface densities (middle), and the Kennicutt-Schmidt relation
(right), i.e., the relation between SFR surface density and gas
surface density, at the final simulation redshift. The Kennicutt-Schmidt relation is characterized by a
power-law behavior and a suppression of star formation below gas
surface densities $\lesssim 10^2 \Msunpcinvsq$. The suppression is a
result of the star formation recipe, which limits star formation to
densities above $500 \cmci$. At the final simulation redshift, such
densities are realized both in the central region and in the disks,
and the galaxy shows a spatially extended morphology of star
formation. In contrast, star formation in the galaxy
before disk formation is limited to the central region (see Figure~\ref{Fig:DisksDensities}). The Kennicutt-Schmidt
power law behavior can be understood by writing $\Sigma_{\rm SFR} =
\tau_{\rm gas}^{-1}\Sigma_{\rm gas}$, where $\tau_{\rm gas} = \rho/
\dot{\rho}_{\star}$ is the gas consumption time. Using $\tau_{\rm gas}
= \tau_{\star}$ (Equation~\ref{Eq:StarFormationTimeScale}), we find,
$\Sigma_{\rm SFR}= \tau_\star^{-1} \Sigma_{\rm gas}$. This relation
is shown with the dotted line in Figure~\ref{Fig:DiskProfiles}, and
the results from the simulation are in close agreement with it. 
\par

\begin{figure*}
\begin{center}
  \includegraphics[trim = 0 0 0 0mm, width = 0.24\textwidth]{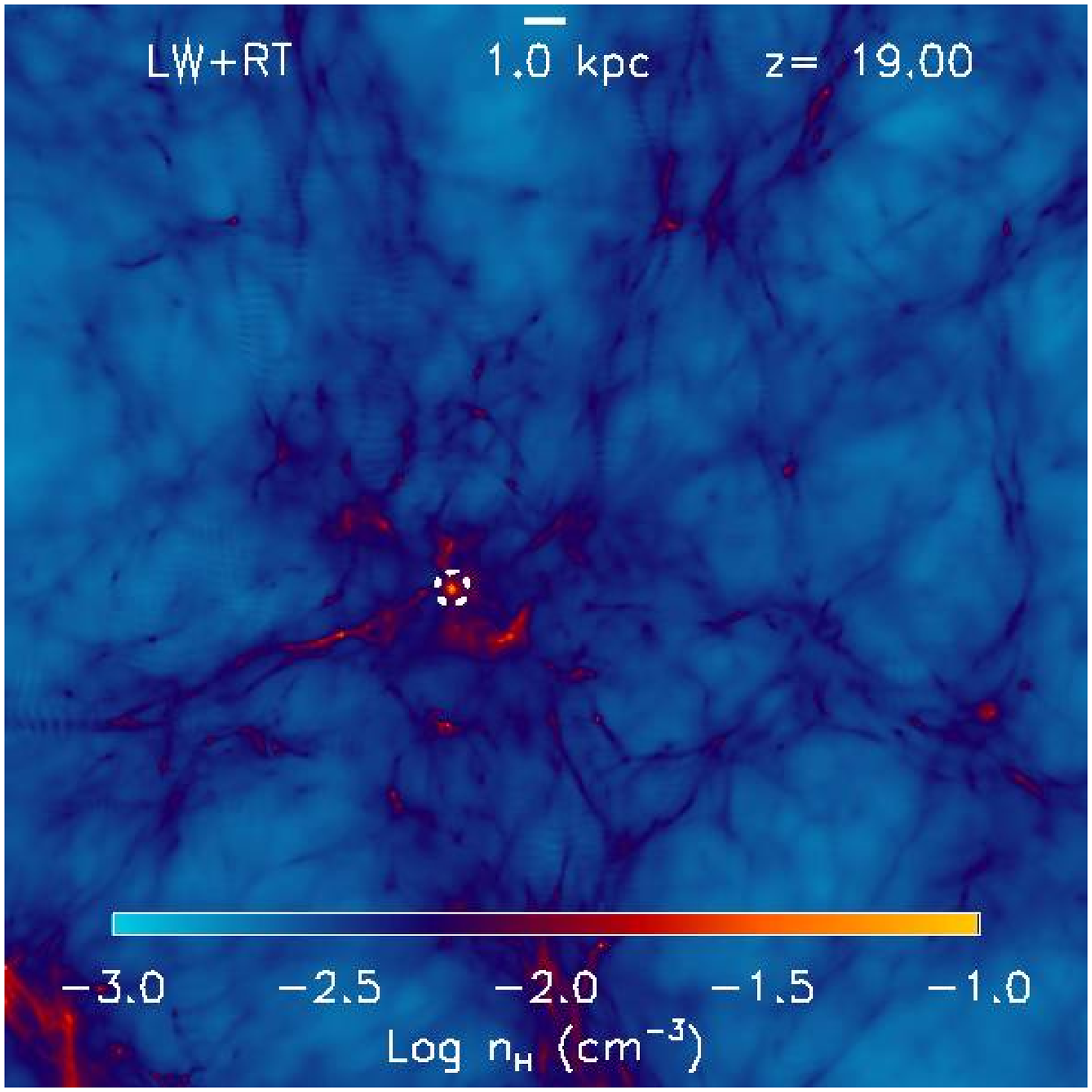}
  \includegraphics[trim = 0 0 0 0mm, width = 0.24\textwidth]{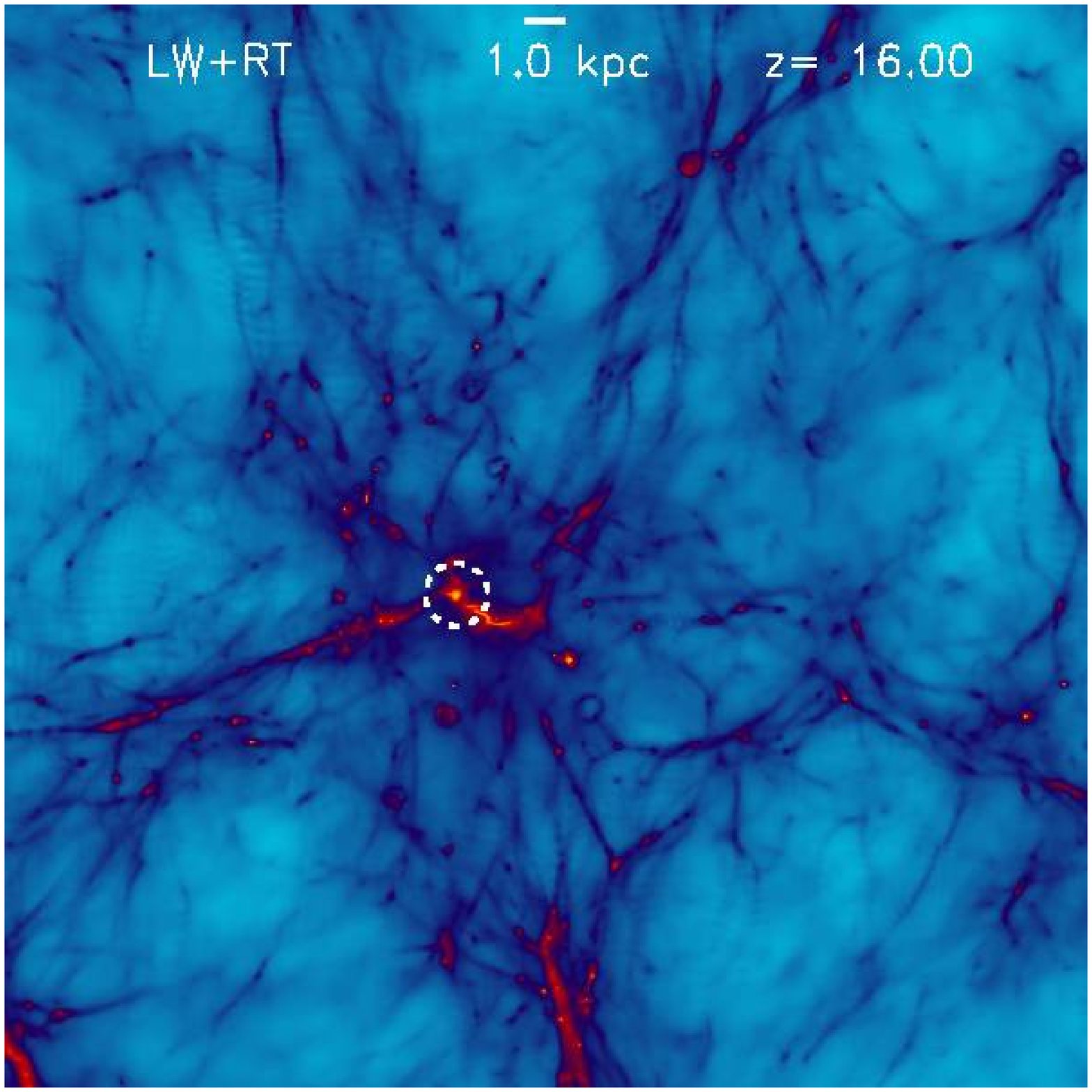}
 \includegraphics[trim = 0 0 0 0mm, width = 0.24\textwidth]{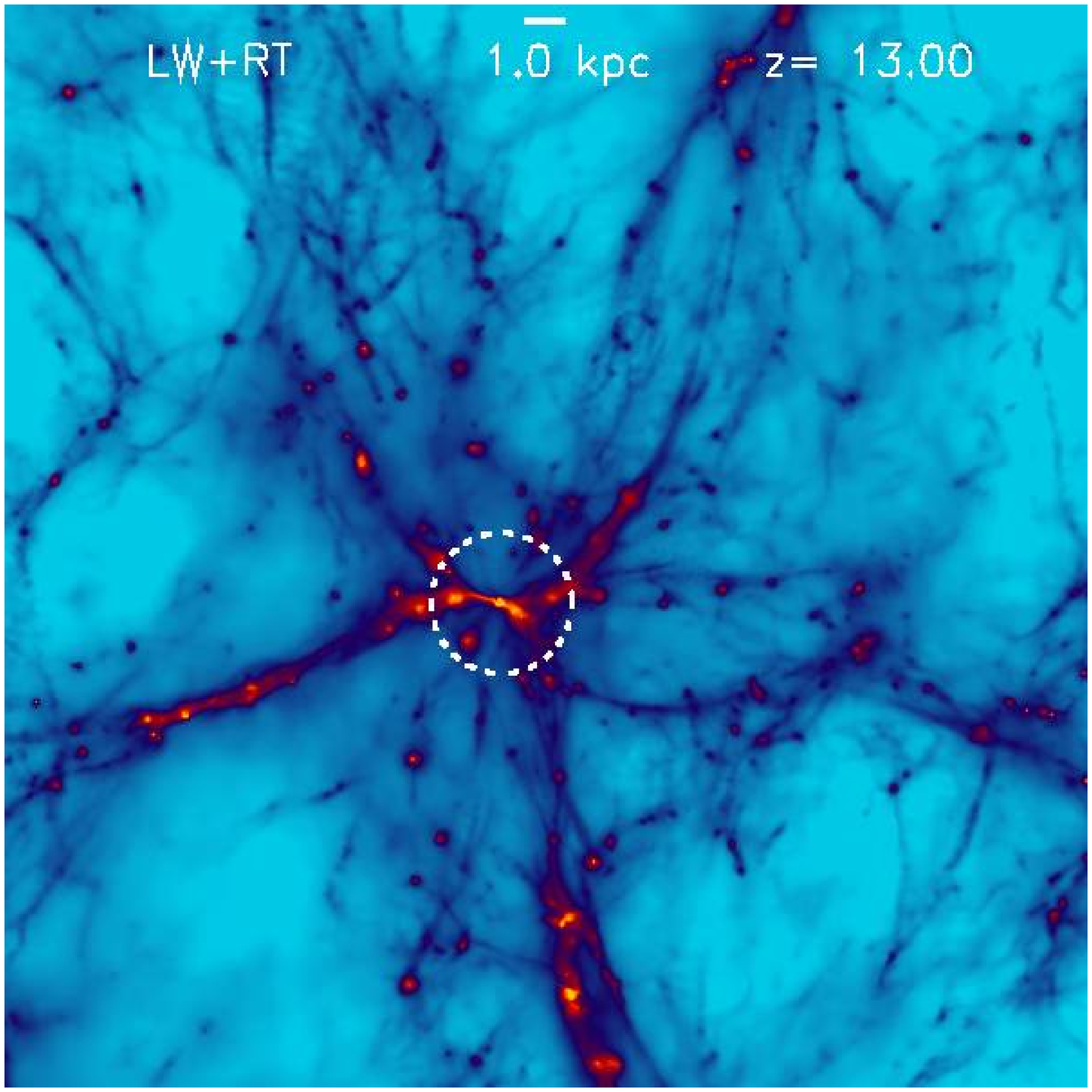}
  \includegraphics[trim = 0 0 0 0mm, width = 0.24\textwidth]{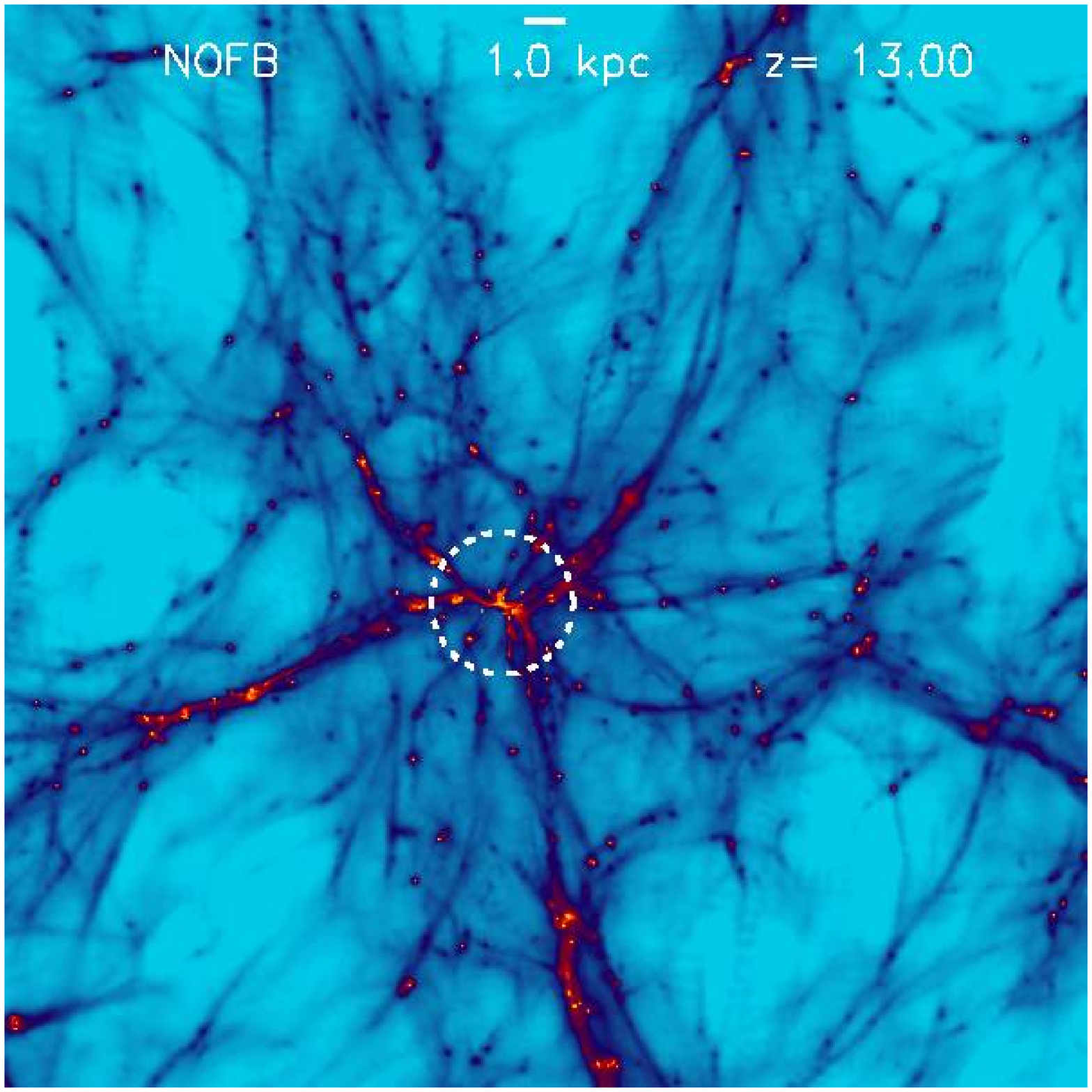}\\
  \includegraphics[trim = 0 0 0 0mm, width = 0.24\textwidth]{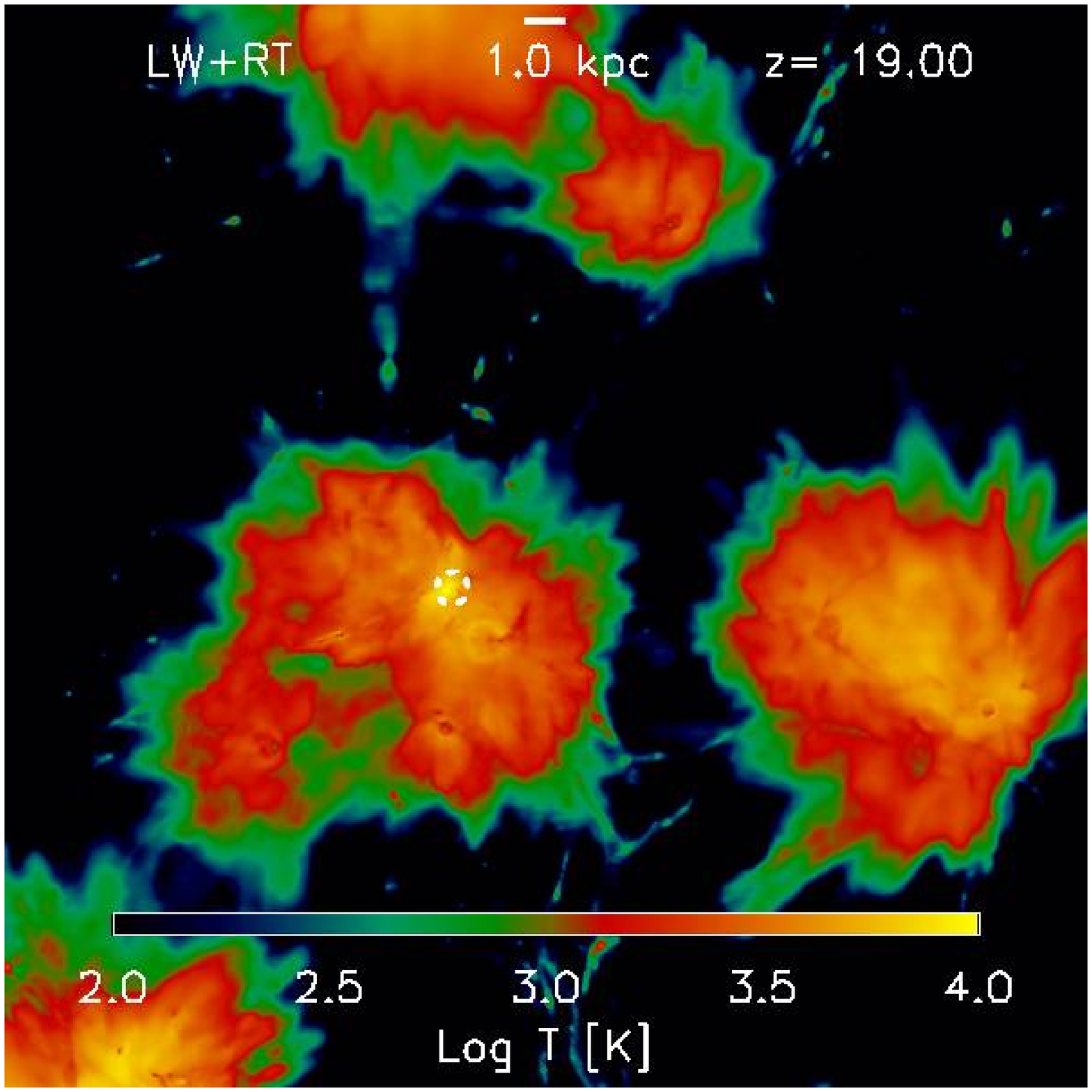}
  \includegraphics[trim = 0 0 0 0mm, width = 0.24\textwidth]{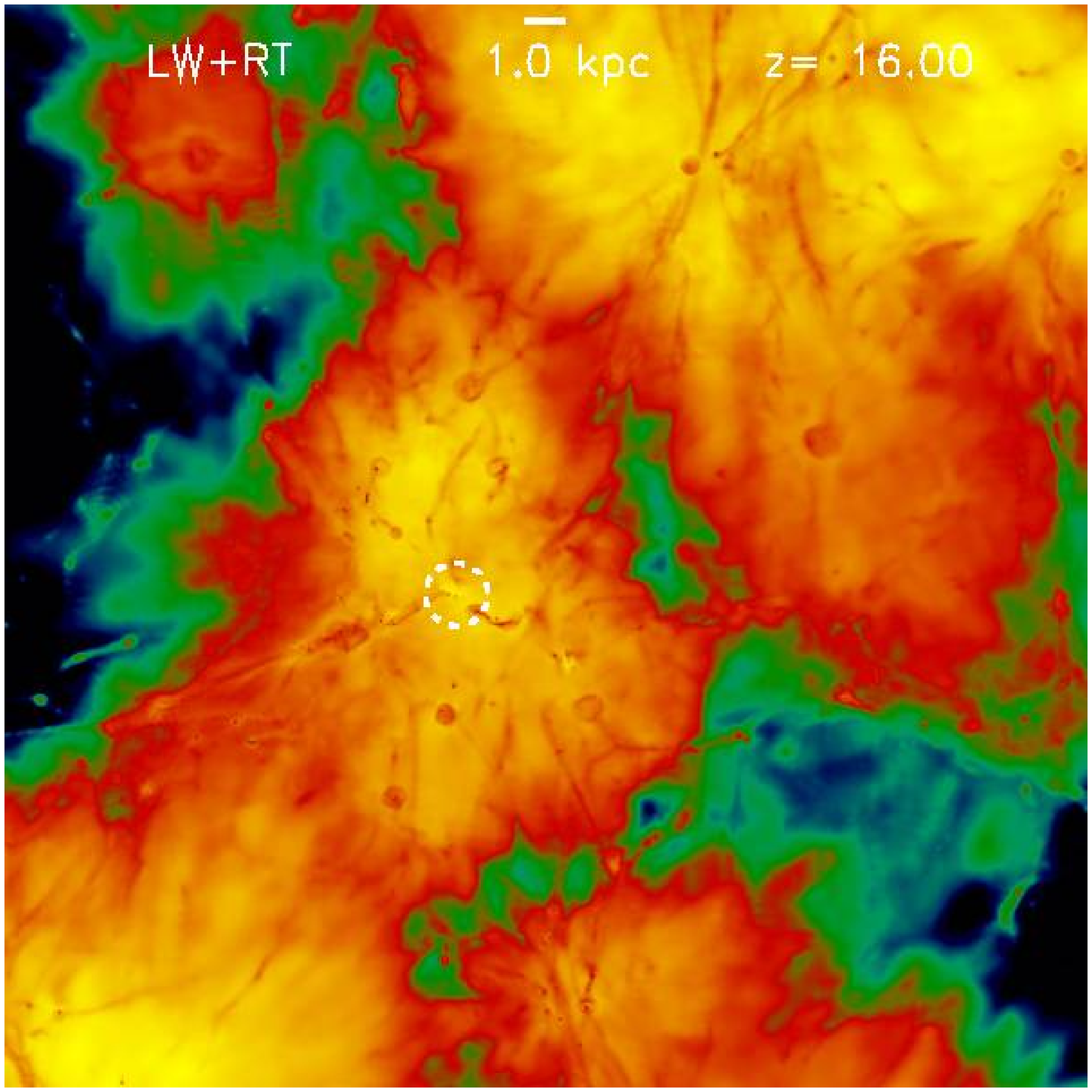}
 \includegraphics[trim = 0 0 0 0mm, width = 0.24\textwidth]{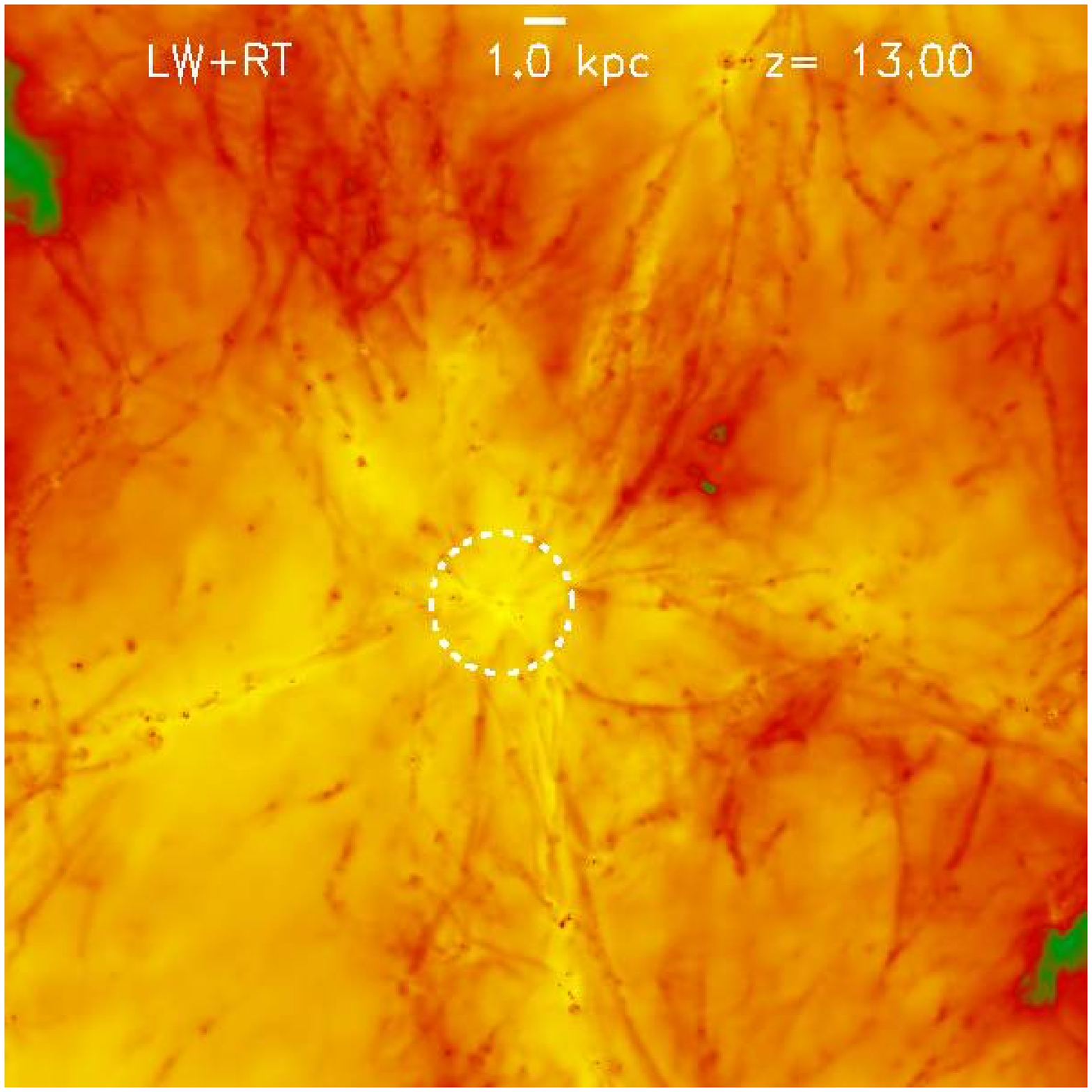}
  \includegraphics[trim = 0 0 0 0mm, width = 0.24\textwidth]{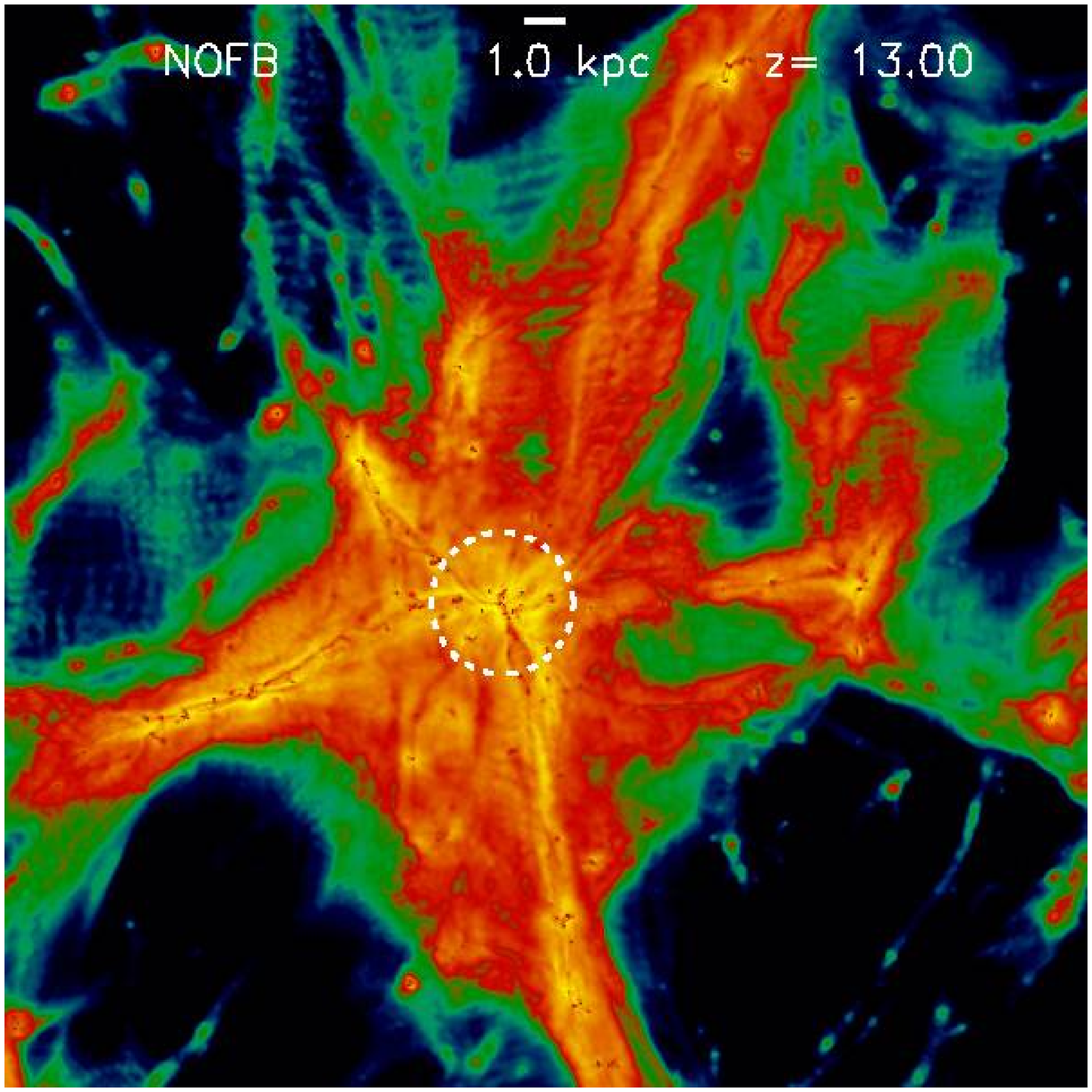}
  \caption{Evolution of gas densities (top) and temperatures (bottom)
    in simulation {\it LW+RT} at three representative redshifts $z = 19, 16$, and $13$ (from left to
    right). For comparison, the right-most panels show the gas
    densities and temperatures in simulation {\it NOFB} at $z =
    13$. The cubical slices of fixed physical dimensions are centered on the comoving position of the most-bound 
    particle in the dwarf galaxy at $z = 10$. The dashed circles are centered on the dwarf galaxy progenitor and have a radius equal to the virial radius of the progenitor. Photoheating raises the gas temperature to $\sim 10^4
    \K$. The implied increase in gas pressure evaporates gas from
    inside low-mass halos and smooths gas density fluctuations in the
    IGM.} \label{Fig:images}
\end{center}
\end{figure*}

Taken at face value, the large Toomre Q values that indicate stability
against disk fragmentation seem inconsistent with the nonzero star
formation rates of the disks. However, our simulations do not have
sufficient resolution or the physical detail required to study
fragmentation of the disks into individual stars. We have therefore
employed a phenomenological model for star formation according to
which stars form from gas with densities larger than the adopted
threshold density for star formation, an approach employed in most
galaxy formation simulations. It remains open if, at higher resolution
or greater physical detail, the disks in our simulation would fragment
to form stars, or if they would remain stable, possibly feeding a
central massive black hole (e.g., \citealp{Eisenstein:1995};
\citealp{Koushiappas:2004}; \citealp{Lodato:2006}). Addressing these
issues in cosmological simulations such as the simulations here is computationally
challenging. Simulations of isolated disk galaxies have shown that
star formation is slower in disks that are more stable as quantified 
by the smallest Toomre Q value in the disk (\citealp{Li:2005}). 
It may therefore be that star formation in
high-redshift low-mass disk galaxies is less efficient than implied by our
simulations.

\section{Reionization and Radiative Feedback from the First Stars}
\label{Sec:Reionization}
\par
Figure~\ref{Fig:images} shows projections of the gas density (top) and
temperature (bottom) in cubical slices through the refinement region
in simulation {\it LW+RT} at three representative redshifts $z = 19,
16$, and $13$ (from left to right). For comparison, the figure also
shows the gas density and temperature at $z = 13$ in the corresponding
slice through simulation {\it NOFB} (right-most panels). In the following, 
we discuss the impact of LW and ionizing radiation on
the properties of the IGM and the formation of low-mass galaxies 
in the neighborhood of the simulated dwarf galaxy.
\par
The feedback processes exerted by radiation are
well-known and are briefly summarized here. LW radiation
photodissociates molecular hydrogen, reducing the ability of low-mass
halos to condense their gas. This suppresses star formation,
providing a negative feedback (e.g., \citealp{Haiman:1997}). Ionizing
radiation photoheats the IGM to $\sim 10^4 \K$, and the implied
Jeans filtering impedes the accretion of gas into halos with
virial temperatures $\lesssim 10^4 \K$ (\citealp{Shapiro:1994}; \citealp{Gnedin:1998}). In
addition, photoheating evaporates gas from low-mass halos and
reduces the ability of gas to cool inside them (e.g.,
\citealp{Efstathiou:1992}; \citealp{Thoul:1996}; \citealp{Wiersma:2009}). On the other hand,
photoionization generates free electrons, which catalyze the formation
of molecular hydrogen, thus increasing the ability of gas inside low-mass
halos to cool and form stars (e.g., \citealp{Ricotti:2002};
\citealp{Oh:2002}). Photoionization therefore provides both a
negative and a positive feedback on star formation. A comprehensive
overview of the effects of radiative feedback can be found in, e.g., \cite{Ciardi:2005}.
\par
\subsection{The Start of Reionization}
\par
\begin{figure*}
\begin{center}
\includegraphics[trim = 0 0 0 0mm, width = 0.49\textwidth]{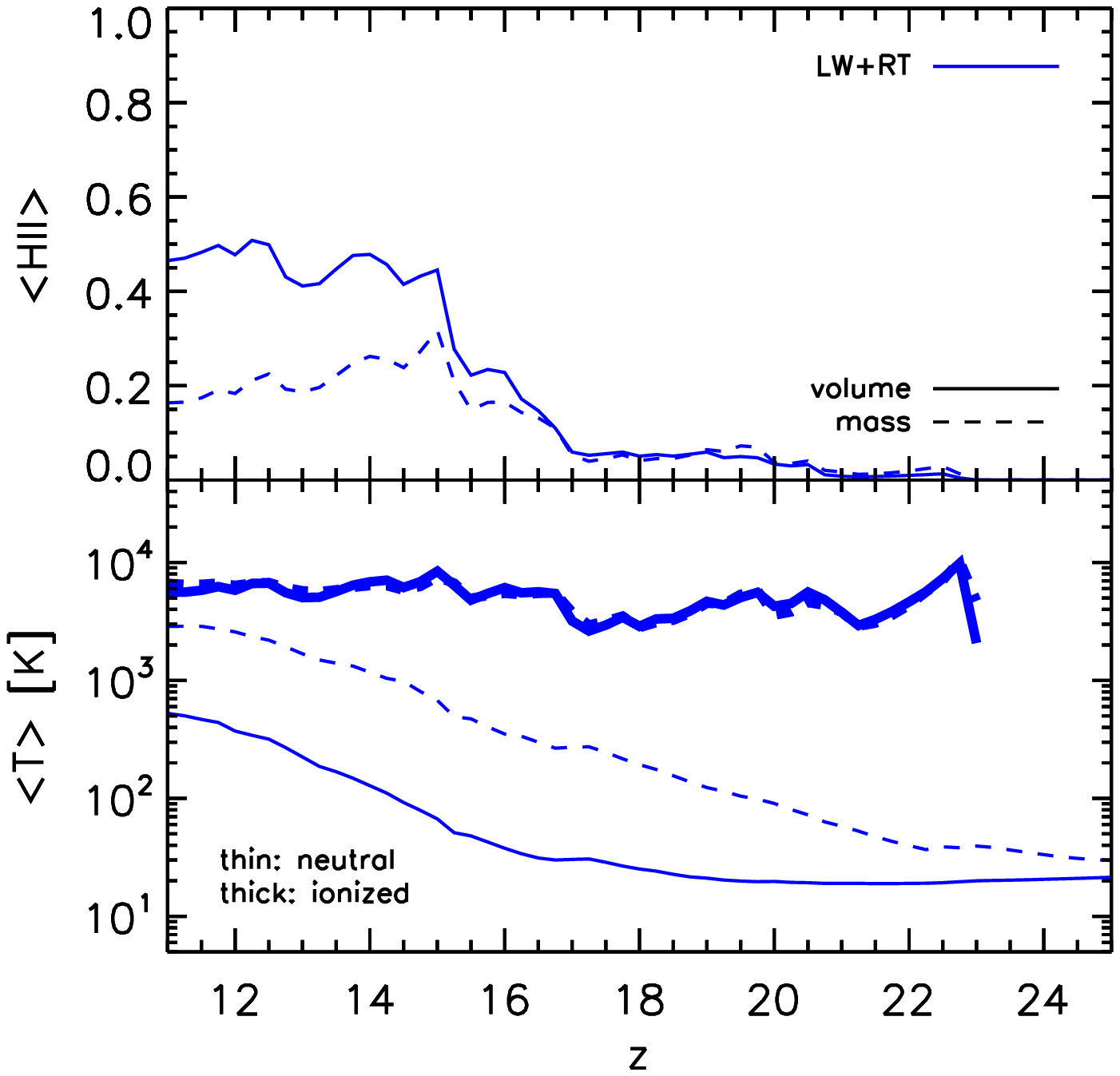}
\includegraphics[trim = 0 0 0 0mm, width = 0.49\textwidth]{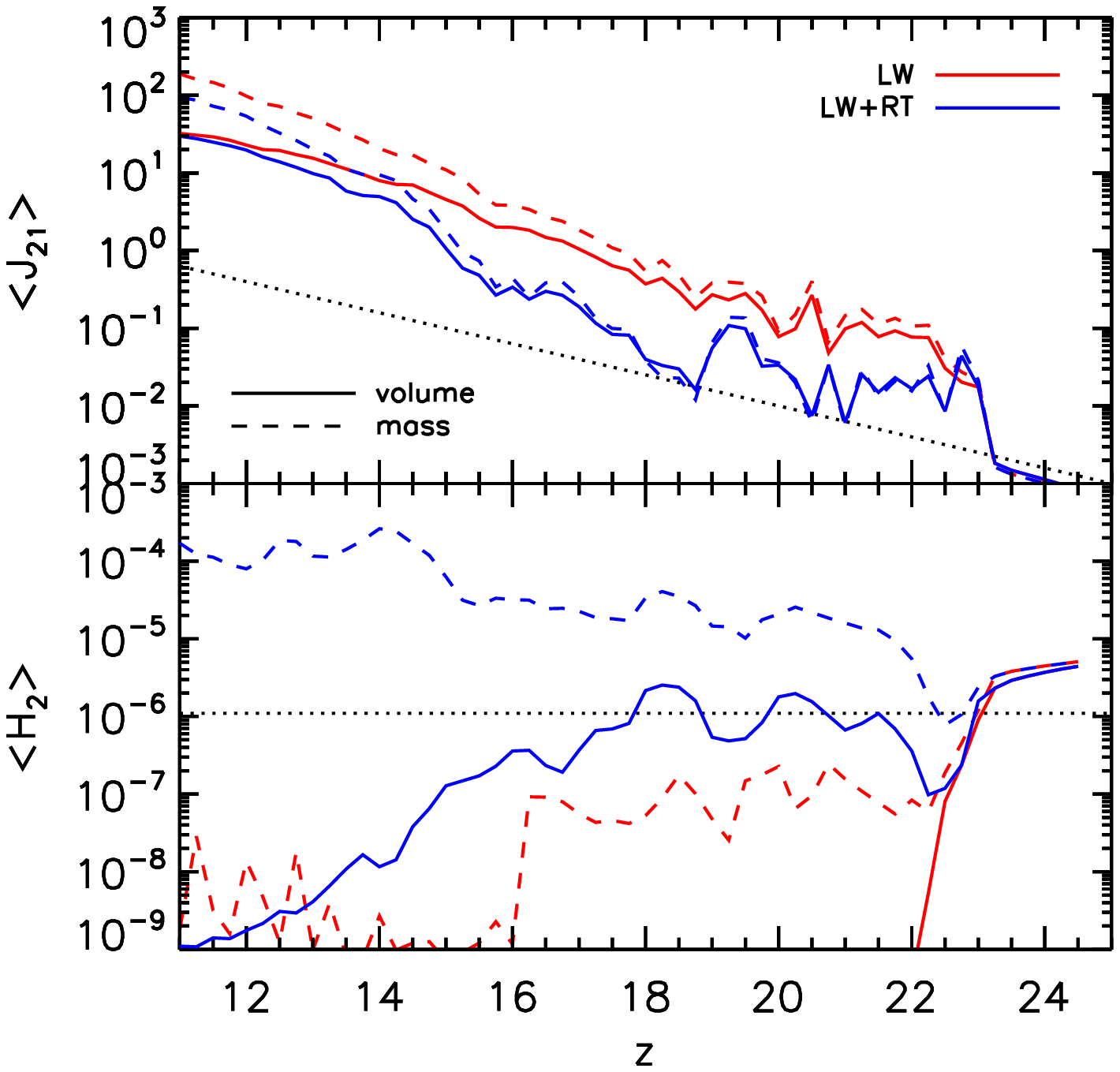}
\end{center}
\caption{Reionization history of the immediate neighborhood of the dwarf
galaxy, defined by the sphere of comoving radius $\approx 100 \kpc$ centered on the comoving position of the dwarf galaxy at $z = 10$, 
the same position used to center the images in Figure~\ref{Fig:images}. In all panels, solid curves show volume-weighted averages, and dashed curves
show mass-weighted averages. {\it Top left:} Mean ionized fractions
$\langle \eta_{\rm HII} \rangle$. {\it Bottom left:} Mean gas temperatures
$\langle T \rangle$. Thick (thin) lines show average temperatures including only gas that is ionized (neutral), 
i.e., gas with $\eta_{\rm HII} > 10^{-3}$ ($\eta_{\rm HII} \le 10^{-3}$). {\it Top right:} Mean normalized LW intensities $
\langle J_{21} \rangle$. The average was computed in log-space to reduce the dynamic range, 
i.e. $\langle J_{21} \rangle \equiv 10^{\langle \log_{10} J_{21} \rangle}$. The dotted line shows the intensity of
the imposed LW background. {\it Bottom right:} Mean molecular hydrogen
fractions $\langle \eta_{\rm H_2} \rangle$. The dotted horizontal line marks the
initial molecular hydrogen fraction, $\eta_{\rm H_2} = 1.1\times
10^{-6}$. The dwarf galaxy highly but not fully reionizes the gas in its overdense neighborhood by the final simulation redshift, $z = 11$.}
\label{Fig:Start}
\end{figure*}

Figure~\ref{Fig:Start} quantifies the ionization and thermal history
(left) as well as the evolution of LW intensities and of the molecular
hydrogen fraction (right) around the dwarf galaxy in simulation {\it
LW+RT} (blue curves). The neighborhood considered here is defined,
for simplicity of geometry and to reduce boundary artifacts, as the 
sphere of comoving radius $5 r_{\rm vir, com} (z=10)$, where $r_{\rm vir, com} (z=10) \approx 34
\ckpc$ is the virial radius of the dwarf galaxy at $z = 10$. We center
the sphere on the comoving position of the most-bound particle of this
galaxy at $z = 10$, and this is the same position on which the slices
in Figure~\ref{Fig:images} are centered. The spherical neighborhood
is contained in the refinement region at all redshifts. Our qualitative discussion below is not sensitive to the
precise definition of the neighborhood adopted here. Volume-weighted (mass-weighted) 
averages are obtained by weighting with the the SPH 
kernel volume (particle mass).
\par
The fraction of the volume ionized (solid) increases with decreasing
redshift until $z \lesssim 15$, after which it remains approximately
constant at $\langle \eta_{\rm HII}\rangle_{V} \approx 0.5$. The fraction 
of the mass ionized (dashed) shows a qualitatively similar behavior, but it slightly decreases after
$z \lesssim 15$ and reaches $\langle \eta_{\rm HII} \rangle_{m} \approx 0.2$ at the final
simulation redshift. The redshift below which the
ionized fractions no longer increase is similar to the redshift at which
the galaxy evolves into an atomically cooling object. The fraction of the mass ionized is
initially close to but slightly larger than the fraction of the volume
ionized, but this changes at $z \lesssim 19$ after which the fraction
of ionized mass becomes increasingly lower than the fraction of the
ionized volume. Reionization thus proceeds from the inside-out, from the dense
halo gas into the diffuse IGM, as expected (e.g., \citealp{Wise:2008a}). The mass- and volume-weighted average gas temperature 
of the ionized gas, which we have defined here as gas with ionized hydrogen fraction $\eta_{\rm HII} > 10^{-3}$, 
fluctuates between $3000$ and $10000 \K$, in good agreement with the results in Figures~12 and 13 in \cite{Wise:2008a}.
\par
Figure~\ref{Fig:Start} shows that the ionization of the region around
the main galaxy in simulation {\it LW+RT} is accompanied by an increase in the average LW
intensities. Before the formation of the first
star, the average LW intensity is close to but slightly
smaller than the intensity of the imposed LW background (dotted line), 
a consequence of self-shielding. Emission of
stellar radiation increases the LW intensity above the
background, and the latter quickly becomes unimportant. At $z \lesssim 16$ the
LW intensities increase more rapidly and approach the intensities
found in simulation {\it LW} that only included LW radiation (red curves), and the LW
radiation efficiently destroys the molecular hydrogen. However, the mass-weighted fraction 
remains significantly larger than the volume-weighted fraction, mostly because of
self-shielding. The comparison with simulation {\it LW} shows that the
inclusion of ionizing radiation promotes the formation of molecular
hydrogen (e.g., \citealp{Haiman:1996}; \citealp{Ricotti:2001}; \citealp{Oh:2002}).
\par
The dwarf galaxy highly but not fully ionizes its neighborhood by $z = 11$. This is
likely due to a combination of reasons. First, as the galaxy evolves into an
atomically cooling object, photoheating is no longer able to 
substantially reduce the gas densities inside it (see Section~\ref{Fig:AssemblyHistory}). 
Ionizing photons emitted by the stellar sources are then efficiently consumed by 
recombinations in the dense gas, thus limiting the ionizing impact on the
surrounding IGM. Second, as we will discuss in 
Section~\ref{Sec:Feedback} below, star formation in the dwarf galaxy 
and its neighboring galaxies exerts a strong negative feedback 
on neighboring low-mass galaxies, suppressing star formation 
inside them. These galaxies thus cannot significantly contribute to reionizing the gas. Finally, the
considered volume lacks more massive galaxies, which would be robust
against the feedback from the dwarf galaxy and could potentially help 
reionizing the gas. This is an artifact of the refinement region being too small 
to contain these rarer halos. Note also that the galaxy resides in a 
highly overdense region, and our computation of the mean ionized fraction 
includes contributions from the self-shielded neutral gas in halos inside this region.
\subsection{Feedback on Galaxy Formation}
\label{Sec:Feedback}

\begin{figure}
\begin{center}
\includegraphics[trim = 0 20 50 20mm, width = 0.49\textwidth]{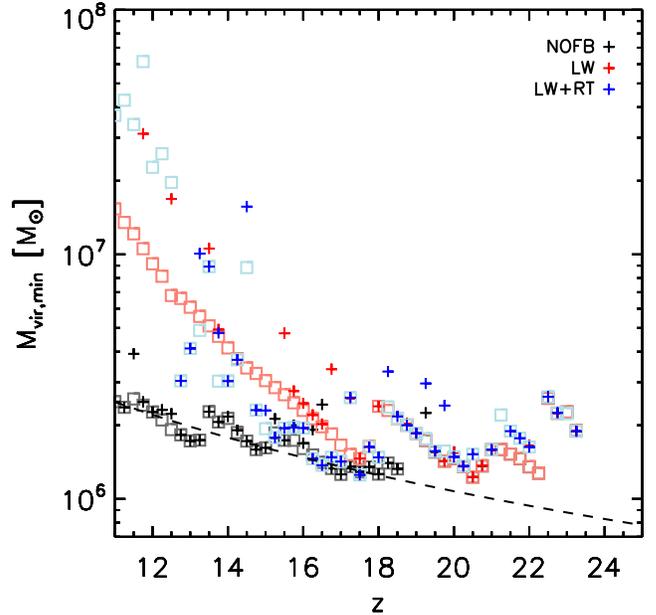}
\end{center}
\caption{Minimum collapse mass, i.e., the mass of the lowest mass halo
 inside the refinement region that forms stars for the first time
 (crosses). The minimum collapse mass is shown both as inferred from 
simulation {\it LW+RT} (blue), which included both 
dissociating and ionizing radiation, and,
 for comparison, in the simulations without radiation ({\it NOFB};
 black), and dissociating radiation only ({\it
 LW}; red). We also show the virial mass of the lowest mass halo
 that forms stars, including halos that have previously hosted star
 formation, in the same three simulations (squares with matching but
 lighter colors). The dashed curve shows a virial mass corresponding
 to a virial temperature $T_{\rm vir} = 2200 \K$
 (Equation~\ref{Eq:VirialTemp} with $\mu = 1.22$), which provides a good
 description of the evolution of the collapse mass in simulation {\it NOFB}. 
 The minimum collapse mass is raised in simulation {\it LW} due to 
 photodissociation of molecular hydrogen by LW photons and in simulation {\it RT+LW} additionally due to
 the Jeans filtering of the IGM. In the latter simulation, the enhanced molecular hydrogen fraction
 in fossil \ion{H}{2} regions can lead to a reduction in the minimum collapse mass below that in simulation {\it LW}.}
\label{Figure:MinimumMass}
\end{figure}

\begin{figure*}
\begin{center}
\includegraphics[trim = 30 0 0 0mm, width = 0.32\textwidth]{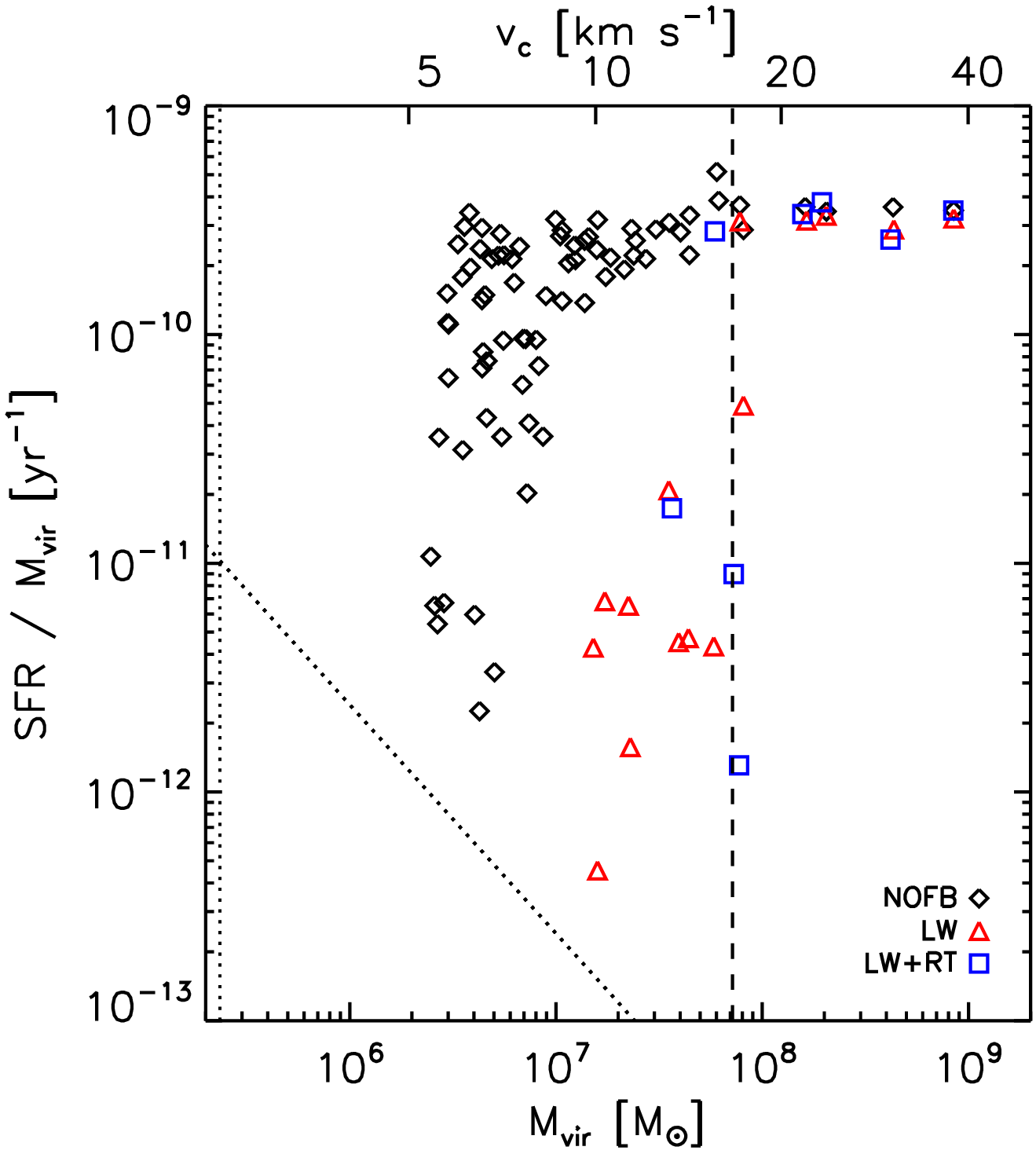}
\includegraphics[trim = 30 0 0 0mm, width = 0.32\textwidth]{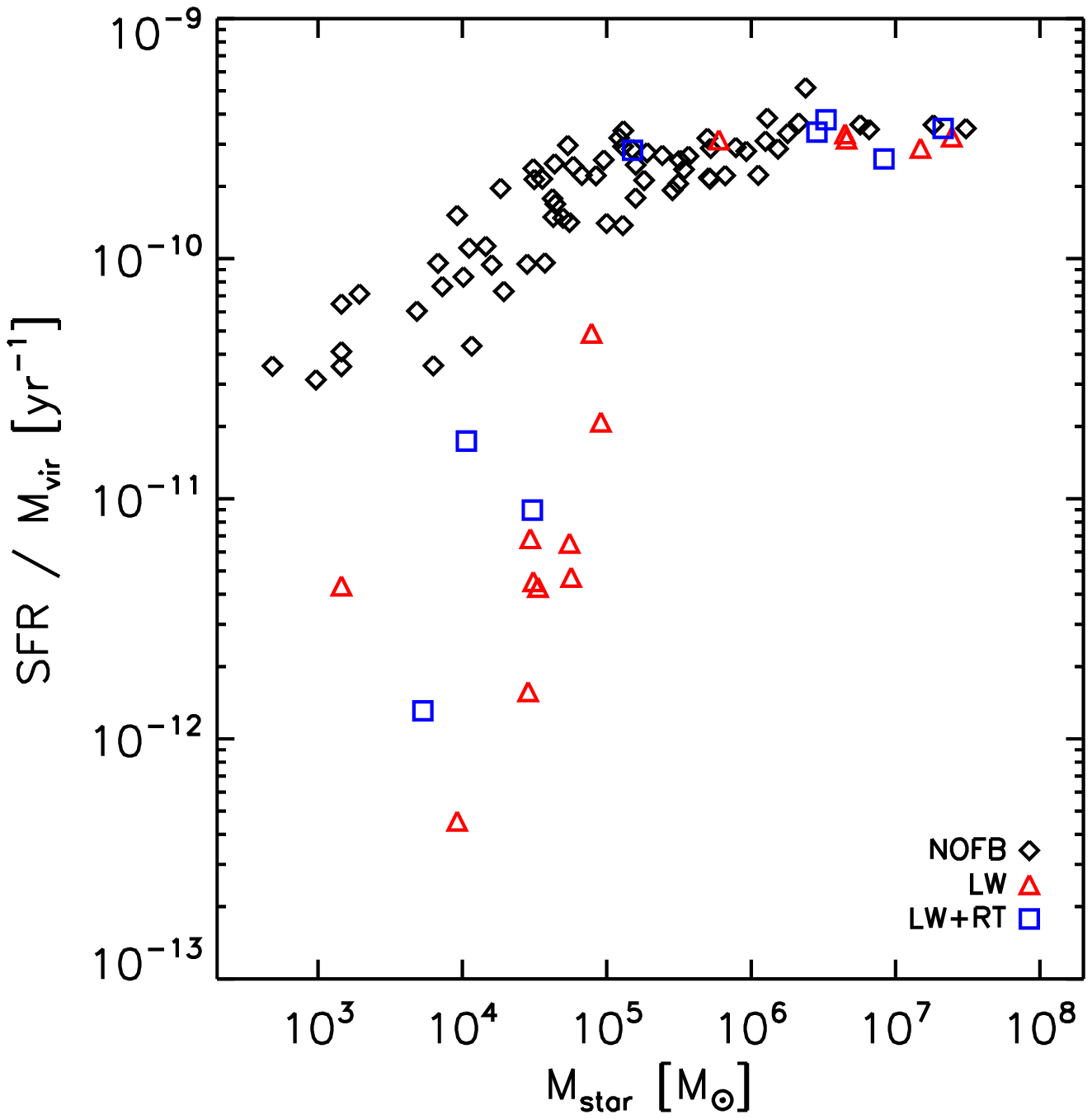}
\includegraphics[trim = 30 0 0 0mm, width = 0.32\textwidth]{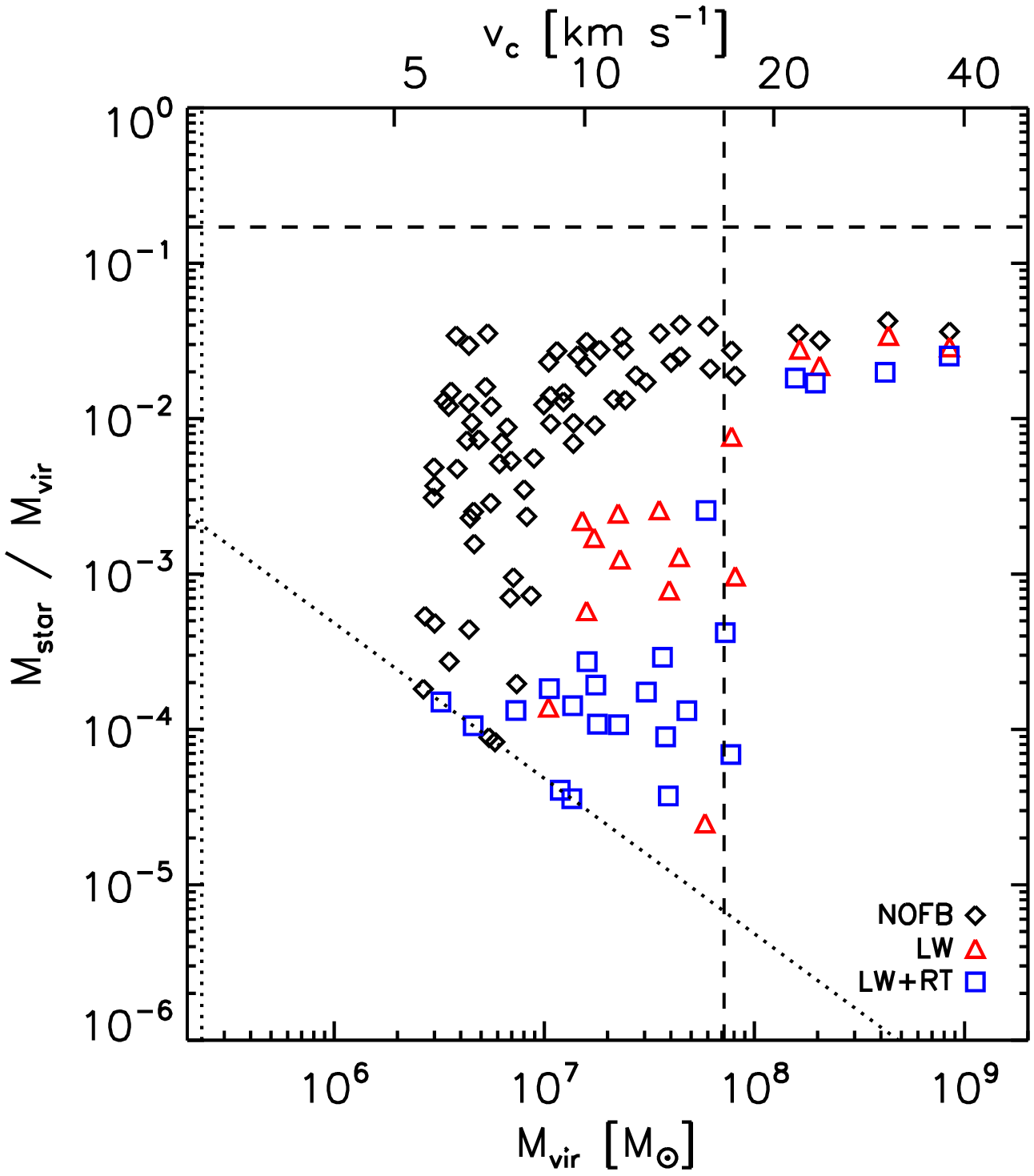}\\
\includegraphics[trim = 30 0 0 0mm, width = 0.32\textwidth]{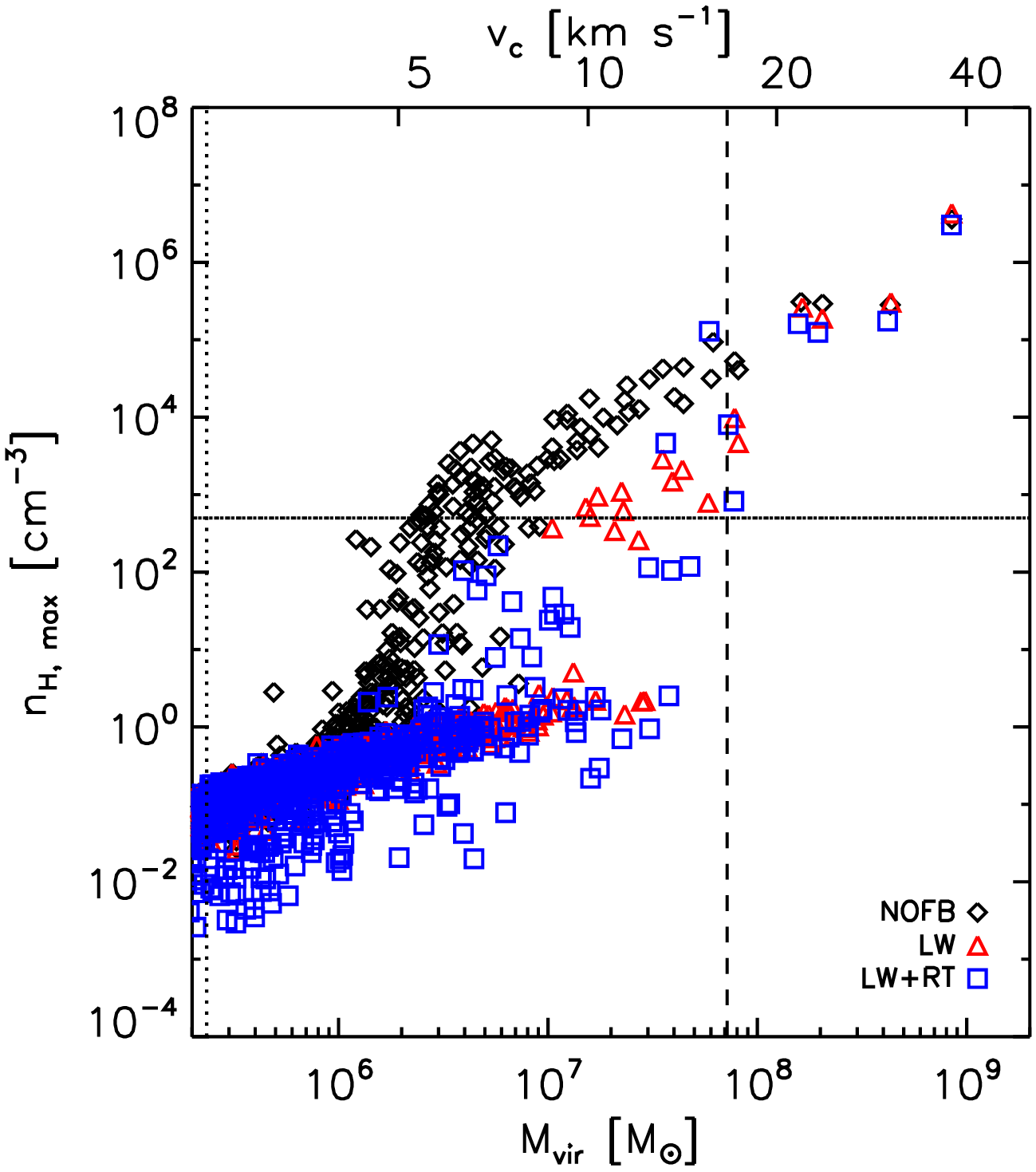}
\includegraphics[trim = 30 0 0 0mm, width = 0.32\textwidth]{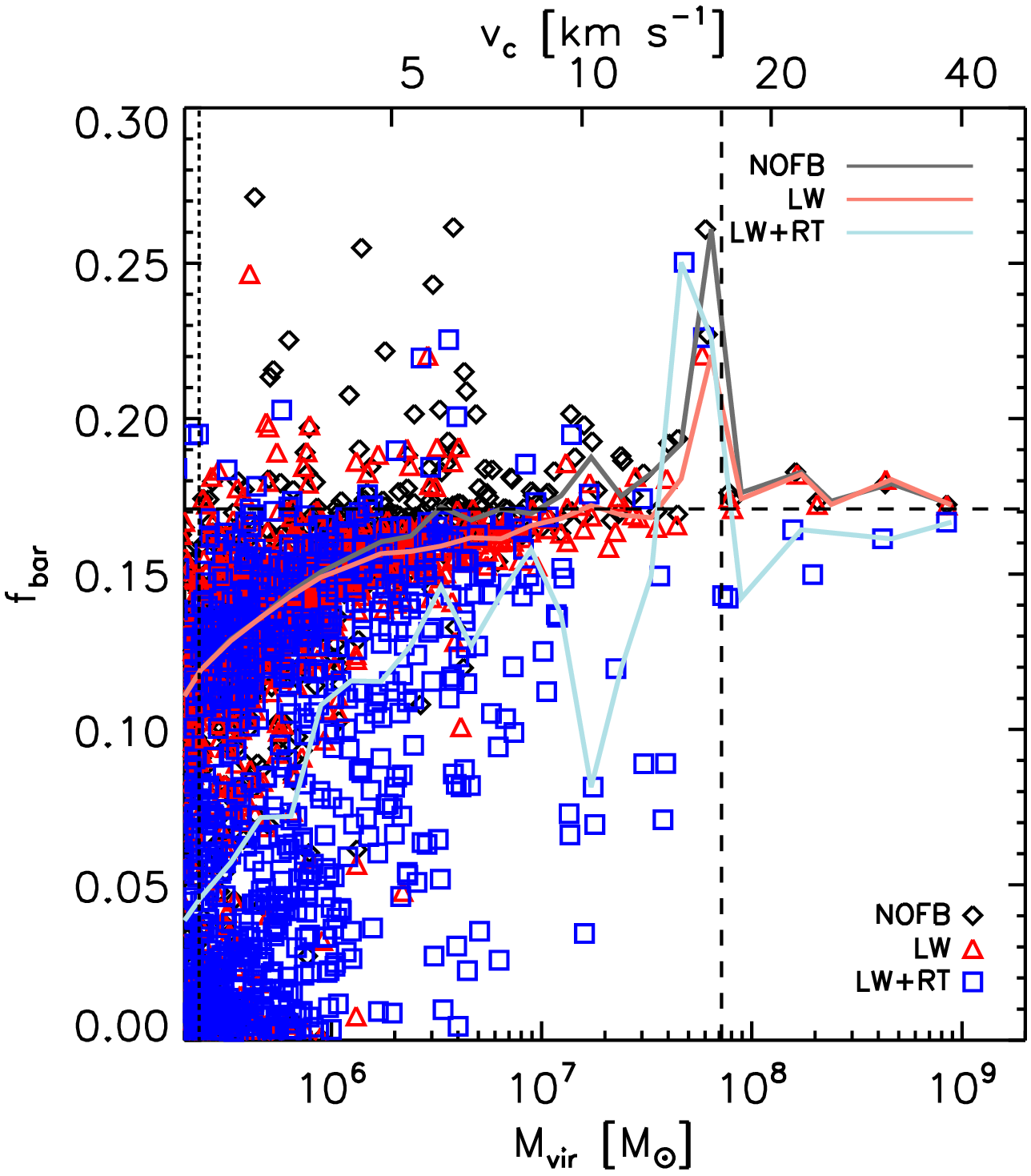}
\includegraphics[trim = 30 0 0 0mm, width = 0.32\textwidth]{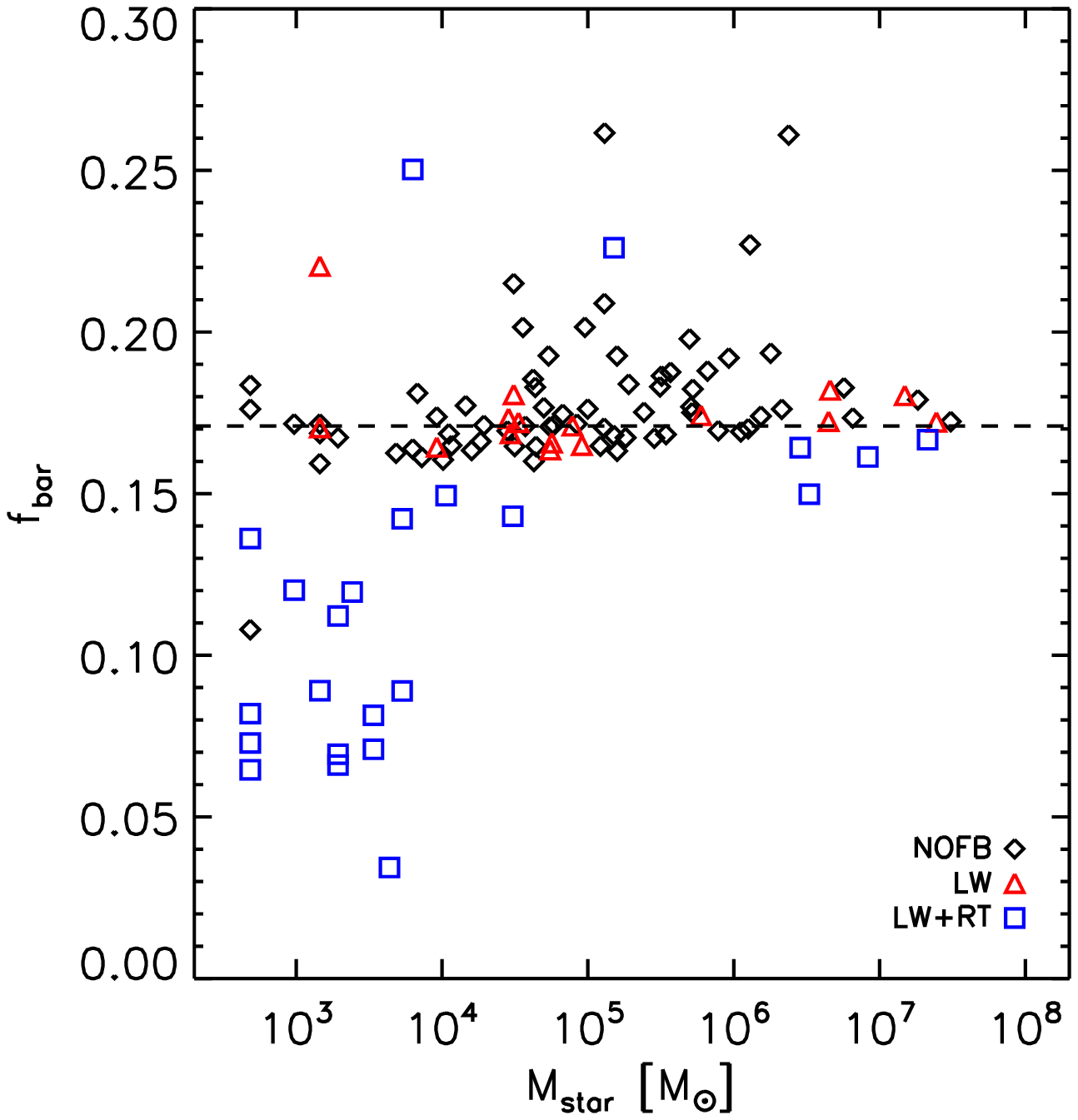}
\end{center}
\caption{Properties of halos at $z = 11$, the final simulation redshift, in the refinement region
 in the three simulations without radiation ({\it NOFB}; black diamonds), with only LW
 radiation ({\it LW}; red triangles), and with LW and ionizing radiation
 ({\it LW+RT}; blue squares). Each symbol represents a single FOF halo. Dashed and dotted vertical lines, if shown, mark the
 virial mass of a halo with virial temperature $T_{\rm vir} = 10^4
 \K$ (Equation~\ref{Eq:VirialTemp} with $\mu = 0.6$), and the mass $100~m_{\rm DM}$. A horizontal dashed line, if
 shown, marks the universal baryon fraction. The top axis, if shown, labels the virial velocity 
 $v_{\rm c} = (GM_{\rm vir} / r_{\rm vir})^{1/2} = 17.0\, (M/10^8 \Msun)^{1/3} [(1+z)/10]^{1/2} \kms $
 (e.g., Equation~3.11
   in \citealp{Loeb:2010}). {\it Top left:} star
 formation rate, normalized by virial mass, as a function of virial mass. {\it Middle left}: SFR, normalized by virial mass, 
 as a function of stellar mass. {\it Top right:} stellar mass
 fraction. The diagonal dotted line marks the stellar mass fraction
 implied by the presence of a single star particle. 
 {\it Bottom left:} maximum gas density. The horizontal line marks the star formation threshold density. {\it Bottom middle:} 
 baryon fraction as a function of virial
 mass. The curves of matching but lighter colors show medians to help guide
 the eye. {\it Bottom right:} baryon fraction as a function of stellar mass. All quantities have been computed considering 
 only matter inside the virial radius $r_{\rm vir}$. Photodissociation of molecular hydrogen by LW radiation suppresses 
 star formation completely only in minihalos with masses significantly below the atomic cooling limit. Photoheating, on the other hand, 
 suppresses star formation also in more massive minihalos. The baryon fraction in simulation {\it LW+RT} is lowest in non-starforming halos with with masses 
 $\lesssim 10^6 \Msun$, in which it is strongly reduced by radiative feedback from external sources, and in star-forming halos with masses $10^7-10^8 \Msun$,
 in which it is additionally reduced by radiative feedback from internal sources.}
\label{Fig:Feedback}
\end{figure*}

Figure~\ref{Figure:MinimumMass} shows the evolution of the minimum
virial mass of star-forming halos in the refinement region, as found
in simulation {\it LW+RT} in which stars emitted both LW and
ionizing radiation ({\it LW+RT}; blue). The corresponding evolutions
in the simulation in which stars emitted LW but not ionizing radiation 
({\it LW}; red), and in the simulation in which no radiation was present 
({\it NOFB}; black), are also shown. We
computed the minimum mass both by considering halos that form stars
for the first time (crosses), which is hereafter referred to as
the minimum collapse mass, and by considering all halos, including
those that have previously formed stars (boxes with matching but lighter
colors). Our definition of the minimum collapse mass is insensitive to our
choice of the threshold density for star formation $n_{\rm H, SF} = 500
\cmci$, at least in the absence of feedback from star formation,
because the transition from densities $n_{\rm H} \gtrsim 10 \cmci$ to 
$ n_{\rm H}\lesssim 10^3 \cmci$ which bracket the adopted
star formation threshold density then occurs within a narrow range of halo
masses (see the bottom left panel in Figure~\ref{Fig:Feedback}). 
\par  

In the absence of LW or ionizing radiation, the minimum collapse mass is
consistent with the mass of halos at virial temperature
$T_{\rm vir} \approx 2200 \K$ (Equation~\ref{Eq:VirialTemp} with $\mu = 1.22$) 
independent of redshift. This mass is consistent with but slightly larger than the masses
$\sim 5 \times 10^5 - 10^6\Msun$ at the time of gas collapse 
in previous simulations of minihalos (e.g.,
\citealp{Yoshida:2003}; \citealp{Oshea:2007}; \citealp{Wise:2008a};
\citealp{Greif:2008}). The small difference is
probably a numerical artifact of our limited resolution, which is
lower than those realized in the previous works. Differences
are also expected because the mass of the halo that forms the first
star is subject to statistical uncertainties related to the low number
of investigated halos (e.g., \citealp{Greif:2008};
\citealp{Oshea:2008}), and because of differences in the strength of
dynamical heating from tidal interactions and mass accretion 
(\citealp{Yoshida:2003}; \citealp{Oshea:2007}). The inclusion of LW radiation increases the
minimum collapse mass. This increase in the minimum collapse mass due to photodissociation by LW radiation  
has been investigated in detail in a number of previous works (e.g.,
\citealp{Machacek:2001}; \citealp{Mesinger:2006};
\citealp{Oshea:2008}; \citealp{Chalence:2012}).
\par
The inclusion of both LW and ionizing radiation increases the minimum
collapse mass in a similar manner but often by a smaller factor,
demonstrating a positive feedback from the enhanced formation of molecular
hydrogen in ionization fronts and fossil \ion{H}{2} regions. Eventually, however, 
the negative feedback from photoheating outweighs the positive
feedback, and there is no new star-forming halo inside the
refinement region below $z \lesssim 13$. Star formation can still
proceed in halos that have previously formed stars, but the masses of
these halos are significantly larger than those in simulation {\it
NOFB}. This is the result of the Jeans filtering of the IGM, 
which impedes the accretion of gas on low-mass halos. The increased scatter 
in the minimum collapse mass in the presence of photoionization 
is an expression of the local nature of feedback 
from photoheating, which is limited to inside the \ion{H}{2} regions.
\par
The top panels of Figure~\ref{Fig:Feedback} show the specific SFRs, i.e., the SFRs
divided by virial mass, both as a function of virial mass (left) 
and of stellar mass (middle), and the stellar mass
fractions (right) for the halos inside the high-resolution region at the final simulation
redshift, $z = 11$. The bottom panels of Figure~\ref{Fig:Feedback} show the maximum gas densities inside
the virial radius (left) and the baryon mass fractions as function of virial (middle)
and stellar mass (right) for these halos. 
\par
In the simulation without radiation ({\it NOFB}), 
the specific SFR $\sim 2 \times 10^{-10} \yri$ is nearly
independent of halo mass in the range $\sim 10^7-10^9 \Msun$. 
The specific SFR is reduced in a fraction of the halos with
masses $\lesssim 10^7 \Msun$ just above the minimum collapse mass,
which we attribute primarily to the effects of dynamical heating during the gravitational collapse
of these halos (e.g., \citealp{Yoshida:2003}). Indeed, the fact that
the transition to high central gas densities occurs within a finite
range of halo masses shows that dynamical heating plays a
non-negligible role, and in the lowest mass halos this may be amplified
by the limited mass resolution we afford. Note that
some of the halos show an increased baryon fraction $f_{\rm bar}
\gtrsim 0.2$ as a result of dynamical interaction and ongoing mergers with other 
halos. 
\par
The inclusion of LW radiation implies a complete suppression of star
formation only in halos with masses $\lesssim 2\times 10^7 \Msun$,
corresponding to virial temperatures significantly below $10^4 \K$
(vertical dashed line). This is in qualitative agreement with the
results from previous high-resolution simulations of the collapse of
minihalos in the presence of a LW radiation background. These works
demonstrated that as the minihalo mass approaches the atomic cooling
limit, the presence of LW radiation cannot prevent the build-up of
molecular hydrogen, because it is catalyzed by the elevated electron
fraction inside central structure formation shocks (e.g.,
\citealp{Wise:2007a}; \citealp{Oshea:2008}), and also because of
self-shielding (e.g., \citealp{Ahn:2007}; \citealp{Susa:2007}). In
contrast, the additional inclusion of ionizing radiation can
potentially suppress star formation in all halos below the atomic
cooling limit by evaporating the gas from the halo centers. However, as we have 
already discussed above in Figure~\ref{Figure:MinimumMass}, 
inspection of the stellar mass fractions reveals that in
simulation {\it LW+RT}, star formation is possible down
to lower halo masses than in simulation {\it LW}, a consequence of  
the positive feedback from ionization ahead of ionization fronts 
and recombinations of HII regions on the formation of molecular hydrogen 
(\citealp{Ricotti:2002}; \citealp{Ricotti:2008}).
\par
The simultaneous inclusion of both LW and ionizing radiation reduces
the average baryon fractions in nearly the full range of simulated
halo masses. The reduction of the baryon fraction is strongest, on
average, for the lowest-mass halos with masses $\lesssim 10^6 \Msun$,
and for halos in the intermediate mass regime, with masses in the
range $10^7-10^8\Msun$. Because halo masses $\lesssim 10^6 \Msun$ are
below the minimum collapse mass, halos in this mass range do not form
stars, and hence their baryon fraction is reduced with respect to that in simulation 
{\it NOFB} due to the effects of radiation from external sources, and due to Jeans
filtering in the photoheated IGM. More massive halos may accrete gas
despite external feedback and Jeans filtering, which explains why the
baryon fraction increases, on average, in the halo mass range $\sim
10^6-10^7 \Msun$. Halos in the range $10^7-10^8\Msun$ are efficiently
forming stars, and therefore their baryon fractions are strongly
reduced by radiative feedback from internal sources. Finally, halos with masses
$\gtrsim 10^8\Msun$, corresponding to virial temperatures $\gtrsim
10^4 \K$, are robust not only against Jeans filtering and feedback
from external sources, but also against the feedback from internal
sources. The baryon fraction of these halos is reduced primarily 
because there was not yet enough time for accretion to compensate
for the mass loss in past episodes of gaseous outflows.
\par
\cite{Ricotti:2008} investigated radiative feedback from the first
galaxies using cosmological simulations of comoving size $1-2 \Mpch$
including processes such as chemical enrichment not treated
here. Utilizing resolution similar to that realized here, they
demonstrated that the positive feedback due to enhanced molecular
hydrogen formation near ionization fronts and inside recombining HII
regions can be very strong and compensate for the negative feedback due
to photodissociation of molecular hydrogen (see also, \citealp{Ricotti:2002}). 
Our simulations may underestimate this positive feedback because we do not capture the
spectral hardening of the ionizing radiation, implying less
ionization to stimulate the formation of hydrogen ahead of
ionization fronts. \cite{Ricotti:2008} further demonstrated that the
stellar mass fractions of the lowest mass halos are characterized by a
large scatter at fixed halo mass. Our simulations do not find as large
a scatter, which may be a consequence primarily of the larger star
particle mass and statistical limitations caused by the small size of
the refinement region in our simulations, although other differences
between the simulations may contribute. We do however find a similar
large scatter in the baryon fraction at low halo masses, and we also
agree with \cite{Ricotti:2008} that this scatter is driven by
radiative feedback from both external and internal sources.

\section{Prospects for Observations with JWST}
\label{Sec:Prospects}
\begin{figure*}
\begin{center}
\includegraphics[trim = 0 0 90 -10mm, width = 0.49\textwidth]{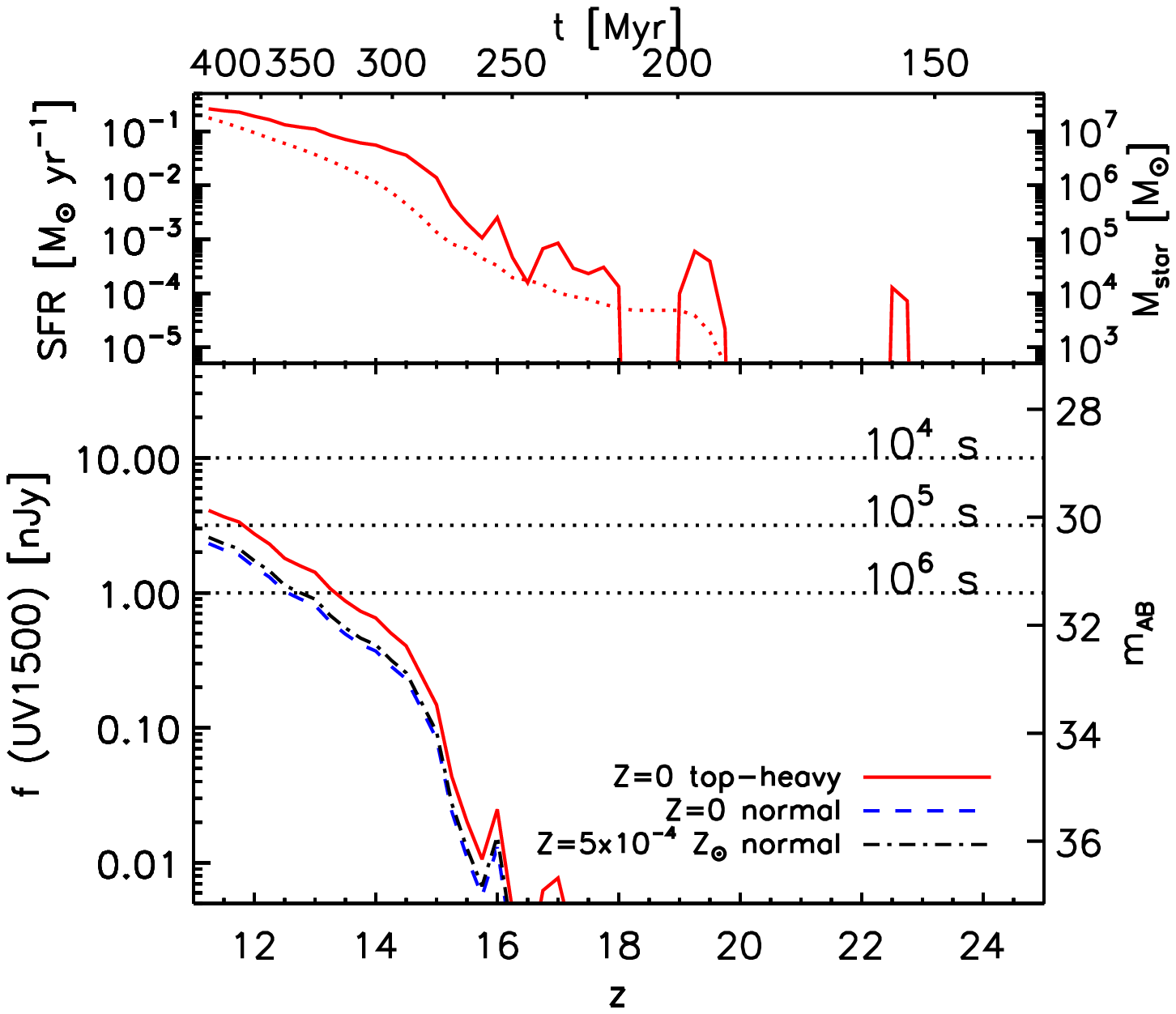}
\includegraphics[trim = 0 0 90 -10mm, width = 0.49\textwidth]{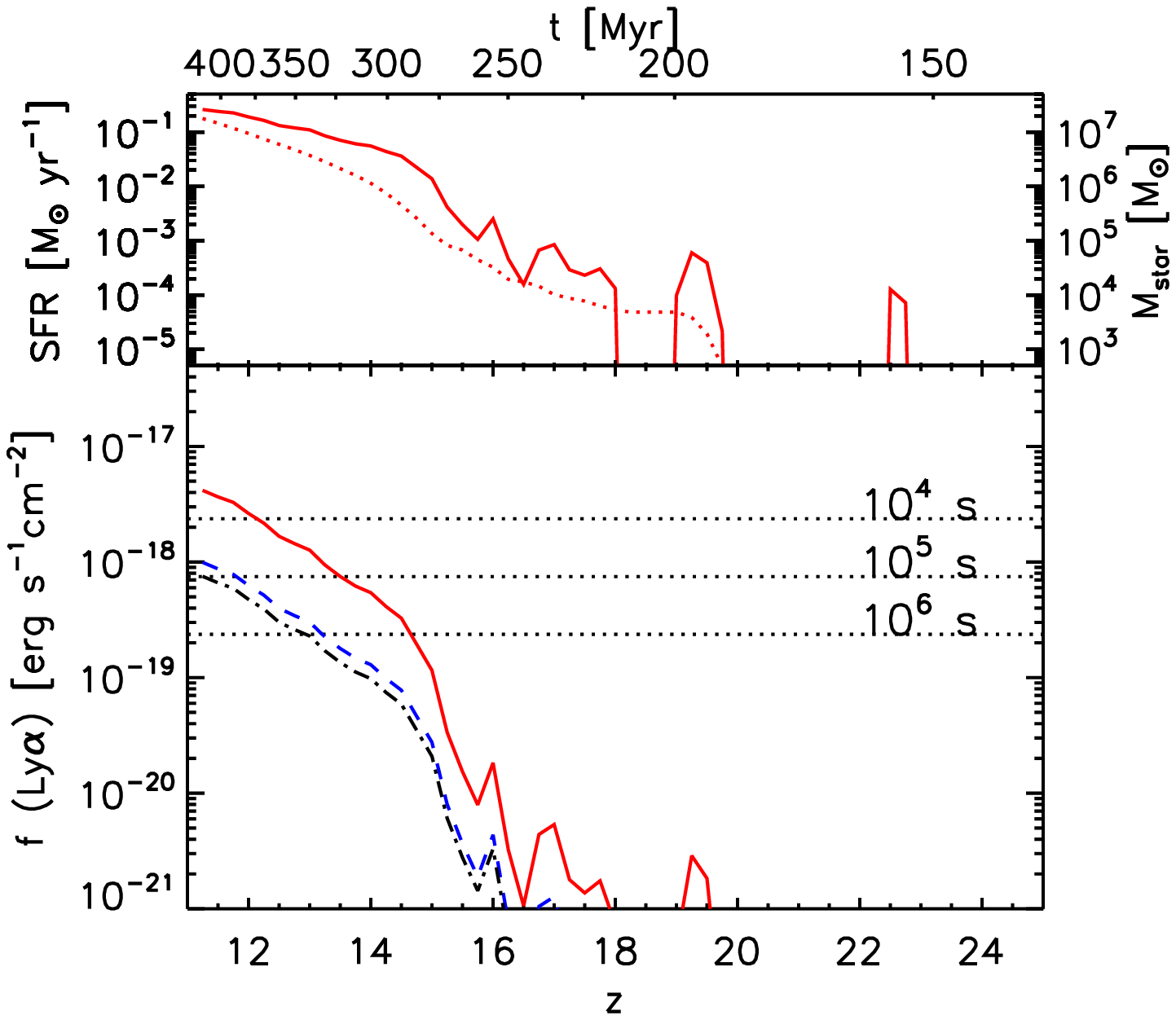}\\
\end{center}
\caption{Flux of the combined stellar and nebular UV1500 continuum (left), and 
in the Ly$\alpha$ recombination line (right) as implied
by the SFRs of the dwarf galaxy in simulation {\it LW+RT} including LW and ionizing radiation (solid curves in the top panels, 
which additionally show the stellar mass in that galaxy with dotted curves and labels on the right-hand axis). 
The estimate is based on the \cite{Schaerer:2003} stellar
population synthesis models with zero metallicity and top-heavy IMF (red solid). For comparison, we also show the expected flux 
assuming metal-free stars with a normal IMF (dashed blue), and stars
with metallicities $Z=5\times 10^{-4} \Zsun$ and normal IMF
(dot-dashed black). The flux estimates assume that all the ionizing photons are absorbed inside the
galaxy, i.e., $f_{\rm esc} = 0$. The line flux estimates can be rescaled to non-zero escape fractions by multiplication with $(1-f_{\rm esc})$. 
All flux estimates scale linearly with the SFR. Dotted horizontal lines show the sensitivity
limits for observations with {\it JWST}, assuming exposures of $10^4$,
$10^5$, and $10^6\s$ (top to bottom) and S/N=10 (\citealp{Panagia:2005}; \citealp{Gardner:2006}). Dwarf galaxies with SFRs $\sim 0.1 \Msunyri$ such as simulated here will be among the faintest galaxies {\it JWST} will detect in deep exposures of the $z \gtrsim 10$ universe.}
\label{Fig:JWST}
\end{figure*}

In this section we estimate the fluxes of
stellar and recombination radiation expected from the dwarf galaxy simulated 
here. We use these estimates to discuss the
detectability of the first galaxies with the upcoming {\it JWST} (e.g., \citealp{Gardner:2006}). 
We focus on the non-ionizing UV continuum at wavelengths around $1500 \Ang$ (hereafter UV1500), 
and on the Ly$\alpha$ recombination line. We have presented initial estimates of the
expected flux in \cite{Pawlik:2011a}. The discussion here improves on
our earlier work by basing the flux estimates on the
dwarf galaxy simulation {\it LW+RT} presented in the current work. This 
simulation, which tracked the emission of LW and ionizing radiation from massive metal-free stars, 
provides us with direct predictions of the SFRs in low-mass high-redshift galaxies
evolving under the radiative feedback from the first stars.
\par
{\it JWST} will image high-redshift galaxies in the UV1500 continuum using
NIRCam. {\it JWST} will also perform spectroscopic observations of these
galaxies in the Ly$\alpha$ line using NIRSpec. Observations of the UV
continuum are routinely used to infer the SFR
density in the high-redshift universe (e.g., \citealp{Bunker:2004}; \citealp{Sawicki:2006}; 
\citealp{Bouwens:2009}; \citealp{Finkelstein:2010}), a key quantity
which not only constrains the nature of the stellar populations (e.g.,
\citealp{Dunlop:2012}; \citealp{Bouwens:2010}), but also allows one to
address the capability of galaxies to reionize the universe (e.g., 
\citealp{Stiavelli:2004}; \citealp{Bouwens:2009}; \citealp{Finkelstein:2010}). Observations of
the Ly$\alpha$ line, on the other hand, have enabled, e.g.,
spectroscopic confirmation of galaxy candidates well into the epoch of
reionization (e.g., \citealp{Rhoads:2004}; \citealp{Iye:2006}; \citealp{Lehnert:2010}; \citealp{Ono:2012}). 
\par
{\it JWST} will further enable spectroscopic observations of high-redshift
galaxies in the He1640 line, using NIRSpec, and in the H$\alpha$ line,
using MIRI. The intrinsic strength of the H$\alpha$ line is weaker by a
factor of about 10 than that of the Ly$\alpha$ line (e.g.,
\citealp{Schaerer:2003}). Unlike the Ly$\alpha$ line, the H$\alpha$ line is, however, not affected 
by resonant scattering (e.g., \citealp{Loeb:1999}; \citealp{Santos:2004}), rendering it a potentially competitive
probe of high-redshift galaxy formation. The He1640 line, 
on the other hand, is highly
sensitive to the metallicity and IMF of the stellar populations, and a
high ratio of luminosities in the He1640 to Ly$\alpha$ or H$\alpha$
lines has been suggested a smoking gun for the existence of massive
metal-free stars (e.g., \citealp{Tumlinson:2001};
\citealp{Oh:2001}; \citealp{Bromm:2001b}). A detection of the He1640 line would therefore put
strong constraints on current theories of metal enrichment and star
formation in the high-redshift universe, but is extremely challenging
because of its weak intrinsic strength (e.g., \citealp{Zackrisson:2011}; 
\citealp{Cai:2011}; \citealp{Inoue:2011}). Because we find that both the H$\alpha$ and He1640 line fluxes 
expected from dwarf galaxies such as simulated here are generally too weak to be observed with {\it JWST}, 
we do not further discuss these emission lines. However, the reader may refer to our earlier discussion of the observability of the first 
galaxies in H$\alpha$ and He1640 in \cite{Pawlik:2011a}.
\par
We use the \cite{Schaerer:2003} population synthesis models for
constant star formation to convert the SFRs of the simulated dwarf
galaxy in simulation {\it LW+RT}, which 
included both LW and ionizing radiation, into intrinsic luminosities in the
Ly$\alpha$ recombination line, and in the UV
continuum. The \cite{Schaerer:2003} models require us to specify the
IMF and the metallicities of the stellar populations, neither of which
is predicted by our simulations and hence has to be assumed. To bracket 
plausible scenarios we repeat our analysis for three models (see the discussion 
in \citealp{Pawlik:2011a}). The first model assumes that stellar
populations consist of metal-free very massive stars, which we describe 
by employing the \cite{Schaerer:2003} model with
zero metallicity and an IMF with Salpeter slope in the stellar mass
range $50-500\Msun$ (hereafter, top-heavy IMF). This is the same model
employed to compute the ionizing and LW luminosities of the stellar
bursts in our simulations. The second model assumes an IMF with
Salpeter slope in the range $1-100 \Msun$ (hereafter, normal IMF) and
zero metallicity. The third and final model assumes the same IMF as
the second model, but a non-zero (but low) metallicity $Z = 5 \times
10^{-4}\Zsun$.
\par
\par
\par
We convert the line luminosities into observed line fluxes using\footnote{This differs from \cite{Pawlik:2011a}, where we converted 
the line flux into an equivalent flux density (Equation~4 in that work).} 
$ f (\lambda_{\rm o}) = L (\lambda_{\rm e}) / [4 \pi d_{\rm L}^2(z)]$, 
where $L (\lambda_{\rm e})$ is the intrinsic Ly$\alpha$ line luminosity, and 
$d_{\rm L}(z)$ the luminosity distance to redshift $z$. The continuum luminosities are converted into 
observed flux densities using an analogous equation (Equation~5 in \citealp{Pawlik:2011a}). The 
line luminosities and the nebular contribution to the UV continuum
luminosities are derived assuming that all ionizing photons are
absorbed, i.e., that the escape fraction is $f_{\rm esc} = 0$, thus 
maximizing the flux in the recombination lines. The
line luminosities can be rescaled to the case of non-zero escape
fractions by multiplication with $(1 - f_{\rm esc})$. We ignore the 
effects of resonant scattering on the observed Ly$\alpha$ line fluxes. 
Ly$\alpha$ RT simulations show that such effects can be very important, but they often
depend sensitively on the specific structure of the investigated galaxies and hence are difficult
to generalize (e.g., \citealp{Verhamme:2008}; \citealp{Dijkstra:2012}). 
Finally, our flux estimates scale linearly with the SFRs. 
\par
\par
The curves in the bottom part of each panel in Figure~\ref{Fig:JWST}
show the resulting UV continuum fluxes (left), and the line
flux in Ly$\alpha$ (right) according to the three population synthesis models. For reference, the SFRs on which these flux estimates are based 
are displayed at the top of each panel (compare with the 
corresponding panel in  Figure~\ref{Fig:AssemblyHistory}). We also show the total stellar masses at the top of each panel
(right-hand axis). Figure~\ref{Fig:JWST} shows that the galaxy simulated here would be
observable out to redshifts $z \lesssim 13$ in both the UV continuum
and the Ly$\alpha$ line in deep surveys with exposures $10^5-10^6 \s$.
While the flux in the UV continuum is insensitive to the properties of
the stellar populations, the Ly$\alpha$ flux is significantly larger
in the case of metal-free stellar population with top-heavy IMF than in the other two cases. At redshifts higher than $z \gtrsim 13$, the reduction in the SFRs, mostly due to
radiative feedback, causes the fluxes to decrease sharply below any
practical detection limit. 
\par
The results presented here are consistent with our earlier results
(\citealp{Pawlik:2011a}), after accounting for the differences in the
SFRs. We caution that the SFRs on which our estimates are based are uncertain for a number
of reasons (see also the discussion in
Section~\ref{Sec:Discussion}). First, our simulations do not account
for feedback from SNe or chemical enrichment, both of which must alter
the SFRs significantly. The star
formation rates are also uncertain because in the absence of
self-regulation by feedback they will depend on the star formation
efficiency (e.g., \citealp{Ricotti:2002}), which is currently not well constrained at the high
redshifts of interest and for which we have assumed a value
appropriate for the local universe (see
Section~\ref{Sec:StarFormation}). As discussed in
Section~\ref{Sec:Disks}, the finite resolution of our simulations
introduces additional uncertainties in the SFRs.  Our work is
a step towards a more general discussion of the
observability of the first galaxies, which ideally should be based on
more sophisticated simulations of a larger galaxy sample (e.g.,
\citealp{Ricotti:2008}). 

\section{Discussion}
\label{Sec:Discussion}
\par
The two nested gas disks in our simulations form only after the halo
has reached a virial temperature significantly larger than $10^4
\K$. In light of the scale-free nature of cold dark matter structure
formation, this relatively late formation of the disks may be
surprising. However, it is consistent with results from previous
zoomed simulations of the first atomically cooling galaxies which have
not exhibited orderly rotation of the gas (e.g., \citealp{Wise:2008c};
\citealp{Greif:2008}; \citealp{Prieto:2013}). On the other hand, the
occurrence of the disks in the dwarf-sized halos fits smoothly in line
with simulations of galaxies more massive than the first atomically
cooling galaxies assembling at lower redshifts (e.g., \citealp{Mashchenko:2008}; 
\citealp{Pawlik:2011a}; \citealp{Romano:2011}; \citealp{Wise:2012}).
\par
The results from these previous works may point at a threshold 
halo mass for the formation of the first disk galaxies 
in the range $\sim 10^8-10^{10}\Msun$ at
$z\sim 10$. This mass scale is similar to the mass scale above which
stellar feedback becomes inefficient. However, the fact that
simulations of the first atomically cooling galaxies have yielded a
turbulent morphology even in the absence of star formation suggests
that the two scales are physically unrelated. Indeed, in our
simulations, the halo mass at the time of disk formation is
insensitive to the inclusion of feedback. Nevertheless, the increased
robustness of dwarf galaxies against stellar feedback helps to
preserve the disks (e.g., \citealp{Kaufmann:2007}). We caution that properties other than
halo mass such as, e.g., the environment, or the merger history, 
may be critical to disk formation in the first galaxies
(\citealp{Prieto:2013}).

\par 
\par
Our results imply that galaxies with SFRs of $\sim 0.1 \Msunyri$ will
be among the faintest galaxies {\it JWST} is likely to detect at $z > 10$,
confirming earlier estimates (e.g., 
\citealp{Haiman:1998}; \citealp{Oh:1999}; \citealp{Tumlinson:2001};
\citealp{Ricotti:2008}; \citealp{Zackrisson:2011}; \citealp{Pawlik:2011a}). According to our
simulations, such galaxies reside in halos with masses of $\sim 10^9
\Msun$ having stellar masses of $\sim 10^7\Msun$. In principle,
{\it JWST} is sufficiently powerful to detect the light from stellar
clusters with masses as low as $10^5-10^6 \Msun$ (e.g.,
\citealp{Johnson:2009}; \citealp{Zackrisson:2011};
\citealp{Pawlik:2011a}). Our simulations do not support the formation
of such massive clusters, which would require local SFRs $\sim 1
\Msunyri$, an order of magnitude higher than found here. However, star formation in our simulations is unresolved, and 
it is possible that star formation in the first galaxies is more clustered or bursty than implied here. 
In this case, {\it JWST} could detect galaxies inside halos less massive
than considered here, or inside halos that are more strongly affected
by feedback than suggested by our simulations. On the other hand, because we do not 
resolve the formation of stars from first principles, 
galaxies could be less efficient star-formers than implied by our simulations, 
and hence be fainter. Such fainter galaxies may still be seen if they are gravitational lensed (e.g.,
\citealp{Zackrisson:2012}). 
\par
Our simulations have ignored a potentially very important physical
process, namely the explosion of massive stars in SNe. SNe can provide
a strong negative feedback by heating and expelling gas from even
relatively massive halos (e.g., \citealp{MacLow:1999}). Previous works have shown
that SN feedback can suppress star formation strongly and lead to 
bursty star formation histories (e.g., \citealp{Stinson:2007}). SN feedback 
further may create a highly spatially
inhomogeneous medium, likely enhancing the fraction of low column
density sight-lines and hence the fraction of escaping ionizing photons
(e.g., \citealp{Yajima:2009}; but see \citealp{Dove:2000}). SN feedback may 
disturb the assembly of disks inside the first galaxies strongly 
(e.g., \citealp{Wise:2012}; but see, e.g., 
\citealp{Mashchenko:2008}). Moreover, SN feedback
may modify the structure of the
dark matter halos significantly (e.g., \citealp{Mashchenko:2008}; \citealp{Governato:2012}; 
\citealp{Brooks:2012}; \citealp{Garrison:2013}).  
\par 
Our simulations also
ignored the chemical enrichment of the gas by the metals synthesized
in stars. Simulations that track the production and transport of metals suggest
that the transition between metal-free and metal-enriched stellar
populations may occur early in the history of the universe (e.g.,
\citealp{Tornatore:2007}; \citealp{Maio:2010}; \citealp{Wise:2012}). Significant uncertainties, however, remain as to the
efficiency of the mixing of metals with the primordial gas and the
level of spatial homogeneity of metal enrichment (e.g.,
\citealp{Scannapieco:2002}; \citealp{Ricotti:2008}). This, together with the fact that not all stars are
expected to explode in SNe and enrich the gas but may instead collapse
directly into black holes (e.g., \citealp{Heger:2003}), leaves open the
possibility of the formation of metal-free stars in select regions of
the universe down to relatively low redshifts (e.g., \citealp{Tornatore:2007}; 
\citealp{Trenti:2009}; \citealp{Johnson:2010}; \citealp{Fumagalli:2011}; \citealp{Simcoe:2012}). However, 
even if all stars would collapse into black
holes without SNe, feedback from accretion onto the
black holes may still affect the evolution of the galaxies 
in a manner not captured by our simulations (e.g., \citealp{Ricotti:2004}; 
\citealp{Alvarez:2009}; \citealp{Jeon:2012}).
\par
\section{Summary}
\label{Sec:Summary}
We have presented cosmological smoothed particle hydrodynamics simulations of a dwarf galaxy assembling in a halo reaching
$10^9 \Msun$ at $z = 10$. The simulations were identical to our
earlier simulations of such a galaxy in that they followed the
non-equilibrium chemistry and cooling of primordial gas. They improved
on our earlier simulations by including the formation of massive
metal-free stars. To investigate the radiative feedback from these stars, 
we compared a simulation in which star
particles emitted both molecular hydrogen dissociating and hydrogen/helium 
ionizing radiation and a simulation in
which star particles emitted only dissociating radiation with a simulation
inside which star particles remained dark.
\par
Our main results are:
\begin{itemize}

\item Dissociating and ionizing radiation exert a
      strong negative feedback by suppressing star formation in the 
      main minihalo progenitor of the dwarf galaxy, but have little effect on star formation as soon as 
      the progenitor evolves into an atomically cooling galaxy. 

\item Radiative feedback suppresses the central dark matter densities
      in the dwarf galaxy main progenitor minihalo relative to the densities found in the simulation
      without radiation. The dark matter density profile of the dwarf
      galaxy is singular isothermal independent of the inclusion of
      radiation shortly after the minihalo has evolved into an atomic cooling halo.

\item The dwarf galaxy halo hosts two nested disks below $z
      \lesssim 12.5$. The formation history and structure of the disks
      are insensitive to the inclusion of dissociating and ionizing
      radiation. These results are consistent with a picture in which the first
      disk galaxies form inside dark matter halos with masses $\gtrsim 10^9
      \Msun$ at $z \gtrsim 10$.

\item The inclusion of dissociating and ionizing radiation lowers the
      baryon fractions inside the minihalos in the neighborhood of the
      dwarf galaxy. The baryon fractions are lowest in minihalos with
      masses $\lesssim 10^6 \Msun$, a consequence of Jeans filtering and photoevaporation from external ionizing sources, and
      in minihalos with masses $\sim 10^7-10^8 \Msun$, here primarily a consequence of 
      photoevaporation of gas by internal ionizing sources.

\item Galaxies with star formation rates $\sim 0.1 \Msunyri $ will be
      among the faintest galaxies the upcoming {\it James Webb Space
      Telescope} will detect in deep exposures of the $z \gtrsim 10$
      universe. Our simulations suggest that such galaxies reside
      in halos with masses $\sim 10^9 \Msun$ and have stellar
      masses $\sim 10^7 \Msun$.

\end{itemize}
\par
We caution that our conclusions are subject to statistical
uncertainties implied by the small volume of the high-resolution
region in our simulations. Another major shortcoming of our
simulations is the lack of feedback from supernova explosions. Such
feedback can potentially have a significant impact on the evolution of
low-mass galaxies. Feedback from supernovae may heavily disturb the
assembly of disks, and strongly decrease the star formation rates
inside dwarf galaxies, thus affecting also estimates of their
observability. Our simulations also did not account for the chemical
enrichment of the interstellar and intergalactic gas and the
associated transition from metal-free to metal-enriched stars. The
effects of supernova feedback and chemical enrichment are left to be
investigated in future work.

\acknowledgments
We are grateful to Volker Springel, Joop Schaye, and Claudio Dalla
Vecchia for letting us use their versions of {\sc gadget} as well as
their implementations of FOF and {\sc subfind}. We thank the referee for 
the constructive comments which improved the discussion of the present work. 
AHP thanks Joop Schaye, Claudio Dalla Vecchia, Alireza Rahmati, Milan Rai{\v
  c}evi{\'c}, Jacob Hummel, and Marcel Haas for helpful
discussions. The simulations presented here were carried out at the
Texas Advanced Computing Center (TACC). This work also benefited from
the use of supercomputer facilities of the National Computing
Facilities Foundation (NCF). This research is supported by NASA
through Astrophysics Theory and Fundamental Physics Program grant
NNX09AJ33G and through NSF grant AST-1009928. AHP receives funding
from the European Union's Seventh Framework Programme (FP7/2007-2013)
under grant agreement number 301096-proFeSsOR.

\appendix

In this appendix we discuss our implementation of photoionization and photoheating by stellar sources 
(Appendix~\ref{sec:photoionization}). We describe the implementation of the radiative transfer (RT; Appendix~\ref{sec:rt}), 
the treatment of multi-frequency radiation (Appendix~\ref{sec:grey}), the 
computation of the photoionization and photoheating rates (Appendix~\ref{sec:rates}), and the coupling of the 
RT to the hydrodynamical evolution (Appendix~\ref{sec:coupling}). We also discuss tests of our 
RT implementation (Appendix~\ref{sec:tests}), specifically addressing the coupling of the 
RT with the solver for the chemical and thermal evolution (Appendix~\ref{sec:tests:rt}), 
and with the hydrodynamics (Appendix~\ref{sec:tests:rhd}). We emphasize that the approximations and choices 
of numerical parameters employed in this work and described below are specific to the current work.

\section{Implementation}
\label{sec:photoionization}
\subsection{\traphic}
\label{sec:rt}
The RT makes use of the RT code 
\traphic\ (\citealp{Pawlik:2008}; \citealp{Pawlik:2011b}) implemented in a 
customized copy of version 3 of {\sc gadget} (\citealp{Springel:2005}; \citealp{Schaye:2010}). In simulations 
with \traphic, the RT equation is
solved by tracing a finite number of discrete photon packets emitted by ionizing source particles 
through the simulation box. This is done directly on the irregular grid 
defined by the SPH particles, at the speed of light, and in a 
photon-conserving manner (\citealp{Abel:1999b}). In addition to the description 
of \traphic\ given below, the reader may refer to the original publications for further details.
\par
Each photon packet carries photons of
characteristic frequency $\nu$. We denote the total number of
frequency bins used to discretize the radiation spectrum by $N_\nu$. The
transport of photons during a single time step proceeds by a
succession of emission and transmission steps that move photon packets
from individual particles, either SPH or star particles, to a number of
$\tilde{N}_{\rm ngb}$ SPH neighbors. The neighbors of a given particle
are particles located inside the neighbor sphere, which is a sphere
centered on the particle and contains $\tilde{N}_{\rm ngb}$ neighboring particles. Therefore, $\tilde{N}_{\rm ngb}$ determines the spatial
resolution at which the RT is carried out. We choose $\tilde{N}_{\rm
ngb} = 48 $, which reflects a compromise between keeping high spatial resolution and controlling particle discreteness noise
(\citealp{Pawlik:2008}). 
\par
Star particles emit photon packets to their $\tilde{N}_{\rm ngb}$ neighboring SPH
particles inside $N_{\rm EC}$ emission cones, 
each subtending a solid angle of $4 \pi / N_{\rm EC}$. Emission cones 
tessellate the sky and are used to accomplish the isotropic emission of photon packets to the SPH
neighbors despite the possibly highly anisotropic distribution of the
SPH particles (see the Appendix in \citealp{Pawlik:2008}). Photon packets 
emitted inside emission cones containing
multiple SPH neighbors are split in proportion to the inverse
squared distance between the source and the neighbors to account for
the dilution of the radiation field with distance from the source. The 
central axes of the emission cones define the initial propagation directions
of the emitted photon packets. The parameter $N_{\rm EC}$ determines the angular sampling of the sky
as seen from any given ionizing source. We choose $N_{\rm EC}= 8$. Even though the orientation of the 
emission cone tessellation is randomly rotated at each RT time step, our choice of a small number of emission cones may 
imply increased random scatter in the distribution of the photoionized and photoheated gas, especially around 
halos that contain only a few star particles (see Appendix~\ref{sec:parameters} for an illustration). 
\par
SPH particles that receive photon packets transmit these packets along
the associated propagation directions to their downstream
neighbors. Individual photon packets are transmitted only to the SPH
neighbors located inside transmission cones, which are regular cones
with solid angles subtending $4 \pi / N_{\rm TC}$ steradians centered around the propagation
directions and with their apex attached to the transmitting
particle, where $N_{\rm TC}$ is a parameter specified below. The use of transmission cones 
prevents uncontrolled diffusion of photon packets
on the set of irregularly distributed SPH particles 
and keeps the photon transport directed. Photon
packets that are emitted inside transmission cones containing multiple
SPH neighbors are split equally between these neighbors.
\par
The parameter $N_{\rm TC}$  sets the angular resolution at which the RT is
performed. Because the transmission cones are defined locally at the
positions of the transmitting SPH particles and because photon packets
are only distributed among the subset of the $\tilde{N}_{\rm ngb}$
neighbors that fall inside these transmission cones, the angular
resolution is independent of the distance from the sources. In this
work we adopt $N_{\rm TC}= 128$. In test simulations of the RT around
multiple sources inside a static cosmological density field 
presented in \cite{Pawlik:2008} we found that the results had converged
at a lower angular resolution of $N_{\rm TC}= 32$. Note that 
virtual particles are created to accomplish the emission and transport of photon
packets in cones that do not contain
SPH neighbors, as described in \cite{Pawlik:2011b}. 
\par
\par
Photon packets received by SPH particles are merged, which allows one to
control the number of photon packets inside the simulation box and
to avoid the scaling of the computational cost with the number of
sources. The merging is done by binning photon packets in solid angle
using a set of $N_{\rm RC}$ tessellating reception cones with solid
angles $4 \pi / N_{\rm RC}$ attached to each SPH particle. The merging
is done separately for each frequency bin and hence it limits the number
of photon packets to be transmitted next to at most $N_{\rm RC} \times
N_\nu$. Accordingly, the computational cost of the RT
scales with $N_{\rm RC}$, but not with the number of ionizing sources.
In this work we set $N_{\rm RC} = 8$, motivated by test simulations similar to 
those presented in \cite{Pawlik:2011b}. 
\par
Photon packets are associated a clock that is used to control the
speed at which they travel (see \citealp{Pawlik:2008}). In this work we set this speed equal to the physical speed of light. However,
for computational efficiency we allow photon packets to travel only a
single inter-particle distance during individual RT time steps. In the
limit of small RT time steps, transporting photon packets at the speed
of light but limited by a single inter-particle distance is equivalent
to solving the time-dependent RT equation (see \citealp{Pawlik:2008} for a similar argument
in the case of the time-independent RT). The
hybrid approach employed here prevents the photons from traveling faster than
the speed of light but, depending on the size of the RT time step and
the inter-particle distance, may imply that photons travel at an
effective speed that is lower than the speed of light, possibly
resulting in an artificially delayed propagation of ionization fronts. We will
discuss the effects of this approximation in Appendix~\ref{sec:parameters}.
\par
\subsection{Absorption of Ionizing Photons by Hydrogen/Helium}
\label{sec:grey}
A fraction $1-\exp [-\tau (\nu)]$ of each photon packet in frequency bin $\nu$ emitted or
transmitted from particle $i$ at position $\mathbf{r}_i$ to SPH
neighbor $j$ at position $\mathbf{r}_j$, separated by the propagation
distance $d_{ij}$, is absorbed. The optical depth $\tau(\nu)$ is the sum $\tau(\nu)
= \sum \tau_\alpha (\nu)$ of the optical depths $\tau_\alpha (\nu)$ of
each absorbing species $\alpha \in \{\rm HI, HeI, HeII\}$, and
$\tau_\alpha(\nu) = \int_{\mathbf{r}_i}^{\mathbf{r}_j} dr\
\sigma_\alpha (\nu) n_\alpha (\mathbf{r}) \approx \sigma_\alpha (\nu)
n_\alpha (\mathbf{r}_j) d_{ij}$. The number of
photons absorbed by species $\alpha$ is $\delta N_{\rm abs, \alpha}
(\nu) = [w_\alpha(\nu)/\sum_{\alpha} w_\alpha(\nu)] \delta N_{\rm abs} (\nu)$,
where the weights $w_\alpha(\nu) = \tau_\alpha (\nu)$
(\citealp{Pawlik:2011b}) and $ \delta N_{\rm abs} (\nu) = \sum_\alpha
\delta N_{\rm abs, \alpha} (\nu)$ is the total number of absorbed
photons.
\par
In this work we choose, for reasons of computational efficiency, 
to transport hydrogen-ionizing radiation using a single frequency bin, i.e., we set
$N_\nu = 1$. In this bin, we adopt  frequency-averaged
photoionization cross-sections $\langle \sigma_{\gamma \alpha}
\rangle$ for the absorption by the individual species $\alpha$ using
the grey approximation (e.g., \citealp{Mihalas:1984}),
\begin{equation}
 \langle \sigma_{\gamma \alpha} \rangle \equiv  \int_{\nu_\alpha}^\infty~d\nu \frac{4 \pi J_{\nu}(\nu)}{h_{\rm P}
   \nu} \sigma_{\gamma \alpha} (\nu) \times \left [\int_{\nu_{\rm HI}}^\infty~d\nu \frac{4 \pi J_{\nu}(\nu)}{h_{\rm P}\nu} \right ]^{-1},
\label{Eq:Crosssection}
\end{equation}
where $J_{\nu}(\nu)$ is the mean intensity of the
radiation field, and $h_{\rm P} \nu_\alpha$ is the ionization
potential. We assume that the mean intensity $J_{\nu}(\nu)$ is characterized by a
black body spectrum of temperature $T_{\rm BB} = 10^5\K$, appropriate for the emission 
of radiation by the first stars (e.g., \citealp{Schaerer:2003}), which 
gives $\langle\sigma_{\gamma \rm HI} \rangle = 1.63 \times 10^{-18} \cmsq$,
$\langle\sigma_{\gamma \rm HeI} \rangle = 2.28 \times 10^{-18} \cmsq$,
and $\langle\sigma_{\gamma \rm HeII} \rangle = 6.19 \times 10^{-20}
\cmsq$. 
We have employed the fits to the photoionization cross-sections from \cite{Verner:1996}.
\par
At fixed mean intensity $J_{\nu}(\nu)$, the grey approximation implies
photoionization rates identical to those inferred from a full
multi-frequency treatment. However, because of the use of only a single frequency bin, our
simulations ignore the hardening of the radiation spectrum with 
distance from the source caused by the preferential absorption
of lower energy photons with larger absorption cross sections (see, e.g., the discussion
in \citealp{Pawlik:2011b}).
\par 

\subsection{Photoionization and Photoheating Rates}
\label{sec:rates}
At the end of each RT time step the photoionization
and photoheating rates are computed. We determine the photoionization
rates directly from the total number of photons $N_{\rm abs, \alpha} (\nu) =
\sum \delta N_{\rm abs, \alpha} (\nu)$ absorbed by a given SPH particle
during the RT time step $\Delta t_{\rm r}$ in frequency bin $\nu$,
\begin{equation}
\Gamma_{\gamma \alpha} = \frac{\sum_\nu N_{\rm abs,
\alpha}(\nu)}{\eta_\alpha N_{\rm H} \Delta t_{\rm r}},
\end{equation} 
where $N_{\rm H} \equiv m_{\rm gas} X_{\rm H} / m_{\rm H}$ is the
number of hydrogen atoms associated with the SPH particle of mass
$m_{\rm gas}$. This ensures that the same number of photons that
have been removed from the simulation is used to determine the
ionization balance and temperature of the ionized gas, i.e., photon
conservation (\citealp{Abel:1999b}). 
\par
The heating rate per atom due to photoionization of species $\alpha$, assuming a single frequency bin in the 
grey approximation, is $\mathcal{E}_{\gamma \alpha} = \Gamma_{\gamma \alpha} \langle \epsilon_\alpha \rangle$, where
\begin{equation}
\langle \epsilon_\alpha \rangle = \int_{\nu_\alpha}^\infty~d\nu \frac{4 \pi
    J_{\nu}(\nu)}{h_{\rm P} \nu} \sigma_{\gamma \alpha} (\nu) (h_{\rm P}\nu -
  h_{\rm P}\nu_\alpha) \times \left[ \int_{\nu_\alpha}^\infty~d\nu \frac{4 \pi J_{\nu}(\nu)}{h_{\rm P} \nu} \sigma_{\gamma \alpha} (\nu) \right]^{-1}
\label{Eq:AverageExcessEnergy}
\end{equation}
is the average energy of the absorbed ionizing photons in excess of the photoionization threshold. Because we assume 
that the mean intensity $J_{\nu}(\nu)$ is characterized by a black
body spectrum of temperature $T_{\rm BB} = 10^5\K$, the average
excess energies for photoionization of hydrogen and helium are $\langle
\epsilon_{\rm HI}\rangle = 6.32 \eV$, $\langle \epsilon_{\rm
HeI}\rangle = 8.70 \eV$, and $\langle \epsilon_{\rm HeII}\rangle =
7.88 \eV$.
\par

\subsection{Radiation-Hydrodynamical Coupling}
\label{sec:coupling}
The radiation-hydrodynamical evolution of the gas is followed by
invoking a series of subcycles to compute the dynamics, radiative
transfer, chemistry, cooling, and heating of the gas. The
dynamical evolution of the gas is followed on the gravito-hydrodynamical time steps set
by the standard time integration scheme of the {\sc gadget} code (Section~\ref{sec:gravity}). 
The RT is performed by subcycling the smallest (among all particles) gravito-hydrodynamical 
time step  $\Delta t_{\rm s}$, with RT time steps of size $\Delta t_{\rm r}$. 
Unless stated otherwise, we adopt $\Delta t_{\rm r} = \min [\Delta t_{\rm s}, 0.1 \Myr]$. We assume
that the chemical abundances and temperatures do not change during a
single RT time step. 
\par
Chemistry, heating, and cooling of the gas are computed at the end
of each RT time step. This is done by subcycling the RT time step using time steps 
computed by the implicit non-equilibrium solver described in Section~\ref{sec:cooling}. 
Note that because photoionization rates are obtained from the
number of absorptions computed under the assumption that species
fractions and temperatures remain constant during the time $\Delta
t_{\rm r}$, not all photons that have been absorbed always end up consumed
in the computation of the chemistry and thermodynamics of the gas. This is a consequence
of the evolution of the chemical composition and temperature
during the subcycling. To ensure photon conservation 
we reinsert the remaining photons into the
RT at the beginning of the next RT
time step, and set them to be propagated along their original
directions.
\par
The change in chemical abundances and temperature affects the dynamics of the particles only
at the end of their gravito-hydrodynamical time steps, at which time the particle
entropies are updated and the sizes of the new gravito-hydrodynamical time steps
are determined. In {\sc gadget}, particles evolve along a hierarchy of
individual particle gravito-hydrodynamical time steps that may be much longer than
the smallest such time step, the latter also being the time step at which the
radiative cooling and heating is computed. This means that the
radiation-hydrodynamical response of the gas to photoionization
heating may occur with a delay. Recently, \cite{Durier:2011} have shown, 
in the context of thermal feedback from supernova explosions, that the lack of a prompt
response to localized energy injection may lead to a strong violation
of the conservation of energy when using the standard {\sc gadget}
time integration scheme employed also here. Our simulations may suffer from 
similar numerical artifacts.

\section{Tests}
\label{sec:tests}

\begin{figure*}
\begin{center}
\includegraphics[trim = 70 0 70 0mm, width = 0.45\textwidth]{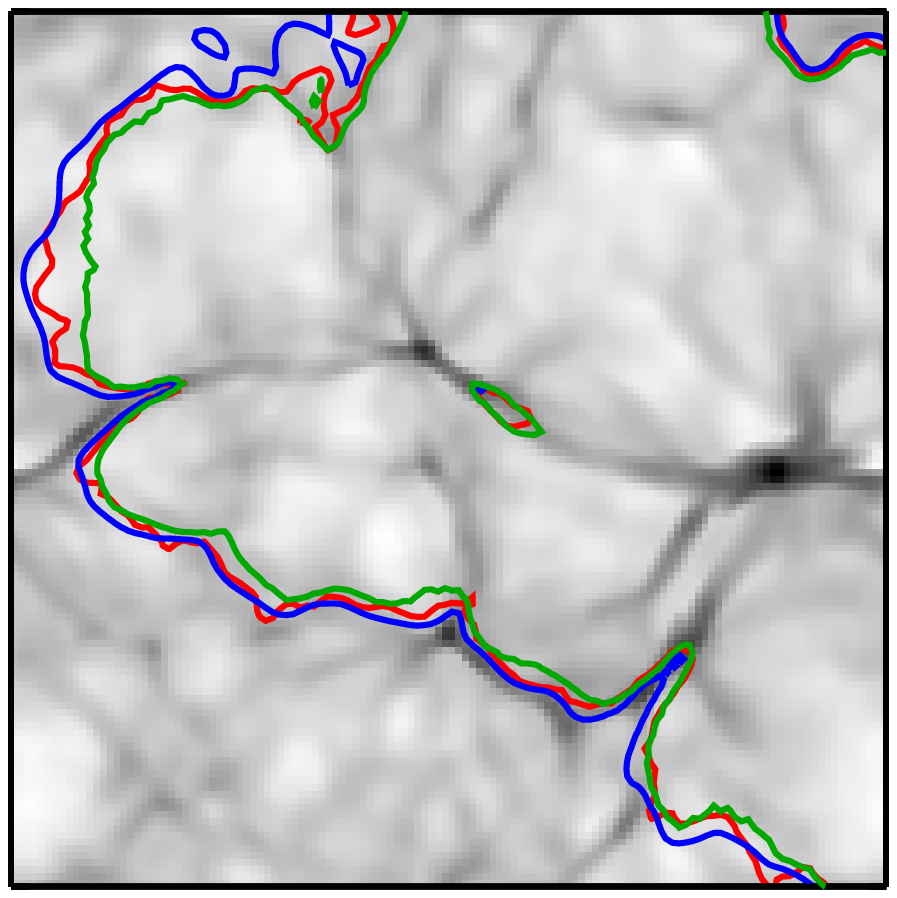}
\includegraphics[trim = 70 0 70 0mm, width = 0.45\textwidth]{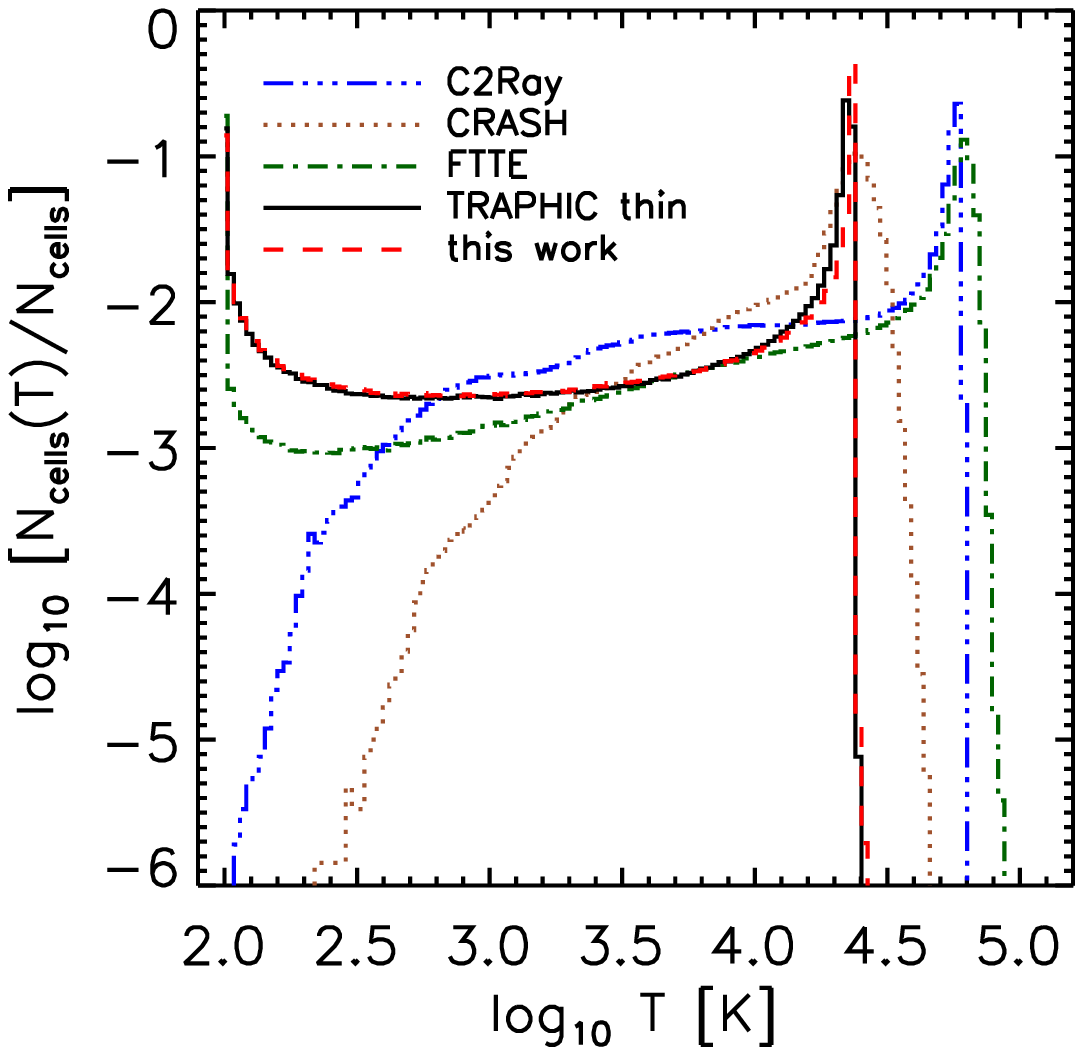}\caption{Test 4: cosmological reionization by multiple stellar sources (\citealp{Iliev:2006}). 
  Left: thin slice through the center of the simulation box
  at time $t = 0.2 \Myr$, showing contours of neutral hydrogen fraction $\eta_{\rm HI} = 0.5$,
  on top of the density field (grey image). The ionization fronts
  obtained in this work (red) are in very good agreement with the reference
  results obtained using {\sc c2ray} (blue; \citealp{Mellema:2006})
  and {\sc crash} (green; \citealp{Ciardi:2001}; \citealp{Maselli:2003}; 
  \citealp{Maselli:2009}). Right: histograms of the gas temperature at $t =
  0.2 \Myr$. The results obtained in this work are in excellent
  agreement with those presented in \cite{Pawlik:2011b} (histogram labeled \traphic\
  thin for reference with our earlier work) obtained with a different solver for the chemical and thermal
  evolution. They are in very good agreement with the reference
  solutions obtained with {\sc c2ray}, {\sc crash}, and {\sc ftte}
  (\citealp{Razoumov:2005}) as reported in \cite{Iliev:2006}, if one
  accounts for the differences in the atomic physics and the
  approximations employed.}
\label{fig:test4:slices}
\end{center}
\end{figure*}

We have previously described a number of tests of our implementation
of \traphic\ in {\sc gadget} for the RT on static
density fields. In \cite{Pawlik:2008}, we carried out monochromatic
RT simulations of increasing complexity, assuming that the
gas is composed only of hydrogen and has a fixed temperature. By
comparing with reference solutions such as those published in
\cite{Iliev:2006}, we demonstrated the ability of \traphic\ to
accurately capture the evolution of ionization fronts, and to reproduce 
the ionized fractions. We also
showed that \traphic\ is able to produce sharp shadows behind opaque
absorbers, which gives rise to the typical ``butterfly'' shape of the ionized regions (e.g.,
\citealp{Abel:1999b}). In \cite{Pawlik:2011b}, we extended our
implementation of \traphic\ to enable the transport of multi-frequency
radiation, to account also for the ionization of helium, and to
compute the evolution of the gas temperature due to radiative cooling
and photoionization heating. Results from test simulations with this
new implementation were in excellent agreement with reference
solutions, such as those in \cite{Iliev:2006}.
\par
The implementation of \traphic\ in {\sc gadget} that we use in this
work differs from that described in \cite{Pawlik:2008} and
\cite{Pawlik:2011b} in two main respects. First, we have substituted the explicit
solver for the evolution of the chemistry and temperature of
primordial atomic gas in the presence of photoionizations used and
tested in \cite{Pawlik:2011b} with the implicit solver used in
\cite{Pawlik:2011a} and described in Section~\ref{sec:cooling}. This
solver accounts for additional species, such as molecular hydrogen,
and employs different rates for the atomic and molecular
physics. It has been described and extensively tested in a number of
publications (e.g., \citealp{Johnson:2006};
\citealp{Greif:2010}). However, it has not yet been tested in
combination with \traphic. In Appendix~\ref{sec:tests:rt} we therefore repeat
the RT test simulation on a static cosmological density
field from \cite{Pawlik:2011b}. The results of this test are in 
excellent agreement with our previous results, demonstrating 
the validity of our implementation. 
\par
Second, the current work is the first to employ \traphic\ in
radiation-hydrodynamic simulations that account for the feedback of
photoionization heating on the gas dynamics. We have performed the
radiation-hydrodynamical tests from \cite{Iliev:2009}, and we have
achieved excellent agreement with the published reference results in
all of them. In Appendix~\ref{sec:tests:rhd} we provide a discussion of one of
the tests relevant to the present work, the simulation of the
evaporation of a minihalo by an internal ionizing source. 
\par
Finally, in Appendix~\ref{sec:parameters}, we investigate how our adoption of a limited
photon propagation speed and a finite angular sampling affects the early evolution of the minihalo in 
simulation {\it LW+RT} analyzed in the main text.

\subsection{Radiative Transfer and Chemical and Thermal Evolution}
\label{sec:tests:rt}
In this section we repeat Test~4 of the cosmological RT code
comparison project (\citealp{Iliev:2006}) that we have previously discussed in
\cite{Pawlik:2011b}, using a chemistry solver different from the one employed here. 
The test involves the simulation of \ion{H}{2} regions around multiple ionizing sources in a static
cosmological density field. It was designed to capture important
aspects of state-of-the-art simulations of hydrogen
reionization, such as the delayed propagation in or trapping of ionization fronts 
by dense gas.
\par
The setup of this test is identical to that of Test 4 in
\cite{Pawlik:2011b}, to which we refer the reader for a detailed
description. Briefly, the initial conditions are provided by a
snapshot (at redshift $z \approx 8.85$) from a cosmological $N$-body and
gas-dynamical uniform-mesh simulation. The simulation box is $L_{\rm
box} = 0.5\cMpch$ on a side, where $h = 0.7$, and is uniformly divided
into $N_{\rm cell} = 128^3$ cells. We Monte Carlo sample this 
density field to replace the mesh cells with $N_{\rm SPH}=N_{\rm
cell}=128^3$ SPH particles. The gas consists purely of atomic hydrogen
and is assumed to be initially neutral at temperature $T = 100\K$. The ionizing sources are chosen to
correspond to the 16 most massive halos in the box, and are assumed
to have blackbody spectra with temperature
$T_{\rm BB} = 10^5\K$. 
\par
For comparison with \cite{Pawlik:2011b} we solve the
time-indepen\-dent RT equation with an angular resolution of $N_{\rm
c} = 32$, set the number of neighbors to which 
sources emit radiation to $\tilde{N}_{\rm ngb}=32$, employ a time step
$\Delta t_{\rm r} = 10^{-4} \Myr$, and transport photons only over a
single inter-particle distance per time step. We transport radiation using a single
frequency bin, employing the grey photoionization cross-section
$\langle \sigma_{\gamma \rm HI}\rangle = 1.63 \times 10^{-18}
\cms$. We assume that each photoionization adds $\langle \epsilon_{\rm
HI} \rangle = 6.32 \eV$ to the thermal energy of the gas, as appropriate for the adopted blackbody spectrum.
\par
In Figure~\ref{fig:test4:slices} we show results of this simulation and compare them with those presented in
\cite{Pawlik:2011b} and also with the reference results reported in
\cite{Iliev:2006}. The agreement is very good if one accounts for the differences in the atomic
physics and the approximations employed (see \citealp{Pawlik:2011b} for a detailed discussion).

\subsection{Radiation-Hydrodynamical Coupling}
\begin{figure*}
\begin{center}
\includegraphics[trim = 15mm 0mm 10mm 0mm, width = 0.32\textwidth]{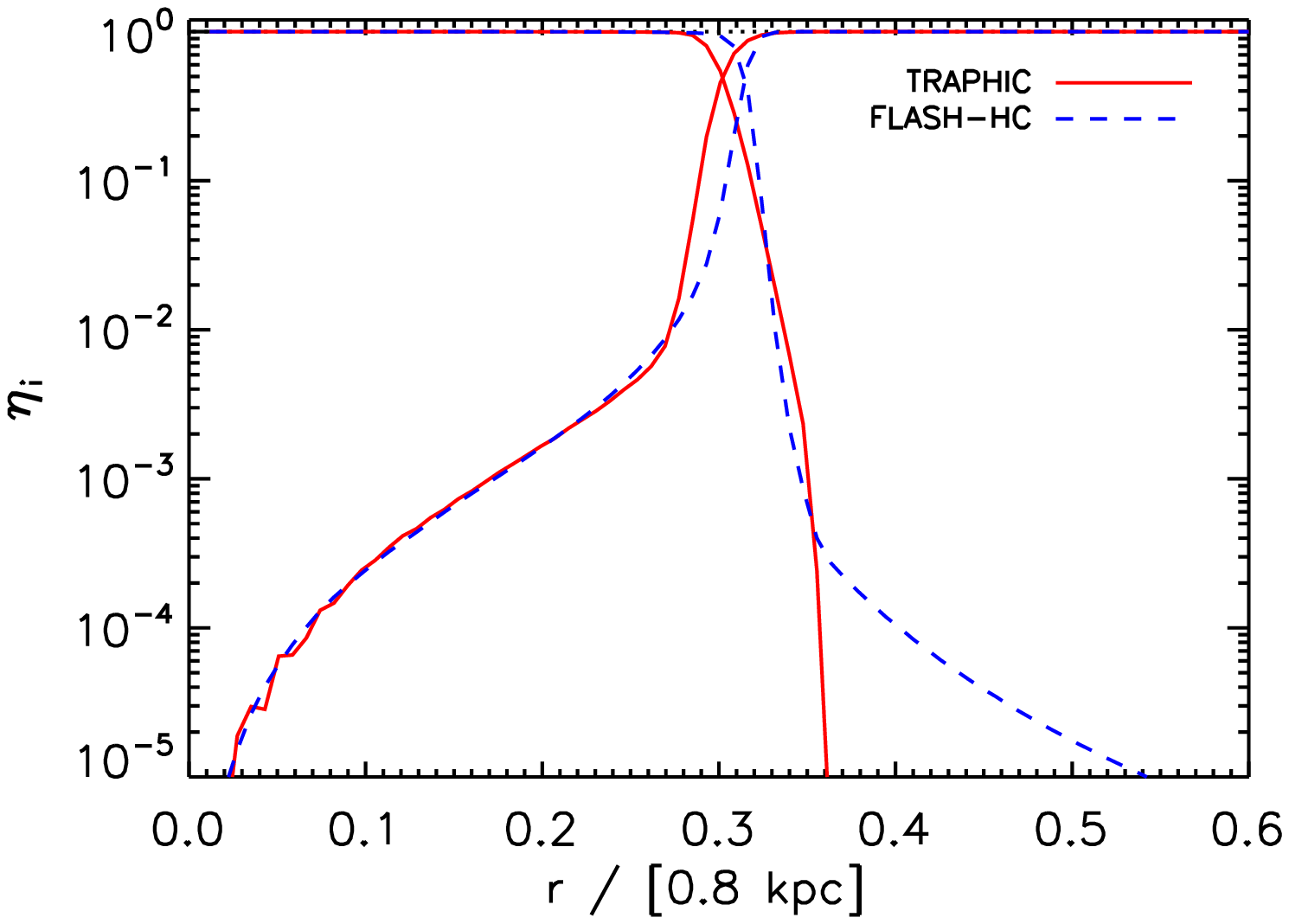}
\includegraphics[trim = 15mm 0mm 10mm 0mm, width = 0.32\textwidth]{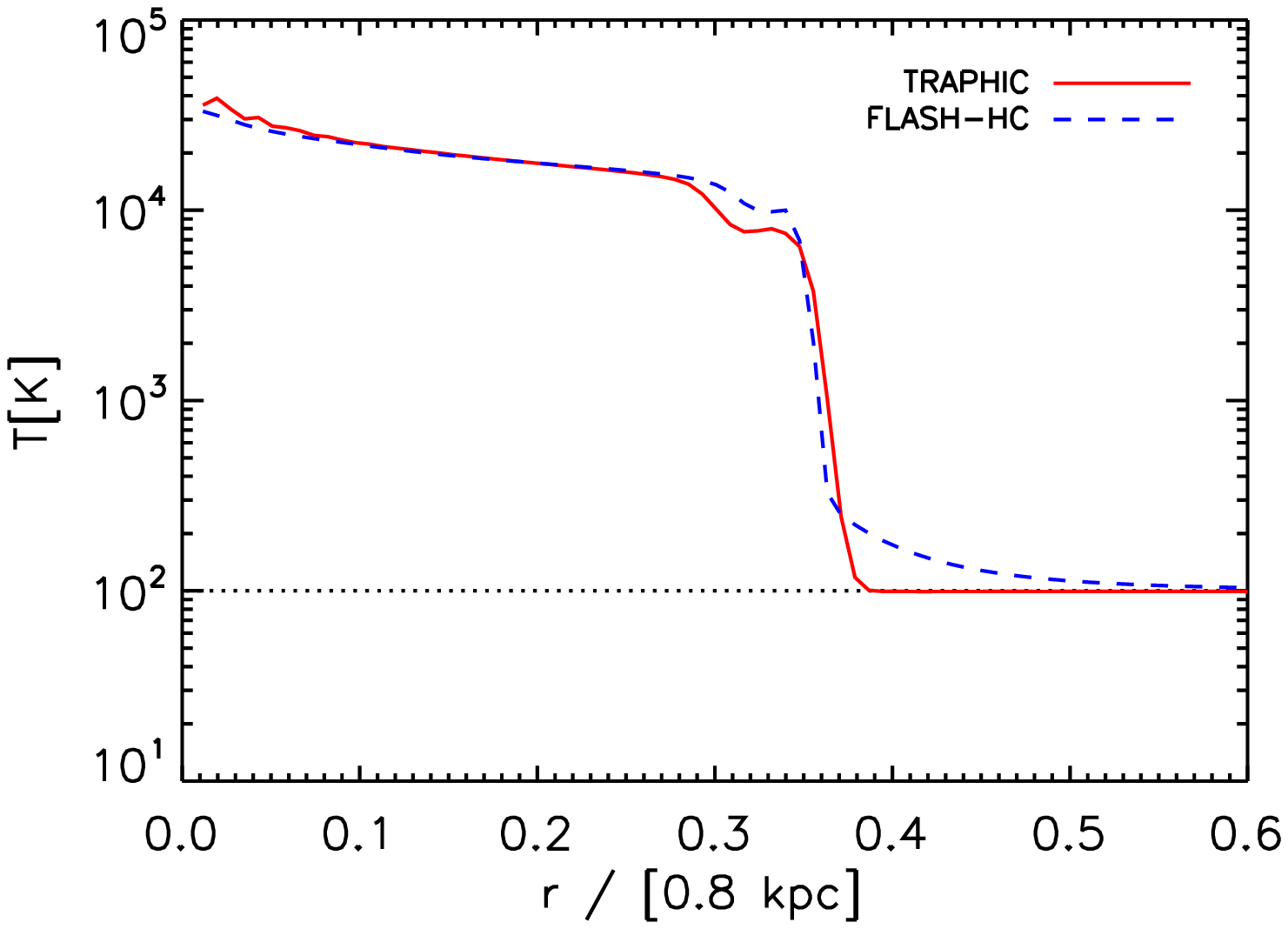}
\includegraphics[trim = 15mm 0mm 10mm 0mm, width = 0.32\textwidth]{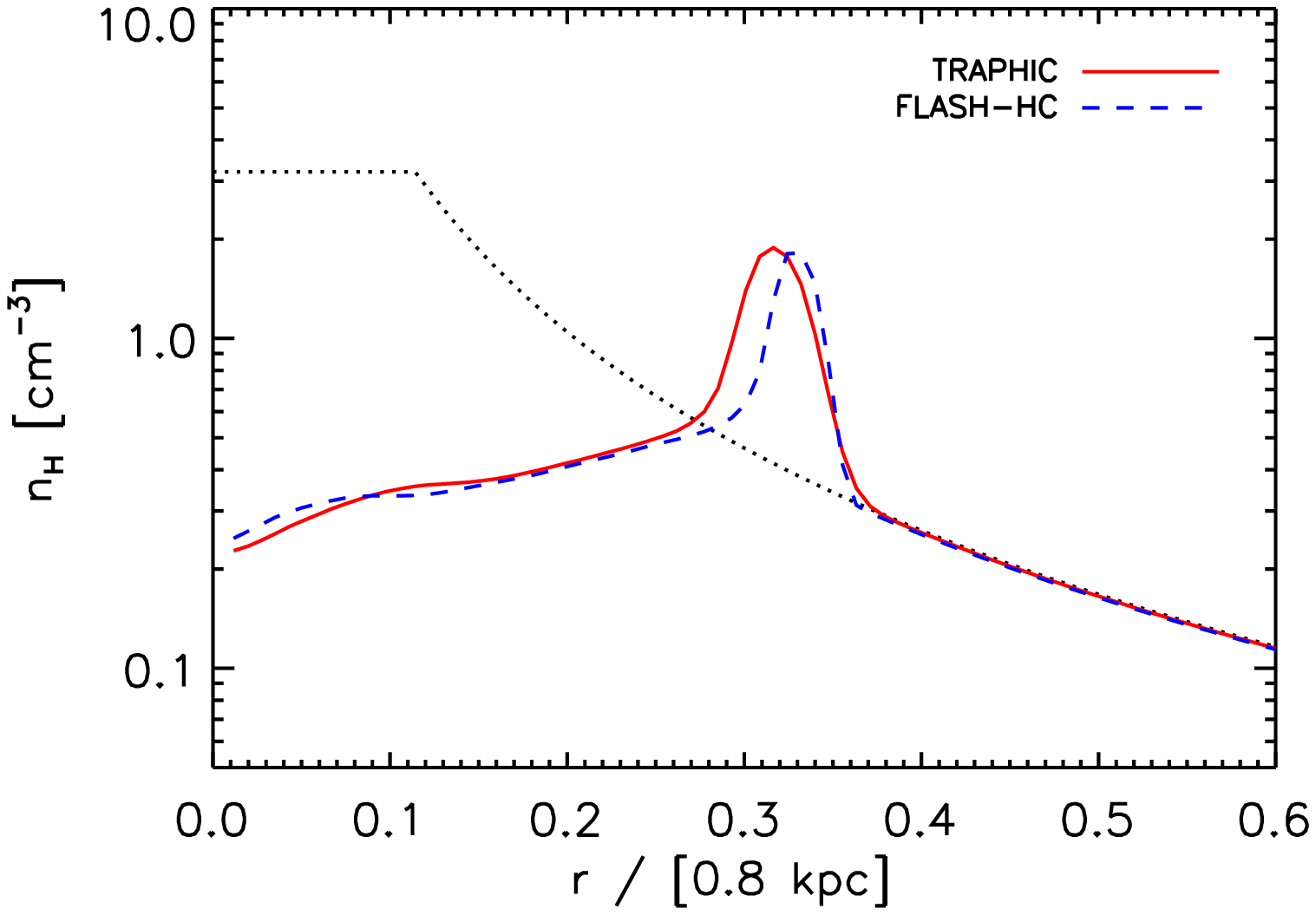}
\caption{Test 6: profiles of the neutral and ionized fractions (left), temperature (middle), 
  and gas density (right) at $t = 10 \Myr$. The results obtained 
  with \traphic\ (red solid) are compared with results obtained with {\sc flash-hc} as published in \citealp{Iliev:2009} (blue dashed). 
  The initial profiles at $t = 0$ are also shown (black dotted). The agreement between the two results is very good. The small differences in the 
  ionized fraction and temperature at $r/0.8 \kpc \gtrsim 0.35$ 
  are primarily caused by differences in the numerical treatment of the blackbody spectrum. The monochromatic simulation 
  with \traphic\ employed a single frequency bin in the grey approximation, which cannot capture the pre-ionization and 
  preheating ahead of the ionization front found in the multi-frequency simulation with {\sc flash-hc}. The shell of dense gas is slightly broader in the simulation with 
  \traphic\ than in the simulation with {\sc flash-hc}, and its propagation is slightly delayed, leading to a slightly smaller 
  radius at which the densities profile peaks. The small difference in the locations of the density profile peak explains the similarly small 
  difference in the locations of the ionization front between the two simulations. The differences seen here are comparable to or smaller than 
  the differences between the results obtained with a range of other RT codes shown in \cite{Iliev:2009}.}
\label{Fig:Minihalo2}
\end{center}
\end{figure*}

\label{sec:tests:rhd}
In this section we perform Test~6 of the cosmological RT code
comparison project (\citealp{Iliev:2009}). This test consists of the
simulation of the expansion of a \ion{H}{2} region driven by an ionizing
point source at the center of a spherically symmetric region with a
steeply declining gas density profile. In contrast to the test
described in the previous section, the gas distribution is not forced
to remain static but may evolve in response to photoionization
heating from the central source. The test situation shares some of the
main features of the evolution of gas densities inside high-redshift
cosmological minihalos under photoionization from a central massive metal-free
star.
\par 
The physical setup of the test is as follows. The initial gas densities are described by 
a cored isothermal density profile, 
\begin{equation}
 n_{\rm H}(r) = \left\{ 
  \begin{array}{l l}
    n_0 & \quad \text{if $r \le r_0$, }\\
    n_0 (r/r_0)^{-2} & \quad \text{otherwise.}\\
  \end{array} 
  \right.
\end{equation}
We follow \cite{Iliev:2009} and set $n_0=3.2 \cmci$ and $r_0 = 91.5
\pc$, and we assume that the gas, consisting solely of atomic
hydrogen, is initially neutral at temperature $100\K$. The
central source has a constant ionizing luminosity of $10^{50}  \,\textrm{photons} \invs$
throughout the simulation, and is characterized by a
blackbody spectrum with temperature $10^5 \K$. Following \cite{Iliev:2009}, 
all relevant cooling processes are included, except for 
Compton cooling. In this test, gravitational forces are ignored. 
\par 
We implement the
power-law density profile by applying a local radial stretch along
the direction towards the central source to the inter-particle
distances of SPH particles that are initially distributed uniformly
with density $\tilde{n}_{\rm H}$. Denoting the initial particle
positions with $(\tilde{r}, \theta, \phi)$, such a stretching is
expressed by the transformation $(\tilde{r}, \theta, \phi) \to (r,
\theta, \phi)$, where $\theta$ and $\phi$ are the usual two angles
needed to specify a position in spherical coordinates. Mass
conservation then requires the new coordinates $(r, \theta, \phi)$ to
satisfy
\begin{equation}
n_{\rm H}(r) r^2 \sin\theta dr d\theta d\phi =\tilde{n}_{\rm H} \tilde{r}^2 \sin\theta d\tilde{r} d\theta d\phi. 
\end{equation}
Substituting a power-law density profile, $n_{\rm
H}(r) \propto r^n$, the last equation can be integrated to yield $r
\propto \tilde{r}^{3/(3-n)}$. In the present case, $n=2$, and hence $r
\propto \tilde{r}^3$. Therefore, a singular isothermal profile can be
obtained by stretching the initial inter-particle distances between
uniformly distributed particles along the radial directions towards
the center by a factor $\tilde{r}^2$. The initial uniform density field
is generated by placing particles randomly in the simulation box. To
reduce the associated shot noise, the resulting particle distribution is
regularized by evolving the particles under the influence of a
reversed-sign (i.e., repulsive) gravitational force until the particles
settle down into a glass-like quasi-equilibrium
(\citealp{White:1996}). After generating the singular isothermal
density profile, we replace the central region within $r_0$ with a
uniform glass-like particle distribution to introduce the central
core. 
\par
The numerical parameters used in this test are as follows. We place
the ionizing source in the center of a box with linear size $1.6
\kpc$. We sample the gas inside the core with radius $r_0$ using 30000
particles, which corresponds to an equivalent uniform grid resolution
of $160^3$ cells. This resolution is slightly higher than the uniform
grid resolution employed in \cite{Iliev:2006}, which was $128^3$
cells. We use an angular resolution $N_{\rm c} = 8$, a number of
neighbors for the emitting source $\tilde{N}_{\rm ngb}=32$, employ a
RT time step $\Delta t_{\rm r} = 10^{-5} \Myr$, and
transport photons only over a single inter-particle distance per time
step. We limit the sizes of the individual particle hydrodynamical
time steps to be less than $0.1 \Myr$. The radiation is
transported using a single frequency bin and assuming 
a grey photoionization cross-section $\langle
\sigma_{\gamma \rm HI}\rangle = 1.63 \times 10^{-18} \cms$, as well as that
each photoionization adds $\langle \epsilon_{\rm HI} \rangle = 6.32
\eV$ to the thermal energy of the gas. The parameter values are
chosen to make contact with the conditions in the simulations
presented in the main part of this paper, and to ensure numerical
convergence. The test results are not critically sensitive to the
adopted parameter values. However, we note that because of our choice of emitting photon packets
only once per RT time step, the angular sampling depends on the RT time step, and a larger RT time step
would imply an increased scatter in the ionized fraction, 
temperature and density profiles (see Appendix~\ref{sec:parameters}). 
\par
The results of the test are shown and discussed in Figure~\ref{Fig:Minihalo2}. The results 
obtained with \traphic\ are in very good agreement with the reference results 
published in \cite{Iliev:2009}.

\subsection{Effect of Limited Photon Packet Propagation Speed and Finite Angular Sampling}
\label{sec:parameters}
\begin{figure*}
\begin{center}
\includegraphics[trim = 0mm 0mm 0mm 0mm, width = 0.32\textwidth]{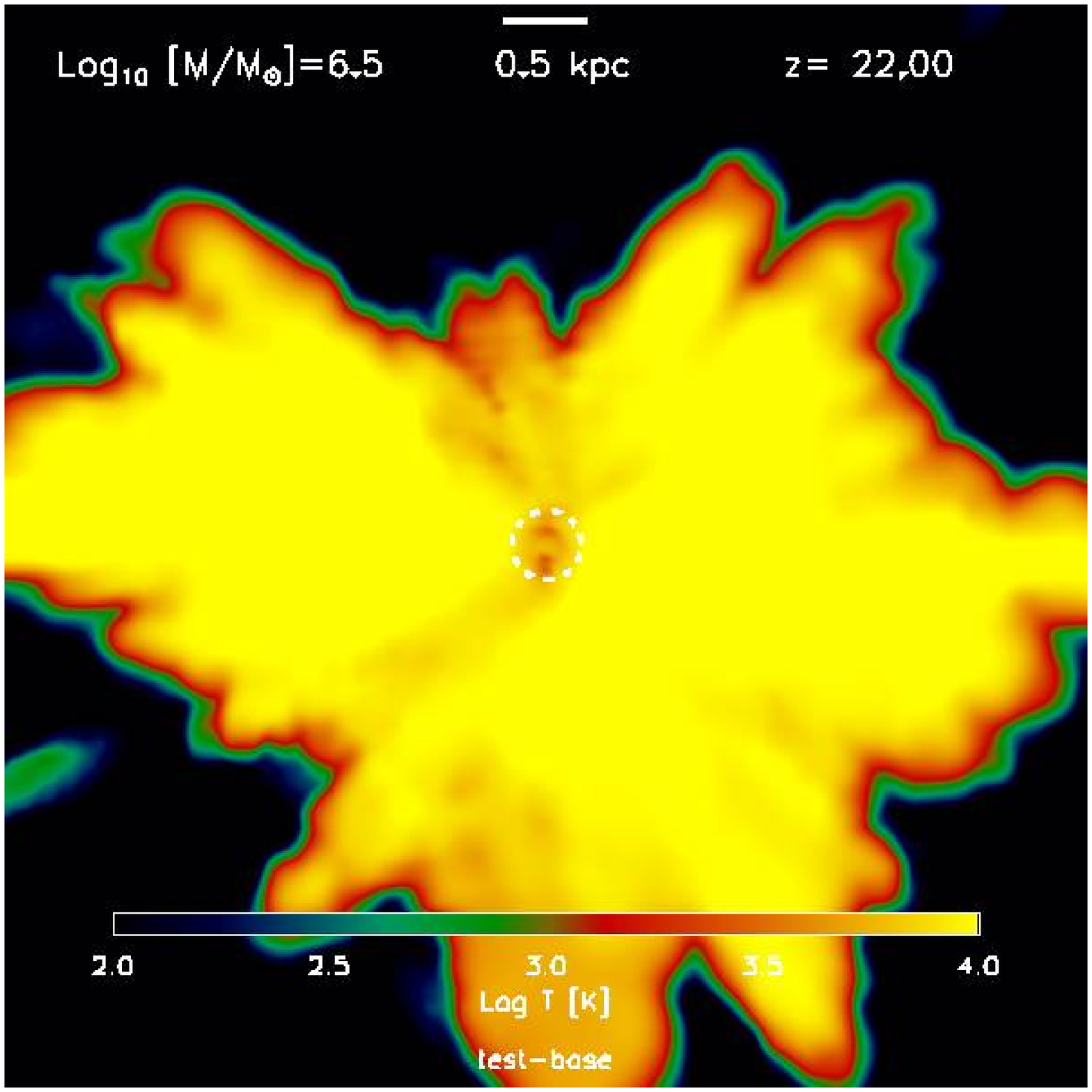}
\includegraphics[trim = 0mm 0mm 0mm 0mm, width = 0.32\textwidth]{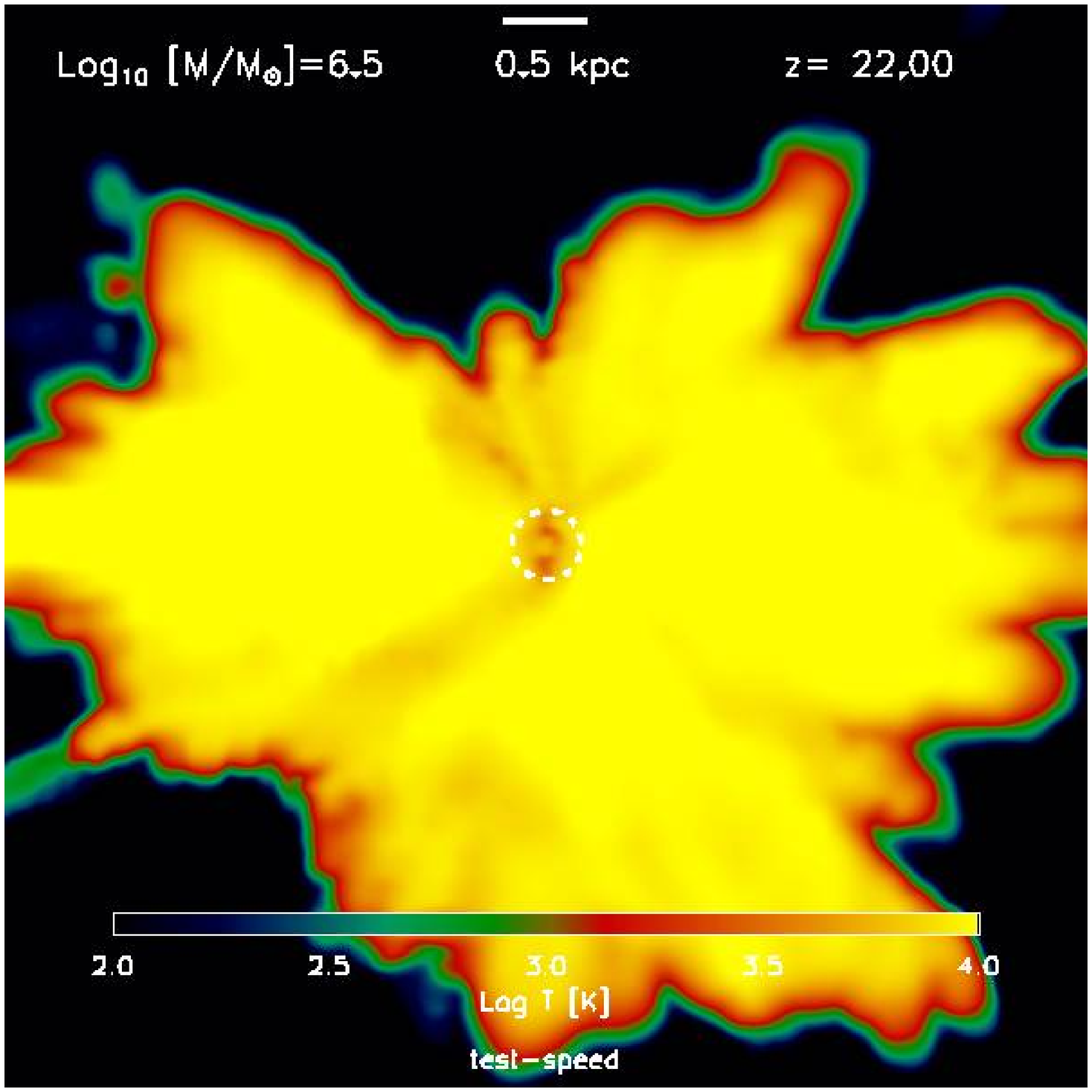}
\includegraphics[trim = 0mm 0mm 0mm 0mm, width = 0.32\textwidth]{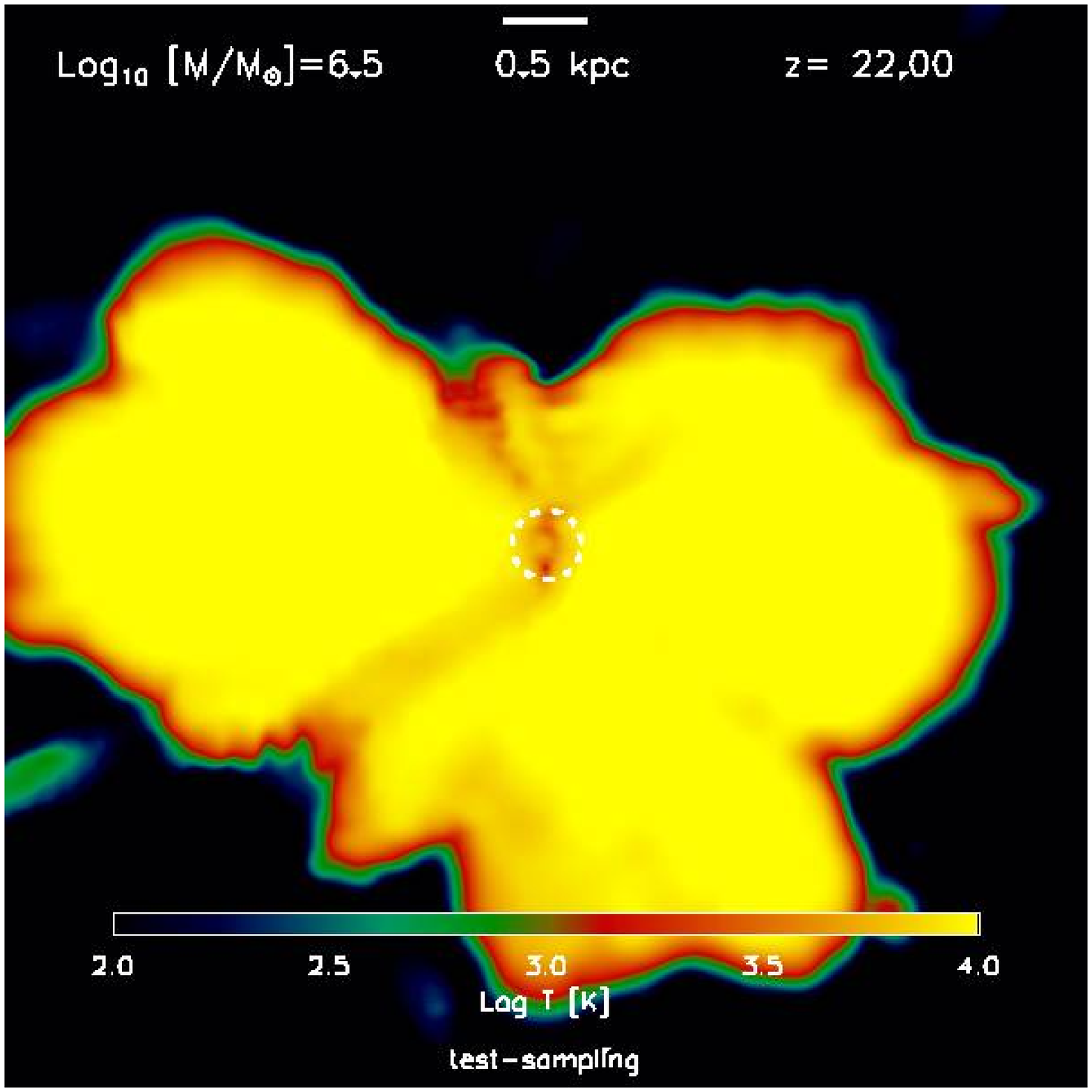}
\caption{Temperature averaged in a thin slice of thickness $0.2 \kpc$
  through the center of the minihalo progenitor at $z = 22$ in three
  simulations that are identical to simulation {\it LW+RT} discussed
  in the main text except for the treatment of the RT. {\it
  Left}: simulation {\it test-base}, with a single emission along $N_{\rm EC} = 8$ directions per RT time step and propagation at the speed of
  light but by at most 1 inter-particle distance, as in simulation {\it
  LW+RT} discussed in the main text. {\it Middle}: simulation {\it test-speed}, with a single emission along $N_{\rm EC} = 8$ directions per RT time step and propagation
  at the speed of light (and no other limitation). {\it Right}: simulation {\it test-sampling} with $10^3$
  emissions along $N_{\rm EC} = 8$ random directions per RT step and propagation at the speed of light (and no
  other limitation). The comparison of simulations {\it test-base} (left) and {\it test-speed} (middle) shows
  that the size of the final photoheated region is insensitive to the
  limit on the distance which photon packets can travel in a given RT
  time step, thanks to the photon-conserving nature of the RT with \traphic. The
  comparison of simulations {\it test-speed} (middle) and {\it test-sampling} (right) shows that an increased
  angular sampling implemented by emitting photons along an increased
  number of random directions reduces artifacts in the shape of the
  photoheated region. Simulation {\it test-sampling} shown in the right panel has been used 
  to generate Figure~\ref{Fig:FirstStar}.}
\label{Fig:parameters}
\end{center}
\end{figure*}

In this section we discuss the effects of our approximations regarding
the limited photon propagation speed and the emission of photon
packets by ionizing sources along a finite set of random directions
per RT time step. To this end we have repeated the RT
around the first stellar burst in the minihalo progenitor of
simulation {\it LW+RT}. The first test simulation that we have
performed, hereafter {\it test-base}, employed RT parameters identical
to those employed in {\it LW+RT}. In particular, in this simulation
ionizing sources emitted photon packets along $N_{\rm EC} = 8$ random
directions per RT time step, and the photon propagation was done at
the speed of light but limited to a single inter-particle distance per
RT time step. To investigate the effect of the limitation in the
propagation speed, simulation {\it test-base} is compared with a
simulation in which photon packets propagate at the speed of light
with no additional limitation regarding the number of inter-particle
distances per RT time step (hereafter {\it test-speed}; and with
ionizing sources emitting photons into $N_{\rm EC} = 8$ 
random directions per RT time step as in {\it test-base}). To investigate the effect of
the angular sampling, simulation {\it test-speed} is compared with a
simulation in which sources emit photon packets along $10^3$ times
more random directions per RT time step (hereafter {\it
test-sampling}; and transporting photons at the speed of light as in {\it test-speed}). The
latter is implemented by repeating the emission of radiation along the
$N_{\rm EC} = 8$ random directions $10^3$ times per RT time step, with
each emission using a different set of $N_{\rm EC} = 8$ random
directions and emitting photon packets that contain by a factor $10^3$
fewer photons.
\par
Figure~\ref{Fig:parameters} shows the gas temperature in a thin slice
through the center of the minihalo progenitor at $z = 22$, at the end
of the stellar burst, in the three test simulations {\it test-base}
(left), {\it test-speed} (middle), and {\it test-sampling}
(right). The final extent of the photoheated region is similar in
simulations {\it test-base} and {\it test-speed} despite the difference
in propagation speeds. This is because the RT with \traphic\ conserves
the number of ionizing photons that are transmitted and absorbed, and
because the final extent of photoheated \ion{H}{2} regions is set primarily by this
number. However, limitations of the speed at which photons are
propagated can lead to differences in extent and shape of the
developing \ion{H}{2} regions early during their evolution (e.g., \citealp{Pawlik:2008}). The
increased angular sampling in simulation {\it test-sampling} reduces
the noise that characterizes the shape of the photoheated region in
simulation {\it test-base}, as expected. However, this noise is already
small in simulation {\it test-base}, which hence yields a photoheated
region similar in shape to that in simulation {\it test-sampling}. We note 
(but do not show) that the evolution of the properties of the minihalo, such as, e.g., 
the maximum gas density, is nearly identical in all three test simulations.


\begin{thebibliography}{}

\bibitem[Abel et al.(1997)]{Abel:1997} Abel, T., Anninos, P., 
Zhang, Y., \& Norman, M.~L.\ 1997, New Astronomy, 2, 181 

\bibitem[Abel et al.(1999)]{Abel:1999b} Abel, T., Norman, M.~L., 
\& Madau, P.\ 1999, \apj, 523, 66 

\bibitem[Abel et al.(2002)]{Abel:2002} Abel, T., Bryan, G.~L., 
\& Norman, M.~L.\ 2002, Science, 295, 93 

\bibitem[Abel et al.(2007)]{Abel:2007} Abel, T., Wise, J.~H., 
\& Bryan, G.~L.\ 2007, \apjl, 659, L87 

\bibitem[Ahn 
\& Shapiro(2007)]{Ahn:2007} Ahn, K., \& Shapiro, P.~R.\ 2007, \mnras, 375, 881 

\bibitem[Ahn et al.(2009)]{Ahn:2009} Ahn, K., Shapiro, P.~R., 
Iliev, I.~T., Mellema, G., \& Pen, U.-L.\ 2009, \apj, 695, 1430 

\bibitem[Ahn et al.(2012)]{Ahn:2012} Ahn, K., Iliev, I.~T., 
Shapiro, P.~R., et al.\ 2012, \apjl, 756, L16 

\bibitem[Alvarez et al.(2006)]{Alvarez:2006} Alvarez, M.~A., Bromm, 
V., \& Shapiro, P.~R.\ 2006, \apj, 639, 621

\bibitem[Alvarez et al.(2009)]{Alvarez:2009} Alvarez, M.~A., Wise, 
J.~H., \& Abel, T.\ 2009, \apjl, 701, L133 

\bibitem[Barkana 
\& Loeb(1999)]{Barkana:1999} Barkana, R., \& Loeb, A.\ 1999, \apj, 523, 54 

\bibitem[Barkana 
\& Loeb(2001)]{Barkana:2001} Barkana, R., \& Loeb, A.\ 2001, \physrep, 349, 125 

\bibitem[Blumenthal et al.(1986)]{Blumenthal:1986} Blumenthal, G.~R., 
Faber, S.~M., Flores, R., \& Primack, J.~R.\ 1986, \apj, 301, 27 

\bibitem[Bouwens et al.(2009)]{Bouwens:2009} Bouwens, R.~J., 
Illingworth, G.~D., Franx, M., et al.\ 2009, \apj, 705, 936 

\bibitem[Bouwens et al.(2010)]{Bouwens:2010} Bouwens, R.~J., 
Illingworth, G.~D., Oesch, P.~A., et al.\ 2010, \apjl, 708, L69 

\bibitem[Bouwens et al.(2012)]{Bouwens:2012} Bouwens, R.~J., 
Illingworth, G.~D., Oesch, P.~A., et al.\ 2012, \apjl, 752, L5 

\bibitem[Bovill 
\& Ricotti(2009)]{Bovill:2009} Bovill, M.~S., \& Ricotti, M.\ 2009, \apj, 693, 1859 

\bibitem[Bovill 
\& Ricotti(2011)]{Bovill:2011} Bovill, M.~S., \& Ricotti, M.\ 2011, \apj, 741, 18 

\bibitem[Boylan-Kolchin et al.(2011)]{Boylan:2011} Boylan-Kolchin, 
M., Bullock, J.~S., \& Kaplinghat, M.\ 2011, \mnras, 415, L40 

\bibitem[Boylan-Kolchin et al.(2012)]{Boylan:2012} Boylan-Kolchin, 
M., Bullock, J.~S., \& Kaplinghat, M.\ 2012, \mnras, 422, 1203

\bibitem[Bovy 
\& Dvorkin(2012)]{Bovy:2012} Bovy, J., \& Dvorkin, C.\ 2012, arXiv:1205.2083 

\bibitem[Bromm et al.(2001)]{Bromm:2001b} Bromm, V., Kudritzki, 
R.~P., \& Loeb, A.\ 2001, \apj, 552, 464 

\bibitem[Bromm et al.(2002)]{Bromm:2002} Bromm, V., Coppi, P.~S., 
\& Larson, R.~B.\ 2002, \apj, 564, 23 

\bibitem[Bromm 
\& Larson(2004)]{Bromm:2004} Bromm, V., \& Larson, R.~B.\ 2004, \araa, 42, 79 

\bibitem[Bromm 
\& Yoshida(2011)]{Bromm:2011} Bromm, V., \& Yoshida, N.\ 2011, \araa, 49, 373 

\bibitem[Brooks 
\& Zolotov(2012)]{Brooks:2012} Brooks, A.~M., \& Zolotov, A.\ 2012, arXiv:1207.2468 

\bibitem[Brown et al.(1989)]{Brown:1989}  Brown P.~N., Byrne G.~D., \& Hindmarsh, A.~C.\ 1989, SIAM J. Sci. Stat. Comput., 10, 1038

\bibitem[Bullock(2010)]{Bullock:2010} Bullock, J.~S.\ 2010, 
arXiv:1009.4505

\bibitem[Bunker et al.(2004)]{Bunker:2004} Bunker, A.~J., Stanway, 
E.~R., Ellis, R.~S., \& McMahon, R.~G.\ 2004, \mnras, 355, 374 

\bibitem[Cai et al.(2011)]{Cai:2011} Cai, Z., Fan, X., Jiang, 
L., et al.\ 2011, \apjl, 736, L28 

\bibitem[Choudhury 
\& Ferrara(2007)]{Choudhury:2007} Choudhury, T.~R., \& Ferrara, A.\ 2007, \mnras, 380, L6 

\bibitem[Ciardi et al.(2001)]{Ciardi:2001} Ciardi, B., Ferrara, A., 
Marri, S., \& Raimondo, G.\ 2001, \mnras, 324, 381 

\bibitem[Ciardi 
\& Ferrara(2005)]{Ciardi:2005} Ciardi, B., \& Ferrara, A.\ 2005, \ssr, 116, 625 

\bibitem[Conroy 
\& Kratter(2012)]{Conroy:2012} Conroy, C., \& Kratter, K.\ 2012, arXiv:1205.3495 

\bibitem[Curtis-Lake et al.(2012)]{Curtis:2012} Curtis-Lake, E., 
McLure, R.~J., Dunlop, J.~S., et al.\ 2012, arXiv:1207.2727 

\bibitem[Dijkstra 
\& Kramer(2012)]{Dijkstra:2012} Dijkstra, M., \& Kramer, R.\ 2012, \mnras, 424, 1672 

\bibitem[Dove et al.(2000)]{Dove:2000} Dove, J.~B., Shull, J.~M., 
\& Ferrara, A.\ 2000, \apj, 531, 846 

\bibitem[Dunlop et al.(2012)]{Dunlop:2012} Dunlop, J.~S., McLure, 
R.~J., Robertson, B.~E., et al.\ 2012, \mnras, 420, 901 

\bibitem[Dunlop(2012)]{D:2012} Dunlop, J.~S.\ 2012, 
arXiv:1205.1543 

\bibitem[Durier 
\& Dalla Vecchia(2011)]{Durier:2011} Durier, F., \& Dalla Vecchia, C.\ 2011, arXiv:1105.3729 

\bibitem[Efstathiou(1992)]{Efstathiou:1992} Efstathiou, G.\ 1992, 
\mnras, 256, 43P 

\bibitem[Eisenstein 
\& Loeb(1995)]{Eisenstein:1995} Eisenstein, D.~J., \& Loeb, A.\ 1995, \apj, 443, 11 

\bibitem[Finkelstein et al.(2010)]{Finkelstein:2010} Finkelstein, S.~L., 
Papovich, C., Giavalisco, M., et al.\ 2010, \apj, 719, 1250 

\bibitem[Finkelstein et al.(2012)]{Finkelstein:2012} Finkelstein, S.~L., 
Papovich, C., Ryan, R.~E., Jr., et al.\ 2012, arXiv:1206.0735 

\bibitem[Fumagalli et al.(2011)]{Fumagalli:2011} Fumagalli, M., 
O'Meara, J.~M., 
\& Prochaska, J.~X.\ 2011, Science, 334, 1245 

\bibitem[Furlanetto et al.(2006)]{Furlanetto:2006} Furlanetto, S.~R., 
Oh, S.~P., \& Briggs, F.~H.\ 2006, \physrep, 433, 181 

\bibitem[Gardner et al.(2006)]{Gardner:2006} Gardner, J.~P., Mather, 
J.~C., Clampin, M., et al.\ 2006, \ssr, 123, 485 

\bibitem[Garrison-Kimmel et al.(2013)]{Garrison:2013} 
Garrison-Kimmel, S., Rocha, M., Boylan-Kolchin, M., Bullock, J., 
\& Lally, J.\ 2013, arXiv:1301.3137 

\bibitem[Glover 
\& Brand(2003)]{Glover:2003} Glover, S.~C.~O., \& Brand, P.~W.~J.~L.\ 2003, \mnras, 340, 210 

\bibitem[Glover 
\& Jappsen(2007)]{Glover:2007} Glover, S.~C.~O., \& Jappsen, A.-K.\ 2007, \apj, 666, 1 

\bibitem[Gnedin 
\& Hui(1998)]{Gnedin:1998} Gnedin, N.~Y., \& Hui, L.\ 1998, \mnras, 296, 44 

\bibitem[Gnedin et al.(2004)]{Gnedin:2004} Gnedin, O.~Y., Kravtsov, 
A.~V., Klypin, A.~A., \& Nagai, D.\ 2004, \apj, 616, 16 

\bibitem[Gnedin et al.(2008)]{Gnedin:2008} Gnedin, N.~Y., Kravtsov, 
A.~V., \& Chen, H.-W.\ 2008, \apj, 672, 765

\bibitem[Gnedin et al.(2011)]{Gnedin:2011} Gnedin, O.~Y., Ceverino, 
D., Gnedin, N.~Y., et al.\ 2011, arXiv:1108.5736 

\bibitem[Governato et al.(2012)]{Governato:2012} Governato, F., 
Zolotov, A., Pontzen, A., et al.\ 2012, \mnras, 422, 1231

\bibitem[Greif 
\& Bromm(2006)]{Greif:2006} Greif, T.~H., \& Bromm, V.\ 2006, \mnras, 373, 128 

\bibitem[Greif et al.(2008)]{Greif:2008} Greif, T.~H., Johnson, 
J.~L., Klessen, R.~S., \& Bromm, V.\ 2008, \mnras, 387, 1021 

\bibitem[Greif et al.(2010)]{Greif:2010} Greif, T.~H., Glover, 
S.~C.~O., Bromm, V., \& Klessen, R.~S.\ 2010, \apj, 716, 510 

\bibitem[Haas et al.(2012)]{Haas:2012} Haas, M.~R., Schaye, J., 
Booth, C.~M., et al.\ 2012, arXiv:1211.3120 

\bibitem[Haiman et al.(1996)]{Haiman:1996} Haiman, Z., Rees, M.~J., 
\& Loeb, A.\ 1996, \apj, 467, 522 

\bibitem[Haiman et al.(1996b)]{Haiman:1996b} Haiman, Z., Thoul, 
A.~A., \& Loeb, A.\ 1996, \apj, 464, 523 

\bibitem[Haiman et al.(1997)]{Haiman:1997} Haiman, Z., Rees, M.~J., 
\& Loeb, A.\ 1997, \apj, 484, 985 

\bibitem[Haiman 
\& Loeb(1998)]{Haiman:1998} Haiman, Z., \& Loeb, A.\ 1998, \apj, 503, 505 

\bibitem[Haiman et al.(2000)]{Haiman:2000} Haiman, Z., Abel, T., 
\& Rees, M.~J.\ 2000, \apj, 534, 11 

\bibitem[Heger et al.(2003)]{Heger:2003} Heger, A., Fryer, C.~L., 
Woosley, S.~E., Langer, N., \& Hartmann, D.~H.\ 2003, \apj, 591, 288 

\bibitem[Hummel et al.(2012)]{Hummel:2011} Hummel, J.~A., Pawlik, 
A.~H., Milosavljevic, M., \& Bromm, V.\ 2012, ApJ, in press (arXiv:1112.5207)

\bibitem[Iliev et al.(2006)]{Iliev:2006} Iliev, I.~T., et al.\ 
2006, \mnras, 371, 1057 


\bibitem[Iliev et al.(2009)]{Iliev:2009} Iliev, I.~T., et al.\ 
2009, \mnras, 400, 1283 

\bibitem[Inoue(2011)]{Inoue:2011} Inoue, A.~K.\ 2011, \mnras, 415, 
2920

\bibitem[Iye et al.(2006)]{Iye:2006} Iye, M., Ota, K., 
Kashikawa, N., et al.\ 2006, \nat, 443, 186 

\bibitem[Jeon et al.(2012)]{Jeon:2012} Jeon, M., Pawlik, A.~H., 
Greif, T.~H., et al.\ 2012, \apj, 754, 34 

\bibitem[Johnson 
\& Bromm(2006)]{Johnson:2006} Johnson, J.~L., \& Bromm, V.\ 2006, \mnras, 366, 247

\bibitem[Johnson et al.(2007)]{Johnson:2007} Johnson, J.~L., Greif, 
T.~H., \& Bromm, V.\ 2007, \apj, 665, 85 

\bibitem[Johnson et al.(2009)]{Johnson:2009} Johnson, J.~L., Greif, 
T.~H., Bromm, V., Klessen, R.~S., \& Ippolito, J.\ 2009, \mnras, 399, 37 

\bibitem[Johnson(2010)]{Johnson:2010} Johnson, J.~L.\ 2010, \mnras, 
404, 1425 

\bibitem[Johnson et al.(2012)]{Johnson:2012} Johnson, J.~L., Dalla 
Vecchia, C., \& Khochfar, S.\ 2012, arXiv:1206.5824 

\bibitem[Kaufmann et al.(2007)]{Kaufmann:2007} Kaufmann, T., Wheeler, 
C., \& Bullock, J.~S.\ 2007, \mnras, 382, 1187 

\bibitem[Klypin et al.(1999)]{Klypin:1999} Klypin, A., Kravtsov, 
A.~V., Valenzuela, O., \& Prada, F.\ 1999, \apj, 522, 82 

\bibitem[Komatsu et al.(2011)]{Komatsu:2011} Komatsu, E., Smith, 
K.~M., Dunkley, J., et al.\ 2011, \apjs, 192, 18 

\bibitem[Koushiappas et al.(2004)]{Koushiappas:2004} Koushiappas, S.~M., 
Bullock, J.~S., \& Dekel, A.\ 2004, \mnras, 354, 292 

\bibitem[Kravtsov(2003)]{Kravtsov:2003} Kravtsov, A.~V.\ 2003, \apjl, 590, L1 

\bibitem[Krumholz 
\& Tan(2007)]{Krumholz:2007} Krumholz, M.~R., \& Tan, J.~C.\ 2007, \apj, 654, 304 

\bibitem[Kuhlen 
\& Faucher-Gigu{\`e}re(2012)]{KuhlenFG:2012} Kuhlen, M., \& Faucher-Gigu{\`e}re, C.-A.\ 2012, \mnras, 423, 862 

\bibitem[Labb{\'e} et al.(2010)]{Labbe:2010} Labb{\'e}, I., 
Gonz{\'a}lez, V., Bouwens, R.~J., et al.\ 2010, \apjl, 716, L103 


\bibitem[Lehnert et al.(2010)]{Lehnert:2010} Lehnert, M.~D., 
Nesvadba, N.~P.~H., Cuby, J.-G., et al.\ 2010, \nat, 467, 940 

\bibitem[Li et al.(2005)]{Li:2005} Li, Y., Mac Low, M.-M., 
\& Klessen, R.~S.\ 2005, \apj, 626, 823 

\bibitem[Lodato 
\& Natarajan(2006)]{Lodato:2006} Lodato, G., \& Natarajan, P.\ 2006, \mnras, 371, 1813 

\bibitem[Loeb 
\& Rybicki(1999)]{Loeb:1999} Loeb, A., \& Rybicki, G.~B.\ 1999, \apj, 524, 527 

\bibitem[Loeb(2010)]{Loeb:2010} Loeb, A.\ 2010, How Did the First 
Stars and Galaxies Form? Princeton Univ. Press, Princeton

\bibitem[Lovell et al.(2012)]{Lovell:2012} Lovell, M.~R., Eke, V., 
Frenk, C.~S., et al.\ 2012, \mnras, 420, 2318 

\bibitem[Machacek et al.(2001)]{Machacek:2001} Machacek, M.~E., 
Bryan, G.~L., \& Abel, T.\ 2001, \apj, 548, 509 

\bibitem[Mac Low 
\& Ferrara(1999)]{MacLow:1999} Mac Low, M.-M., \& Ferrara, A.\ 1999, \apj, 513, 142 

\bibitem[Maio et al.(2009)]{Maio:2009} 
Maio, U., Ciardi, B., Yoshida, N., Dolag, K., \& Tornatore, L.\ 2009, \aap, 503, 25 

\bibitem[Maio et al.(2010)]{Maio:2010} Maio, U., Ciardi, B., 
Dolag, K., Tornatore, L., \& Khochfar, S.\ 2010, \mnras, 407, 1003 

\bibitem[Maselli et al.(2003)]{Maselli:2003} Maselli, A., Ferrara, 
A., \& Ciardi, B.\ 2003, \mnras, 345, 379 

\bibitem[Maselli et al.(2009)]{Maselli:2009} Maselli, A., Ciardi, 
B., \& Kanekar, A.\ 2009, \mnras, 393, 171 


\bibitem[Mashchenko et al.(2006)]{Mashchenko:2006} Mashchenko, S., 
Couchman, H.~M.~P., \& Wadsley, J.\ 2006, \nat, 442, 539

\bibitem[Mashchenko et al.(2008)]{Mashchenko:2008} Mashchenko, S., 
Wadsley, J., \& Couchman, H.~M.~P.\ 2008, Science, 319, 174 

\bibitem[Mayer et al.(2007)]{Mayer:2007} Mayer, L., Kazantzidis, 
S., Mastropietro, C., \& Wadsley, J.\ 2007, \nat, 445, 738 

\bibitem[Mayer(2010)]{Mayer:2010} Mayer, L.\ 2010, Advances in 
Astronomy, 2010, article id. 278434 

\bibitem[McKee 
\& Ostriker(2007)]{McKee:2007} McKee, C.~F., \& Ostriker, E.~C.\ 2007, \araa, 45, 565 

\bibitem[Meiksin(2009)]{Meiksin:2009} Meiksin, A.~A.\ 2009, Rev. Mod. Phys., 81, 1405 

\bibitem[Mellema et al.(2006)]{Mellema:2006} Mellema, G., Iliev, 
I.~T., Alvarez, M.~A., \& Shapiro, P.~R.\ 2006, New Astronomy, 11, 374 

\bibitem[Mesinger et al.(2006)]{Mesinger:2006} Mesinger, A., Bryan, 
G.~L., \& Haiman, Z.\ 2006, \apj, 648, 835 

\bibitem[Mihalas 
\& Weibel Mihalas(1984)]{Mihalas:1984} Mihalas, D., \& Weibel Mihalas, B.\ 1984, New York: Oxford University Press, 1984,  

\bibitem[Moore et al.(1999)]{Moore:1999} Moore, B., Ghigna, S., 
Governato, F., et al.\ 1999, \apjl, 524, L19 

\bibitem[Muratov et al.(2012)]{Muratov:2012} Muratov, A.~L., Gnedin, 
O.~Y., Gnedin, N.~Y., \& Zemp, M.\ 2012, arXiv:1212.0909 

\bibitem[Naoz et al.(2006)]{Naoz:2006} Naoz, S., Noter, S., 
\& Barkana, R.\ 2006, \mnras, 373, L98 

\bibitem[Navarro et al.(1996)]{NavarroEke:1996} Navarro, J.~F., Eke, 
V.~R., \& Frenk, C.~S.\ 1996, \mnras, 283, L72 

\bibitem[Navarro et al.(1997)]{Navarro:1997} Navarro, J.~F., Frenk, 
C.~S., \& White, S.~D.~M.\ 1997, \apj, 490, 493


\bibitem[Oh(1999)]{Oh:1999} Oh, S.~P.\ 1999, \apj, 527, 16 

\bibitem[Oh et al.(2001)]{Oh:2001} Oh, S.~P., Haiman, Z., 
\& Rees, M.~J.\ 2001, \apj, 553, 73 

\bibitem[Oh 
\& Haiman(2002)]{Oh:2002} Oh, S.~P., \& Haiman, Z.\ 2002, \apj, 569, 558 

\bibitem[Okamoto et al.(2008)]{Okamoto:2008} Okamoto, T., Gao, L., 
\& Theuns, T.\ 2008, \mnras, 390, 920 

\bibitem[Ono et al.(2012)]{Ono:2012} Ono, Y., Ouchi, M., 
Mobasher, B., et al.\ 2012, \apj, 744, 83 

\bibitem[O'Shea et al.(2005)]{Oshea:2005} O'Shea, B.~W., Abel, T., 
Whalen, D., \& Norman, M.~L.\ 2005, \apjl, 628, L5 

\bibitem[O'Shea 
\& Norman(2007)]{Oshea:2007} O'Shea, B.~W., \& Norman, M.~L.\ 2007, \apj, 654, 66 

\bibitem[O'Shea 
\& Norman(2008)]{Oshea:2008} O'Shea, B.~W., \& Norman, M.~L.\ 2008, \apj, 673, 14 


\bibitem[Panagia(2005)]{Panagia:2005} Panagia, N.\ 2005, The Initial Mass Function 50 Years Later, Vol. 327, ed. E.
Corbelli \& F. Palle (Dordrecht: Springer), 479

\bibitem[Pawlik \& Schaye (2008)]{Pawlik:2008} Pawlik A.~H., Schaye J., 2008, MNRAS, 389, 651 

\bibitem[Pawlik et al.(2011)]{Pawlik:2011a} Pawlik, A.~H., 
Milosavljevi{\'c}, M., \& Bromm, V.\ 2011, \apj, 731, 54 

\bibitem[Pawlik 
\& Schaye(2011)]{Pawlik:2011b} Pawlik, A.~H., \& Schaye, J.\ 2011, \mnras, 412, 1943 

\bibitem[Petri et al.(2012)]{Petri:2012} Petri, A., Ferrara, A., 
\& Salvaterra, R.\ 2012, \mnras, 422, 1690 


\bibitem[Prieto et al.(2013)]{Prieto:2013} Prieto, J., Jimenez, R., 
\& Haiman, Z.\ 2013, arXiv:1301.5567 

\bibitem[Rai{\v c}evi{\'c} et al.(2011)]{Raicevic:2011} Rai{\v 
c}evi{\'c}, M., Theuns, T., \& Lacey, C.\ 2011, \mnras, 410, 775 

\bibitem[Razoumov 
\& Cardall(2005)]{Razoumov:2005} Razoumov, A.~O., \& Cardall, C.~Y.\ 2005, \mnras, 362, 1413 

\bibitem[Razoumov 
\& Sommer-Larsen(2010)]{Razoumov:2010} Razoumov, A.~O., \& Sommer-Larsen, J.\ 2010, \apj, 710, 1239 

\bibitem[Regan 
\& Haehnelt(2009)]{Regan:2009} Regan, J.~A., \& Haehnelt, M.~G.\ 2009, \mnras, 393, 858 

\bibitem[Rhoads et al.(2004)]{Rhoads:2004} Rhoads, J.~E., Xu, C., 
Dawson, S., et al.\ 2004, \apj, 611, 59 

\bibitem[Ricotti et al.(2001)]{Ricotti:2001} Ricotti, M., Gnedin, 
N.~Y., \& Shull, J.~M.\ 2001, \apj, 560, 580 

\bibitem[Ricotti et al.(2002)]{Ricotti:2002} Ricotti, M., Gnedin, 
N.~Y., \& Shull, J.~M.\ 2002, \apj, 575, 49 

\bibitem[Ricotti(2003)]{Ricotti:2003} Ricotti, M.\ 2003, \mnras, 
344, 1237 

\bibitem[Ricotti \& Ostriker(2004)]{Ricotti:2004} 
Ricotti, M., \& Ostriker, J.~P.\ 2004, \mnras, 352, 547 

\bibitem[Ricotti 
\& Gnedin(2005)]{Ricotti:2005} Ricotti, M., \& Gnedin, N.~Y.\ 2005, \apj, 629, 259 

\bibitem[Ricotti et al.(2008)]{Ricotti:2008} Ricotti, M., Gnedin, 
N.~Y., \& Shull, J.~M.\ 2008, \apj, 685, 21 

\bibitem[Ricotti(2010)]{Ricotti:2010} Ricotti, M.\ 2010, Advances in 
Astronomy, 2010, article id. 271592

\bibitem[Robertson et al.(2010)]{Robertson:2010} Robertson, B.~E., 
Ellis, R.~S., Dunlop, J.~S., McLure, R.~J., 
\& Stark, D.~P.\ 2010, \nat, 468, 49 

\bibitem[Romano-D{\'{\i}}az et al.(2011)]{Romano:2011} 
Romano-D{\'{\i}}az, E., Choi, J.-H., Shlosman, I., 
\& Trenti, M.\ 2011, \apjl, 738, L19 

\bibitem[Ro{\v s}kar et al.(2010)]{Roskar:2010} Ro{\v s}kar, R., 
Debattista, V.~P., Brooks, A.~M., et al.\ 2010, \mnras, 408, 783 

\bibitem[Rydberg et al.(2012)]{Rydberg:2012} Rydberg, C.-E., 
Zackrisson, E., Lundqvist, P., \& Scott, P.\ 2012, arXiv:1206.0007 

\bibitem[Salpeter(1955)]{Salpeter:1955} Salpeter, E.~E.\ 1955, \apj, 
121, 161 

\bibitem[Salvadori 
\& Ferrara(2009)]{Salvadori:2009} Salvadori, S., \& Ferrara, A.\ 2009, \mnras, 395, L6 

\bibitem[Safranek-Shrader et al.(2012)]{Chalence:2012} 
Safranek-Shrader, C., Agarwal, M., Federrath, C., et al.\ 2012, 
ApJ, submitted (arXiv:1205.3835)

\bibitem[Santos(2004)]{Santos:2004} Santos, M.~R.\ 2004, \mnras, 
349, 1137 

\bibitem[Sawala et al.(2012)]{Sawala:2012} Sawala, T., Scannapieco, 
C., \& White, S.\ 2012, \mnras, 420, 1714 

\bibitem[Sawicki 
\& Thompson(2006)]{Sawicki:2006} Sawicki, M., \& Thompson, D.\ 2006, \apj, 648, 299 

\bibitem[Scannapieco et al.(2002)]{Scannapieco:2002} Scannapieco, E., 
Ferrara, A., \& Madau, P.\ 2002, \apj, 574, 590 

\bibitem[Schaerer(2003)]{Schaerer:2003} Schaerer, D.\ 2003, \aap, 397, 527 

\bibitem[Schaye \& Dalla Vecchia(2008)]{Schaye:2008} Schaye, J., \& Dalla Vecchia, C.\ 2008, \mnras, 383, 1210 

\bibitem[Schaye et al.(2010)]{Schaye:2010} Schaye, J., et al.\ 
2010, \mnras, 402, 1536 


\bibitem[Seljak 
\& Zaldarriaga(1996)]{Seljak:1996} Seljak, U., \& Zaldarriaga, M.\ 1996, \apj, 469, 437 

\bibitem[Shang et al.(2010)]{Shang:2010} Shang, C., Bryan, G.~L., 
\& Haiman, Z.\ 2010, \mnras, 402, 1249 

\bibitem[Shapiro 
\& Kang(1987)]{Shapiro:1987} Shapiro, P.~R., \& Kang, H.\ 1987, \apj, 318, 32 

\bibitem[Shapiro et al.(1994)]{Shapiro:1994} Shapiro, P.~R., Giroux, 
M.~L., \& Babul, A.\ 1994, \apj, 427, 25 

\bibitem[Simcoe et al.(2012)]{Simcoe:2012} Simcoe, R.~A., Sullivan, 
P.~W., Cooksey, K.~L., et al.\ 2012, \nat, 492, 79 

\bibitem[Springel et al.(2001)]{Springel:2001a} Springel, V., Yoshida, 
N., \& White, S.~D.~M.\ 2001a, New Astronomy, 6, 79

\bibitem[Springel et al.(2001b)]{Springel:2001b} Springel, V., White, 
S.~D.~M., Tormen, G., \& Kauffmann, G.\ 2001, \mnras, 328, 726 


\bibitem[Springel(2005)]{Springel:2005} Springel, V.\ 2005, \mnras, 
364, 1105 

\bibitem[Stacy et al.(2011)]{Stacy:2011} Stacy, A., Bromm, V., 
\& Loeb, A.\ 2011, \apjl, 730, L1 

\bibitem[Stiavelli et al.(2004)]{Stiavelli:2004} Stiavelli, M., Fall, 
S.~M., \& Panagia, N.\ 2004, \apjl, 610, L1 

\bibitem[Stinson et al.(2007)]{Stinson:2007} Stinson, G.~S., 
Dalcanton, J.~J., Quinn, T., Kaufmann, T., 
\& Wadsley, J.\ 2007, \apj, 667, 170 

\bibitem[Strigari et al.(2008)]{Strigari:2008} Strigari, L.~E., 
Bullock, J.~S., Kaplinghat, M., et al.\ 2008, \nat, 454, 1096 

\bibitem[Susa(2007)]{Susa:2007} Susa, H.\ 2007, \apj, 659, 908 

\bibitem[Tegmark et al.(1997)]{Tegmark:1997} Tegmark, M., Silk, J., 
Rees, M.~J., et al.\ 1997, \apj, 474, 1 

\bibitem[Thoul 
\& Weinberg(1996)]{Thoul:1996} Thoul, A.~A., \& Weinberg, D.~H.\ 1996, \apj, 465, 608 

\bibitem[Toomre(1964)]{Toomre:1964} Toomre, A.\ 1964, \apj, 139, 
1217 

\bibitem[Tornatore et al.(2007)]{Tornatore:2007} Tornatore, L., 
Ferrara, A., \& Schneider, R.\ 2007, \mnras, 382, 945 

\bibitem[Trenti et al.(2009)]{Trenti:2009} Trenti, M., Stiavelli, 
M., \& Michael Shull, J.\ 2009, \apj, 700, 1672 

\bibitem[Tseliakhovich et al.(2011)]{Tseliakhovich:2011} Tseliakhovich, 
D., Barkana, R., \& Hirata, C.~M.\ 2011, \mnras, 418, 906 

\bibitem[Tumlinson et al.(2001)]{Tumlinson:2001} Tumlinson, J., 
Giroux, M.~L., \& Shull, J.~M.\ 2001, \apjl, 550, L1 

\bibitem[Vera-Ciro et al.(2012)]{Vera-Ciro:2012} Vera-Ciro, C.~A., 
Helmi, A., Starkenburg, E., \& Breddels, M.~A.\ 2012, arXiv:1202.6061 

\bibitem[Verhamme et al.(2008)]{Verhamme:2008} Verhamme, A., Schaerer, D., Atek, H., \& Tapken, C.\ 2008, \aap, 491, 89 

\bibitem[Verner et al.(1996)]{Verner:1996} Verner, D.~A., Ferland, 
G.~J., Korista, K.~T., \& Yakovlev, D.~G.\ 1996, \apj, 465, 487 

\bibitem[Wang et al.(2012)]{Wang:2012} Wang, J., Frenk, C.~S., 
Navarro, J.~F., Gao, L., \& Sawala, T.\ 2012, arXiv:1203.4097 

\bibitem[White(1996)]{White:1996} White, S.~D.~M.\ 1996, Cosmology 
and Large Scale Structure, 349 

\bibitem[Wiersma et al.(2009)]{Wiersma:2009} Wiersma, R.~P.~C., 
Schaye, J., \& Smith, B.~D.\ 2009, \mnras, 393, 99 

\bibitem[Wise \& Abel(2005)]{Wise:2005} Wise, J.~H., \& Abel, T.\ 2005, \apj, 629, 615 

\bibitem[Wise 
\& Abel(2007a)]{Wise:2007a} Wise, J.~H., \& Abel, T.\ 2007a, \apj, 671, 1559 

\bibitem[Wise 
\& Abel(2007b)]{Wise:2007b} Wise, J.~H., \& Abel, T.\ 2007b, \apj, 665, 899

\bibitem[Wise 
\& Abel(2008a)]{Wise:2008a} Wise, J.~H., \& Abel, T.\ 2008a, \apj, 684, 1 

\bibitem[Wise 
\& Abel(2008b)]{Wise:2008b} Wise, J.~H., \& Abel, T.\ 2008b, \apj, 685, 40 

\bibitem[Wise et al.(2008)]{Wise:2008c} Wise, J.~H., Turk, M.~J., 
\& Abel, T.\ 2008, \apj, 682, 745 

\bibitem[Wise 
\& Cen(2009)]{Wise:2009} Wise, J.~H., \& Cen, R.\ 2009, \apj, 693, 984 

\bibitem[Wise et al.(2012)]{Wise:2012} Wise, J.~H., Turk, M.~J., 
Norman, M.~L., \& Abel, T.\ 2012, \apj, 745, 50 

\bibitem[Wolcott-Green 
\& Haiman(2011)]{Wolcott:2011} Wolcott-Green, J., \& Haiman, Z.\ 2011, \mnras, 412, 2603 

\bibitem[Wolcott-Green et al.(2011b)]{Wolcott:2011b} Wolcott-Green, 
J., Haiman, Z., \& Bryan, G.~L.\ 2011, arXiv:1106.3523 

\bibitem[Yajima et al.(2009)]{Yajima:2009} Yajima, H., Umemura, M., 
Mori, M., \& Nakamoto, T.\ 2009, \mnras, 398, 715 

\bibitem[Yajima et al.(2011)]{Yajima:2011} Yajima, H., Choi, J.-H., 
\& Nagamine, K.\ 2011, \mnras, 412, 411 

\bibitem[Yoshida et al.(2003)]{Yoshida:2003} Yoshida, N., Abel, T., 
Hernquist, L., \& Sugiyama, N.\ 2003, \apj, 592, 645

\bibitem[Zackrisson et al.(2011)]{Zackrisson:2011} Zackrisson, E., 
Rydberg, C.-E., Schaerer, D., {\"O}stlin, G., 
\& Tuli, M.\ 2011, \apj, 740, 13 

\bibitem[Zackrisson et al.(2012)]{Zackrisson:2012} Zackrisson, E., 
Zitrin, A., Trenti, M., et al.\ 2012, arXiv:1204.0517 

\bibitem[Zeldovich(1970)]{Zeldovich:1970} Zeldovich, Y.~B.\ 1970, \aap, 5, 84 

\bibitem[Zemp et al.(2012)]{Zemp:2012} Zemp, M., Gnedin, O.~Y., 
Gnedin, N.~Y., \& Kravtsov, A.~V.\ 2012, \apj, 748, 54 



\end{thebibliography}
\end{document}